\documentclass[fleqn,letterpaper,11pt]{article}
\usepackage{color,epsfig}
\usepackage[bookmarks,dvips,pdfhighlight=/O,pdfborder={0 0
  0},pdfstartview=FitH]{hyperref}

\usepackage{amsmath,amsfonts,amssymb,amsopn}

\newcommand{\dd}{\text{d}}
\DeclareMathOperator{\im}{Im}

\DeclareMathOperator{\tr}{tr}
\newcommand{\ii}{\text{i}}

\newcommand{\beq}{\begin{equation}}
\newcommand{\eeq}{\end{equation}}
\newcommand{\bce}{\begin{center}}
\newcommand{\ece}{\end{center}}

\newcommand{\lsim}{\lesssim}
\newcommand{\gsim}{\gtrsim}

\newcommand{\tave}[1]{\langle\!\langle{#1}\rangle\!\rangle}

\title{The Chiral Restoration Transition of QCD and Low Mass Dileptons}

\author{R.~Rapp$^{1}$, J. Wambach$^{2}$ and H. van Hees$^{3}$\\
  $^{1}$Cyclotron Institute and Physics Department, Texas A\&M University,\\
  College Station, TX 77843-3366, U.S.A. \\
  $^{2}$ TU Darmstadt, Schlo{\ss}gartentr. 9, D-64289 Darmstadt, Germany\\
  $^{3}$ Institut f{\"u}r Theoretische Physik,
  Justus-Liebig-Universit{\"a}t Giessen, \\
  Heinrich-Buff-Ring 16, D-35392 Giessen, Germany}
\date{\today}

\begin{document}

\maketitle

\vspace{1.0in}

\begin{abstract}
Recent developments in the evaluation of vector-meson spectral functions 
in hot and dense matter are discussed with emphasis on connections to 
the chiral phase transition in QCD.  Model independent approaches 
including chiral low-density expansions, lattice QCD, chiral and QCD 
sum rules are put into context with model predictions for in-medium 
vector-spectral function utilizing effective Lagrangians. Hadronic 
many-body calculations predict a strong broadening (and little mass 
shift) of the $\rho$ spectral function which rapidly increases close to 
the expected phase boundary of hadronic and quark-gluon matter. Pertinent 
dilepton rates appear to degenerate with perturbative quark-antiquark 
annihilation in the Quark-Gluon Plasma, suggestive for chiral symmetry 
restoration. Applications to low-mass dilepton spectra in heavy-ion 
collisions result in quantitative agreement with recent high-quality 
data at the CERN-SPS. Thermal radiation from temperatures around $T_c$ 
consistently reproduces the experimental dilepton excess observed
at masses above 1~GeV as well. The interpretation of dilepton sources 
at high transverse momentum appears to be more involved.   
\end{abstract}

\newpage

\tableofcontents

\newpage

\section{Introduction}
The quest for the elementary entities of matter has always been a
central objective in physics. Of no less interest is the emergence of
the structure (or phases) of matter built from its basic constituents
and their interactions.  The exploration of matter governed by the
strong force is at the forefront of contemporary research in nuclear
physics.  The most common form of strongly interacting matter in the
present-day universe is contained in atomic nuclei, which are bound
states of nucleons, i.e., protons and neutrons.  Nuclear matter as found
in the center of heavy nuclei is characterized by a nucleon (energy)
density of about $\varrho_0=0.16$\,fm$^{-3}$
($\varepsilon_0=0.15$\,GeV/fm$^3$), rendering one table spoon of this
material a mass of about one million kilotons (10$^{12}$\,kg). The
binding of nucleons is strong enough to cause a reduction of nuclear
masses by about 1\% compared to the sum of the rest mass of the
individual nucleons, $m_N\simeq940$\,MeV/$c^2$. But how does the mass
of a nucleon arise?  In the late 1960's it was discovered that the
nucleon itself is a composite object, built of three ``valence'' quarks
of {\it up} ($u$) and {\em down} ($d$) ``flavor''. The bare masses of
$u$ and $d$ quarks are only about 5-10\,MeV/$c^2$, and believed to be
generated by a condensate of (yet to be discovered) Higgs bosons in the
electroweak (EW) sector of the Standard Model of Elementary
Particles. That is, about 98\% of the nucleon's mass is generated 
dynamically by the strong interaction.  Moreover, no individual quarks
have been observed in nature thus far: they are ``confined'' into
hadrons, either baryons or mesons (conglomerates of three valence quarks
or of a quark and antiquark, respectively).  In the 1970's, the quantum
field theory underlying the strong force has been developed, Quantum
Chromodynamics (QCD), based on quarks and gluons as fundamental degrees
of freedom. This theory has been quantitatively confirmed in high-energy
scattering experiments, where the strong coupling constant,
$\alpha_s\simeq0.1$, is relatively small and perturbation theory can be
reliably utilized to obtain quantitative results for observables.
However, at low momentum transfers, $\alpha_s$ becomes large, perturbation 
theory ceases to be applicable and nonperturbative mechanisms take
over. It is in this regime where quark confinement and mass generation
occur, posing formidable challenges for their theoretical 
understanding~\cite{Shuryak:2004}.

\subsection{QCD Vacuum and Chiral Restoration}
It turns out that confinement and mass generation are intimately
connected with the phase structure of strongly interacting matter (see,
e.g., Ref.~\cite{BraunMunzinger:2008tz}). In
fact, even the structure of the QCD vacuum is far from trivial: similar
to the EW sector, it is believed to be filled with condensates, which
are closely related to the origin of hadronic masses.  There are,
however, important differences: the QCD condensates are made of (scalar)
composites of quarks and gluons (rather than elementary fields like the
Higgs boson), and they do not induce a breaking of the gauge symmetry.
In what follows, the so-called ``chiral'' quark-antiquark condensate,
$\langle 0|\bar qq|0\rangle \simeq (-250\,{\rm MeV})^3$, will be of
particular importance. It breaks the (approximate) chiral symmetry of
QCD, which corresponds to the conservation of left and right
``handedness'' of massless quarks (applicable for the light $u$ and $d$
quarks, whose masses are parametrically small, $m_{u,d} \ll |\langle
0|\bar qq|0\rangle|^{1/3}$).  While the quark condensate cannot be
directly observed, its consequences are apparent in the excitations of
the condensed ground state, i.e., in the hadron spectrum. Since chiral
symmetry is a global symmetry (rather than a local one depending on
space-time position), its spontaneous breaking must be accompanied by 
(almost) massless Goldstone bosons. For two quark flavors the latter 
are identified with the three charge states of the pion,
whose mass, $m_\pi\simeq140$\,MeV, is ``abnormally'' small compared to
that of all the other hadrons (e.g., $m_N\simeq940$\,MeV). 
The observed hadron spectrum encodes further
evidences for the spontaneous breaking of chiral symmetry (SB$\chi$S):
chiral multiplets (e.g., $\rho$(770)-$a_1$(1260) or
$N$(940)-$N^*(1535)$), which would be degenerate if the ground state
were chirally symmetric, exhibit a large mass splitting of typically
$\Delta M\simeq500$\,MeV.  The effects of SB$\chi$S seem to (gradually)
cease as one goes up in mass in the hadronic 
spectrum~\cite{Glozman:2007ek}.  This is one of
the indications that SB$\chi$S is a low-energy, strong-coupling
phenomenon which is no longer operative at high momentum transfers where
perturbation theory becomes applicable.

When heating the QCD vacuum its condensate structure is expected to 
change. Loosely speaking, thermally excited hadrons ``evaporate" 
condensed $\bar qq$ pairs which eventually leads to the restoration of 
the spontaneously broken 
chiral symmetry. Numerical computations of the lattice-discretized path
integral for QCD at finite temperature predict chiral symmetry
restoration ($\chi$SR) to occur at a (pseudo-) critical temperature of 
$T_c\simeq$~160-190\,MeV~\cite{Cheng:2006qk,Aoki:2006br}, corresponding 
to an energy density of about $\varepsilon_c\simeq 1$\,GeV/fm$^3$. The 
chiral transition is characterized by a rapid decrease of the $\bar qq$ 
condensate, which, in fact, serves as an order parameter of strongly 
interacting matter. In the limit of vanishing light quark masses and 
for three quark flavors, this transition is of first order, while for 
realistic quark masses as realized in nature (two light quarks $u$ and 
$d$ and a more heavy strange quark, $m_s\simeq 120$~MeV), it is more likely 
a rapid cross-over. Key manifestations of chiral symmetry restoration are 
its (observable) consequences for the hadron spectrum. Chiral partners 
must degenerate
implying massive medium modifications of hadronic spectral functions as
the transition is approached.  This notion is a quite general concept
found, e.g., in solid state physics where phase transitions are 
routinely diagnosed utilizing ``soft-mode spectroscopy".
This applies in particular to a second
order phase transition where the mode associated with an order parameter
becomes massless (soft). But even for bulk matter properties, rapid 
changes in the thermodynamic state variables are directly related to 
changes in the relevant degrees of freedom at the typical thermal scale 
(temperature or Fermi momentum). Interestingly, the chiral transition 
is accompanied by the
dissolution of hadrons into quarks, i.e., the deconfinement transition,
at the same temperature (at least for vanishing net baryon density). 
The reason for the apparent coincidence of the two transitions is not 
understood. The deconfined and chirally restored strongly interacting 
matter is commonly referred to as the Quark-Gluon Plasma (QGP). 
The experimental verification and theoretical understanding of
the mechanisms leading to the QGP are central objectives in modern
nuclear research.

\subsection{Ultrarelativistic Heavy-Ion Collisions}
\label{ssec_urhics}
The only way to produce and study hot and dense strongly interacting
matter in the laboratory is by colliding atomic nuclei at high
energies. Several large-scale experiments at ultrarelativistic
bombarding energies, $E_{\rm lab} \gg m_N$, have been conducted over 
the past $\sim$20 years, most recently at the SPS at CERN
(at center-of-mass energies up to $\sqrt{s}=17.3$~AGeV) and at the
Relativistic Heavy-Ion Collider (RHIC) at Brookhaven (up to
$\sqrt{s}=200$~AGeV)~\cite{Heinz:2000bk,Arsene:2004fa} with the heaviest
available nuclei at A$\simeq$200 (Pb and Au). 

The first question that needs to be answered is whether these reactions 
produce {\em equilibrated matter}, i.e., do the produced particles 
undergo sufficient rescattering to justify the notion of an interacting 
medium characterized by bulk thermodynamic variables?
Extensive and systematic measurements of hadronic observables have lead
to a positive answer. This is extremely exciting as it puts
within grasp the possibility to recreate, at least for a short moment,
the matter which the early universe was made of just a few microseconds
after the Big Bang!  While hadronic measurements are discussed and
interpreted in depth in other contributions of this volume, let us
sketch some of their main features: transverse-momentum ($p_T$) spectra
of different hadron species (pions, kaons, protons, etc.), which
characterize the hadronic fireball just before break-up at its ``thermal
freeze-out'', exhibit a collective explosion reaching an average speed
of about half the speed of light at a final temperature of 
$T^{\rm th}_{\rm fo}\simeq 100$\,MeV. The ratios of the observed hadron 
species point at a significantly higher temperature of 
$T^{\rm ch}_{\rm fo} \simeq 160$\,MeV~\cite{BraunMunzinger:2003zd,
Becattini:2005xt}, i.e., the
chemistry of the fireball (driven by inelastic scattering processes)
appears to freeze out significantly earlier than kinetic equilibrium
(maintained by elastic interactions). This is consistent with the large
difference of empirical elastic (e.g., $\pi\pi\to\rho \to\pi\pi$ or $\pi
N\to \Delta\to\pi N$) and inelastic (e.g., $\pi\pi\to K\bar K$) hadronic
cross sections, with typical values of $\sim$100\,mb vs.  $\sim$1\,mb,
respectively. Since the cross sections determine the relaxation times
according to $\tau \simeq (\varrho_h \sigma v_{\rm rel})^{-1}$
($\varrho_h$: hadron density, $v_{\rm rel}$: relative velocity of the
colliding hadrons), one obtains a clear hierarchy in the underlying
relaxation times, $\tau_{\rm th} \ll \tau_{\rm ch}$.  The interacting
hadronic phase between chemical and thermal freeze-out will play an
important role in the remainder of this article.  More differential
analyses of the flow patterns of the measured hadrons allow to trace
back the matter properties to earlier times in the evolution of the
fireball.  In particular, the magnitude of the ``elliptic flow''
measured at RHIC indicates that the medium thermalizes on a rather short
time scale, $\tau\le$~1-2~fm/$c$ after initial impact, translating into
(energy-) densities of a factor 10 or more above the critical
one\footnote{Elliptic flow characterizes the azimuthal asymmetry in the
  $p_T$-spectra of particles (in the plane transverse to the beam
  axis). In a non-central heavy-ion collision, the initial nuclear
  overlap (interaction) zone is ``almond-shaped''. If the system
  thermalizes before this spatial anisotropy is smeared out (e.g., due
  to free streaming), a larger pressure gradient builds up along the
  ``short'" compared to the ``long" axis of the initial almond.  This
  thermal pressure drives a collective expansion of the ``almond'' which
  is stronger along the short axis and thus results in particle momenta
  with a preference to be aligned with this axis. The magnitude of the
  elliptic flow is thus sensitive to how fast thermalization is
  established.}.  A thermal (hydrodynamic)
description~\cite{Teaney:2001av,Hirano:2002ds,Kolb:2003dz,Nonaka:2006yn}
of the fireball in semi-/central collisions of heavy nuclei at RHIC
appears to be valid for hadrons up to momenta of $p_T\simeq$~2-3~GeV,
comprising approximately 95\% of all produced particles. At high
transverse momenta, $p_T>5$~GeV, hadron production is dominated by hard
scattering, i.e., a primordial parton-parton collision at high momentum
transfer within the incoming nucleons, followed by fragmentation into (a
spray of) hadrons (jets). In central Au-Au collisions at RHIC, a factor 
of $\sim$5 suppression of high-$p_T$ hadron production has been
observed (``jet quenching'')\footnote{Jet quenching is probably also
  present at the SPS but it is quantitatively smaller than at RHIC
  (about a factor of 2 suppression) and masked by a large initial $p_T$
  broadening in the interpenetrating nuclei prior to the hard
  scattering, known as ``Cronin effect''.}.  While these hadrons (or
their parent quarks) do not thermalize, their suppression indicates a
substantial coupling to the created medium, associated with an energy
loss of a fast parton propagating through the fireball.  The (energy-)
density of the medium required to account for this effect is roughly
consistent with the estimate inferred from a hydrodynamic description of
the elliptic flow of low-$p_T$ hadrons.

A second level of questions concerns the relevant degrees of freedom of
the produced matter, i.e., whether there is explicit evidence that
individual partons leave a distinctive footprint in observables. It
turns out that the elliptic flow is again revealing interesting features
in this context: it has been found~\cite{Adare:2006ti,Abelev:2007qg}
that the elliptic-flow coefficient, $v_2^h(K_T)$, of all measured
hadrons, $h$=$\pi$, $K$, $p$, $\Lambda$, $\Sigma$, $\phi$, ...,
exhibits a remarkable universality as a function of transverse kinetic 
energy, $K_T$=$m_T-m_h$ ($m_T$=($p_T^2$+$m_h^2$)$^{1/2}$): when scaled with
the constituent-quark number content, $n_q$, of hadron $h$, all measured
hadron-$v_2$ data appear to collapse on a single curve, 
$v_2^q$($K_t$$\equiv$$K_T/n_q$)=$v_2^h(K_T) / n_q$. 
This has been interpreted as evidence for
a collectively expanding partonic source hadronizing via quark
coalescence. A fully consistent theoretical description of this 
phenomenon has not been achieved yet.

A third level of investigations has to address signals of the
deconfinement and/or chiral restoration transitions. In a rigorous
sense, this requires the assessment of order parameters associated with
these transitions. However, changes in order parameters are not always 
easily observable. This is particularly true in the present context and 
we are led back to the idea of ``mode spectroscopy'',
to be conducted in the environment of a short-lived, rapidly expanding
fireball of a heavy-ion collision. Individual (stable) hadrons emanating
from the collision zone have all long recovered their free (vacuum)
masses by the time they are measured in the detectors. A better
observable are invariant-mass spectra of short-lived resonance decays,
$h\to h_1 h_2$, with a lifetime, $\tau_h$, comparable to, or smaller
than, the lifetime of the interacting fireball, $\tau_{\rm FB}\simeq
10$~fm/$c$.  Such a resonance (e.g., $\Delta\to \pi N$ or
$\rho\to\pi\pi$) has a large probability to decay inside the medium so
that its decay products can carry the information on its invariant-mass,
$m_h^2=(p_1+p_2)^2$, at the point of decay to the detector. In
principle, this would allow to determine the invariant-mass distribution
(or spectral function) of the resonance $h$ in the medium. The problem
is that the decays products, $h_1$ and $h_2$, are likely to undergo
further rescattering in the fireball which destroys the desired
invariant-mass information.  The latter will thus be largely restricted
to the dilute (break-up) stages of the medium in a heavy-ion collision.

\subsection{Dilepton Spectroscopy}
\begin{figure}[!tb]
\begin{minipage}{0.7\linewidth}
\epsfig{file=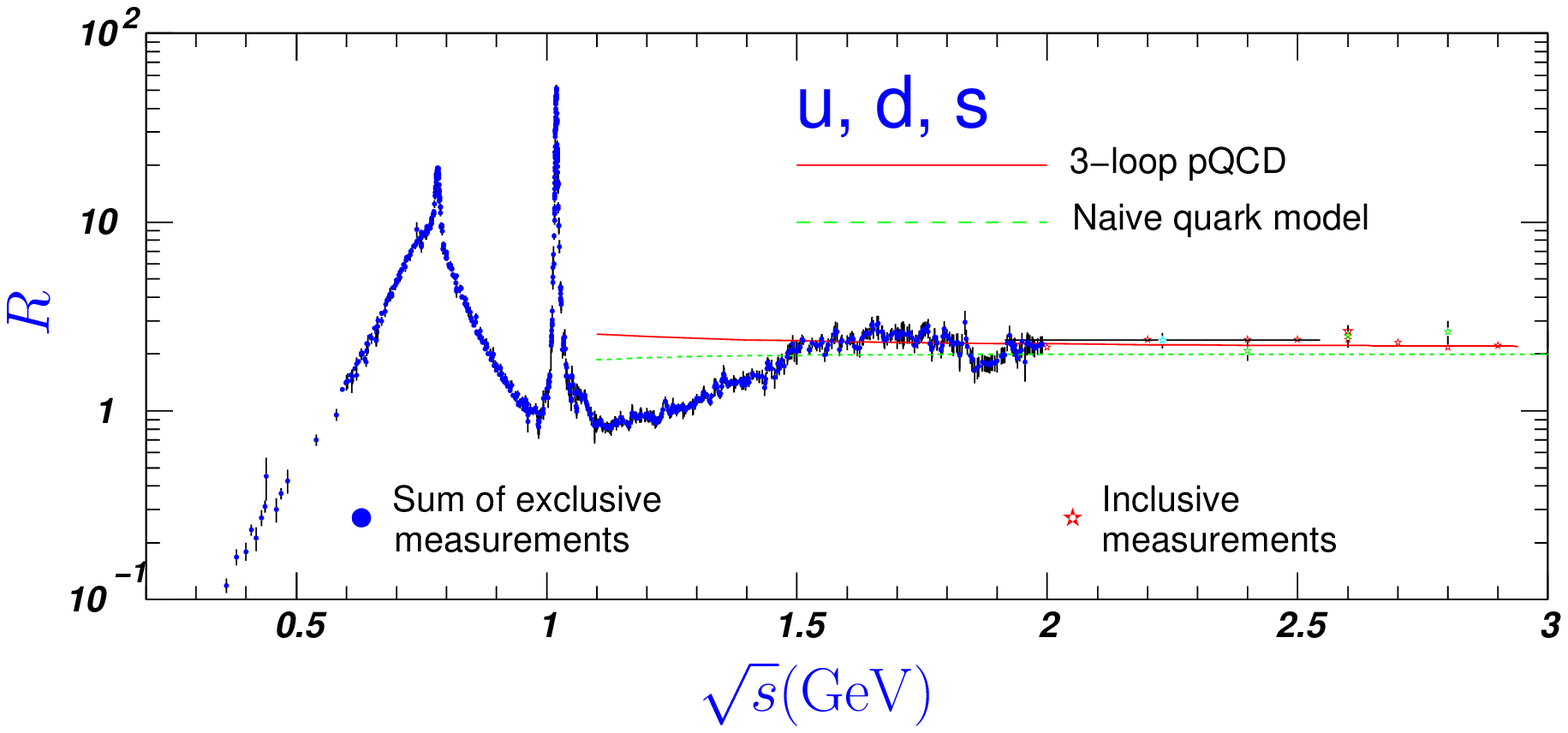,width=0.9\textwidth,height=0.5\textwidth}
\end{minipage}
\hspace{-0.8cm}
\begin{minipage}{0.3\linewidth}
\vspace{0.0cm}
\epsfig{file=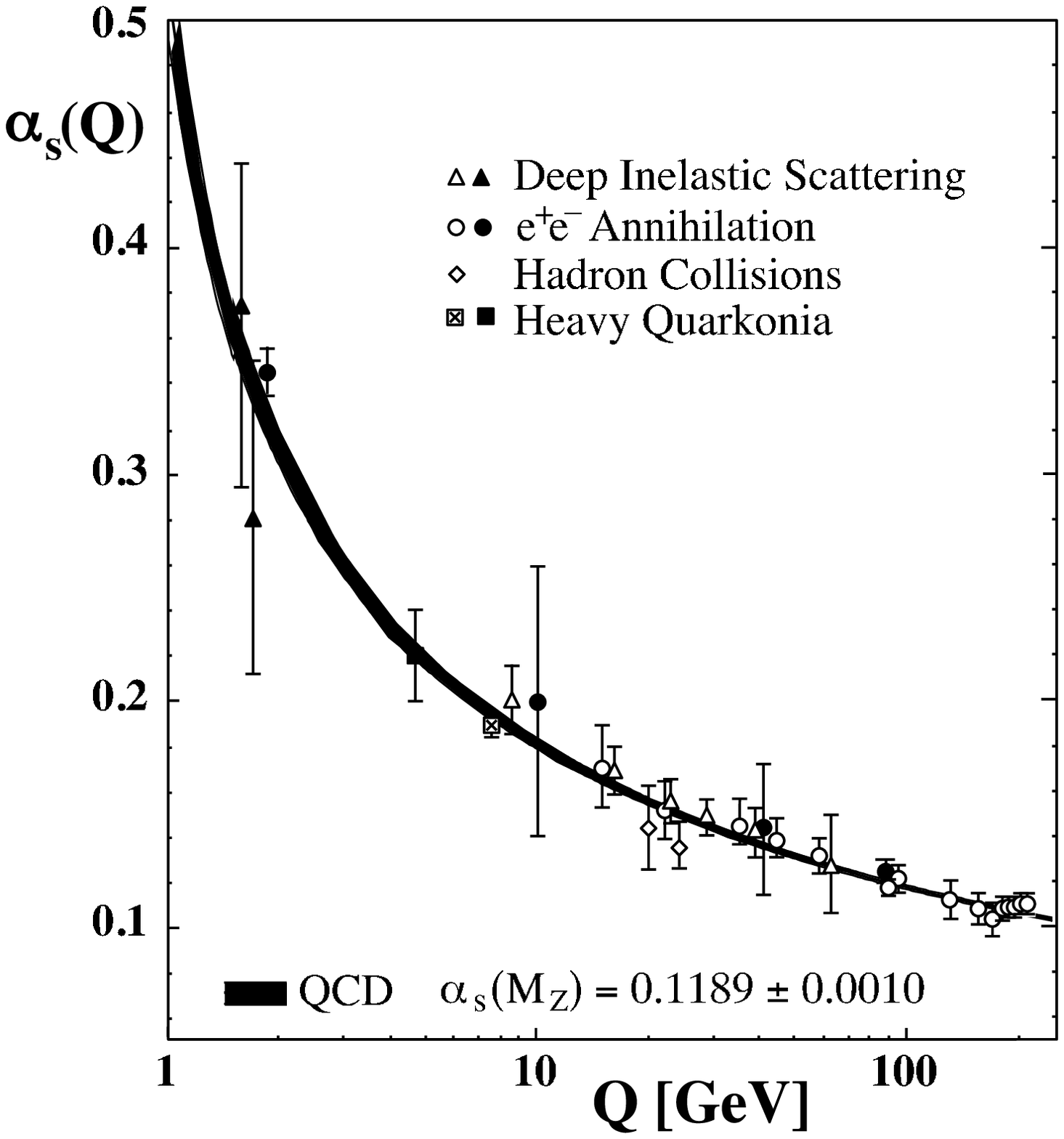,width=1.0\textwidth}
\end{minipage}
\caption{\it Left panel: ratio of cross sections, $R=\sigma_{ee\to{\rm
      hadrons}}/\sigma_{ee\to\mu\mu}$, for electron-positron
  annihilation into hadrons relative to muon-antimuons, as a function of
  center-of-mass energy, $\sqrt{s}$. The experimental data exhibit 
  a nonperturbative resonance regime up to $\sqrt{s}$$\simeq$\,1.1~GeV, 
  exhausted by the light vector mesons $\rho$, $\omega$ and $\phi$, 
  followed by a transition to an almost structureless perturbative regime 
  at $\sqrt{s}$$\gsim$1.5\,GeV. The latter is well described by 
  perturbative QCD (pQCD), especially for $\sqrt{s}$$\ge$2~GeV, where 
  residual ``oscillations'' (due to excited vector resonances) 
  have essentially ceased. The naive quark-model prediction 
  (leading-order pQCD, ${\cal O}(\alpha_s^0)$) 
  is given by $R_{QM}$=$N_c \sum_{q=u,d,s} e_q^2$=2. 
  Right panel: strong coupling constant as a function of momentum 
  transfer~\cite{Bethke:2006ac}; note the increase of
  $\alpha_s$ toward small $Q$ suggestive for the emergence of
  nonperturbative phenomena.}
\label{fig_Pi-em-vac}
\end{figure}

The decisive step to obtain access to hadronic spectral functions in the
hot and dense regions of the medium is provided by electromagnetic (EM)
probes, i.e., photons ($\gamma$) and dileptons (arising from virtual
(timelike) photons, $\gamma^*\to l^+l^-$ with $l$=$e$ or
$\mu$)~\cite{Feinberg:1976ua,Shuryak:1978ij,McLerran:1984ay}. These are
not subject to the strong force and thus suffer negligible final-state
interactions, with a mean free path which is much larger than the typical
size of the fireball, $R_{\rm FB}\simeq 10$~fm.  The natural candidates
for in-medium spectroscopy are the vector mesons ($V$), which carry
the quantum numbers of the photon (spin-parity $J^P$=$1^-$) and thus
directly couple to exclusive dilepton final states, $V\to l^+l^-$. In the
low-mass region ($M$$\le$1\,GeV), which is the region of interest to
study chiral restoration, the prominent vector mesons are $\rho$(770),
$\omega$(782) and $\phi$(1020). In fact, the famous vector dominance
model (VDM)~\cite{Sakurai:1969} asserts that the coupling of a (real or
virtual) photon to {\em any} EM hadronic current exclusively proceeds
via an intermediate vector meson (which is excellently satisfied in the
mesonic sector but subject to corrections in the baryonic sector). Thus,
if VDM holds in hadronic matter, dilepton emission is indeed equivalent
to in-medium vector-meson spectroscopy.  In thermal equilibrium, the
contribution from the (isovector) $\rho$ meson dominates over the
$\omega$ by a factor of $\sim$10 (factor $\sim$5 over the $\phi$,
which, however, is thermally suppressed due to its larger
mass). Furthermore, it can be shown~\cite{Rapp:2004zh} that, in the
context of a heavy-ion collision, low-mass dilepton radiation from the 
hadronic phase dominates over the emission from a putative QGP phase, 
even at collider energies (RHIC and LHC)\footnote{This is due to the much 
  larger three-volume in the hadronic phase; at larger dilepton masses,
  $M$$>$1~GeV, the thermal Boltzmann factor, ${\rm e}^{-q_0/T}$,
  augments the sensitivity to higher temperatures which increases the
  QGP contribution relative to the hadronic one in the inclusive
  dilepton spectrum.}. The excitement (and theoretical activity) in the
field was further spurred by the suggestion of Brown and
Rho~\cite{Brown:1991kk} that the $\rho$-meson mass should drop to
(almost) zero as a consequence of $\chi$SR. Early dilepton measurements
in S(200\,AGeV)-Au collisions at the CERN-SPS by the CERES
collaboration~\cite{Agakishiev:1995xb} found a large enhancement of the
spectrum at invariant masses below the nominal $\rho$ mass, i.e., for
$M$$\simeq$0.2-0.7\,GeV. These data could be well described by a
dropping-mass scenario implemented into relativistic transport models
within a mean-field description~\cite{Li:1995qm,Cassing:1994}. 
Subsequently, more ``conventional'' medium
modifications of the $\rho$ meson were investigated based on its
rescattering on constituents of a hadronic medium, see, e.g.,  
Refs.~\cite{Rapp:1999ej,Alam:1999sc,Bratkovskaya:1999xx,Gale:2003iz} for
reviews. The generic finding of these hadronic many-body calculations
was a strong broadening of the $\rho$ spectral function, which, when
extrapolated to the putative phase transition temperature,
$T_c\simeq175$~MeV, leads to a complete ``melting" of the resonance
structure~\cite{Rapp:1997fs}. The broadening effect could account for a
large part of the low-mass dilepton excess observed in S-Au collisions.
The agreement was even better~\cite{Rapp:1999us} with improved
CERES/NA45 measurements carried out in the heavier Pb(158\,AGeV)-Au
system~\cite{Agakishiev:1997au,Agakichiev:2005ai}.  The connection of
the $\rho$ melting to $\chi$SR appeared to be less direct than in
dropping-mass scenarios. However, in
Refs.~\cite{Rapp:1999if,Rapp:1999us} it was found that the hadronic
dilepton rates following from the ``melted'' $\rho$ close to $T_c$
rather closely resemble the rates computed in a partonic description,
i.e., perturbative quark-antiquark annihilation. In the vacuum, such a
phenomenon is well known from the $e^+e^-$ annihilation into hadrons:
for $M$$\ge$1.5\,GeV the total cross section is well described within
perturbative QCD using quark-antiquark final states, known as
``parton-hadron duality'', cf.~Fig.~\ref{fig_Pi-em-vac}.
It was therefore suggested that the conceptual implication of the $\rho$
melting is a reduction of the ``duality 
threshold"~\cite{Rapp:1999if,Rapp:1999us}, from $M_{\rm dual}$=1.5\,GeV 
in the vacuum to essentially zero around $T_c$. Note
that a ``perturbative'' dilepton rate automatically implies chiral
restoration (i.e., degeneracy of vector and axialvector channels).

The accuracy in the 1995/1996 CERES/NA45 dielectron
data~\cite{Agakishiev:1997au,Agakichiev:2005ai} did not allow for a
decisive experimental discrimination of the dropping-mass and
melting-resonance scenarios. An important step forward was realized with
the NA60 dimuon spectra~\cite{Arnaldi:2006jq} in In(158\,AGeV)-In
collisions at the SPS. Excellent mass resolution and superior statistics
enabled, for the first time, an isolation of the ``excess radiation''
(by subtraction of final-state hadron decays). The shape of the
excess spectrum clearly favors a broadened $\rho$ spectral function over
scenarios involving dropping masses. The original predictions of
hadronic many-body theory~\cite{Rapp:1999us,Rapp:2004zh} are, in fact,
in quantitative agreement~\cite{vanHees:2006ng} with the inclusive mass
spectra in semi-/central In-In collisions.  In the last round of CERES/NA45
data~\cite{Adamova:2006nu} excess spectra have also been extracted 
in Pb-Au collisions (by subtraction of final-state hadron decays using
a statistical model~\cite{BraunMunzinger:2003zd}). 
While the overall data quality does not
reach the level of NA60, the larger collision system and the access to
very small dilepton masses in the dielectron channel (dimuons have a
threshold of 2$m_\mu$=210\,MeV) can provide additional insights.

The dilepton program at the CERN-SPS has thus far reached the highest
level of maturity in the heavy-ion context. It also included a CERES/NA45
measurement in a low-energy Pb(40\,AGeV)-Au run~\cite{Adamova:2003kf},
which produced tantalizing hints for an even larger excess than at
158\,AGeV, but was unfortunately hampered by low statistics.  At much
lower, relativistic bombarding energies (1-2\,AGeV), the DLS
collaboration at the BEVALAC reported a very large dilepton
excess~\cite{Porter:1997rc}, which has recently been confirmed by the
HADES collaboration at SIS~\cite{Agakichiev:2006tg,Agakishiev:2007ts}.  
On the other hand, the dilepton measurements at RHIC are still in 
their infancy (first data indicate substantial excess
radiation~\cite{Afanasiev:2007xw}), but it will become a central
component in future runs~\cite{David:2006sr}. Very interesting results
are also emerging from vector-meson spectroscopy in cold nuclei using
elementary projectiles, i.e., photons~\cite{Trnka:2005ey,Clas:2007mga}
or protons~\cite{Naruki:2005kd}.  It turns out that all of these
observables are closely related, and their broad understanding 
is essential for the determination of the in-medium vector-meson spectral
functions. Of particular importance is the consistency of theoretical 
descriptions
beyond phenomenological applications and the interrelations between
different approaches (including effective hadronic and quark models,
lattice QCD and constraints from sum rules), which will ultimately 
reveal the mechanisms of chiral restoration. In this article, we give
an up-to-date account of these efforts with special emphasis on a
broader picture in the context of $\chi$SR.

\subsection{Outline}
Our article is organized as follows. In Sec.~\ref{sec_chi-sym}, we start 
by recollecting basic features of spontaneous chiral symmetry breaking 
in the QCD vacuum with emphasis on condensate structures and consequences 
for the hadronic excitation spectrum (sub-Sec.~\ref{ssec_cond-hs}),
followed by a discussion of in-medium condensates within the 
landscape of the QCD phase diagram (sub-Sec.~\ref{ssec_phase-dia}).
In Sec.~\ref{sec_vec-mes}, we scrutinize the links of the chain with 
which one hopes to connect thermal dilepton rates and (partial) $\chi$SR.
We first introduce the EM correlation function which is the basic quantity 
figuring into the thermal dilepton rate (sub-Sec.~\ref{ssec_em-corr}).
Model-independent evaluations of medium
effects can be obtained in the low-density limit from current algebra,
in the high-temperature limit from perturbative QCD and, for vanishing
baryon-chemical potential from lattice QCD
(sub-Sec.~\ref{ssec_mod-indep}). A valuable source of model-independent
constraints is provided by chiral and QCD sum rules
(sub-Sec.~\ref{ssec_sum-rules}) which are energy moments of
spectral functions that directly relate to order parameters of QCD
and are generally not restricted in temperature and density. 
For practical applications, effective hadronic models are an indispensable 
tool (sub-Sec.~\ref{ssec_eff-mod}); their reliability, based on the choice 
of interaction vertices and associated parameters, crucially hinges on a
thorough procedure of theoretical and phenomenological constraints; an
important question will also be the fate of the vector dominance model in
the medium. In Sec.~\ref{sec_spectra} the theoretical developments are
tested in recent dilepton production experiments, starting with
elementary reactions off nuclei representative for medium effects in
cold nuclear matter (sub-Sec.~\ref{ssec_nuclei}). The main part of
Sec.~\ref{sec_spectra} is devoted to an analysis of
dilepton spectra in ultrarelativistic heavy-ion reactions
(sub-Sec.~\ref{ssec_hics}), focusing on recent results
obtained at the CERN-SPS by the NA60 and CERES/NA45 collaborations.
The spectral analysis is completed by a
critical assessment of the combined theoretical and
experimental status to date (sub-Sec.~\ref{ssec_appraisal}).  We
finish with concluding remarks in Sec.~\ref{sec_concl}.

\section{Chiral Symmetry, Condensates and Chiral Restoration}
\label{sec_chi-sym}

It is generally accepted that strong interactions are described by 
Quantum Chromodynamics (QCD), introduced in 
1973~\cite{Fritzsch:1973pi,Gross:1973id,Politzer:1973fx}, with
a Lagrangian density given by
\beq
{\cal L}_{QCD}= 
\bar q (i\gamma^\mu D_\mu - {\cal{M}}_q) q
-\frac{1}{4}G^a_{\mu\nu}G_a^{\mu\nu} \quad , 
\qquad
D_\mu=\partial_\mu+ig_s\frac{\lambda_a}{2}A^a_\mu \ ,
\label{LQCD}
\eeq
formulated in terms of elementary quark ($q$) and gluon ($A^a_\mu$) 
fields ($\gamma^\mu$ and $\lambda^a$: Dirac and Gell-Mann matrices,
respectively, ${\cal{M}}_q$=diag$(m_u,m_d,\dots)$: 
current-quark mass matrix). In addition to the local $SU(3)$ color gauge
symmetry, ${\cal L}_{QCD}$ possesses several global symmetries. 
The most relevant one in the present context is Chiral Symmetry,  
which can be exhibited by rewriting ${\cal L}_{QCD}$ in terms of left- 
and right-handed quark fields, $q_{L,R}$=$\frac{1}{2}(1\mp\gamma_5)q$: 
\beq
{\cal L_{\rm QCD}} =\bar q_Li\gamma^\mu D_\mu q_L+
\bar q_Ri\gamma^\mu D_\mu q_R
-(\bar q_L {\cal M}_q q_R+ \bar q_R {\cal M}_q q_L) 
-{1\over 4} G_{\mu\nu}^aG^{\mu\nu}_a  \  . 
\eeq
For small quark masses, i.e., $u$ and $d$ quarks, ${\cal L_{\rm QCD}}$ 
is approximately invariant under rotations  
$q_{L,R} \to {\rm e}^{-i\vec\alpha_{L,R}\cdot {\vec\tau}/2} q_{L,R}$,
where $\vec\alpha_{R,L}$ are 3 real angles and $\tau$ operates in
($u$-$d$) isospin space. Chiral invariance of the QCD Lagrangian
thus refers to the conservation of quark handed-ness and isospin.
Alternatively, one can rewrite the chiral rotations as
$q \to {\rm e}^{-i\vec\alpha_{V}\cdot {\vec\tau}/2} q$ and 
$q \to {\rm e}^{-i\gamma_5\vec\alpha_{A}\cdot {\vec\tau/2}} q$, 
giving rise to conserved isovector-vector and -axialvector
currents, 
\beq
\vec j_{V}^\mu=\bar q\gamma^\mu{\vec\tau\over 2}q \quad , \quad
\vec j_{A}^\mu=\bar q\gamma^\mu\gamma_5{\vec\tau\over 2}q \ . 
\label{jVA}
\eeq



\subsection{Condensates and Hadron Spectrum in Vacuum}
\label{ssec_cond-hs}
As emphasized in the Introduction, the nonperturbative structure of 
the QCD vacuum is characterized by its condensates. A special role is 
played by the quark-antiquark ($q\bar q$) and gluon ($G^2$) condensates. 
Apart from being the condensates involving the minimal number of
quark- and gluon-fields, the former is a main order parameter of 
SB$\chi$S while the latter dominantly figures into the 
energy-momentum tensor of the theory. The vacuum expectation value of 
the latter's trace is given by
\begin{equation}
\langle T^\mu_\mu \rangle = \epsilon - 3 P = -\langle G^2 \rangle + 
m_q \langle \bar qq\rangle  
\label{Tmumu}
\end{equation}  
where $G^2=-(\beta(g_s)/2g_s)~G_a^{\mu\nu} G^a_{\mu\nu}$ involves the
gluon-field strength tensor and the renorma\-li\-zation-group beta function,
$\beta(g_s)$. The latter appears because the nonvanishing vacuum value of
$T^\mu_\mu$ breaks the scale invariance of the classical QCD Lagrangian,
induced by quantum loop corrections.  The small current light-quark
masses, $m_q$$\simeq$5\,MeV, render the contribution of the quark 
condensate to $T^\mu_\mu$ small. The absolute value of the gluon 
condensate is not precisely known, but presumably rather large, around 
1.5\,GeV/fm$^3$$\simeq$(330\,MeV)$^4$. In fact, the magnitude of the 
quark condensate is not small either,
$\langle \bar qq \rangle$$\simeq$(-250\,MeV)$^3$ per light-quark flavor,
and about 50\% of that for strange quarks. This implies that the vacuum
is filled with $\sim$5 quark-antiquark pairs per fm$^3$!  
Also note that the quark condensate maximally violates chiral symmetry
by mixing right- and left-handed quarks, 
$\langle \bar qq \rangle$=$\langle \bar q_Lq_R + \bar q_Rq_L \rangle$, 
implying that a quark propagating through the vacuum can flip its 
chirality by coupling to the condensate.
The intimate relation between chiral symmetry breaking and the associated
Goldstone-boson nature of the pion is highlighted by the Gell-Mann-Oakes
Renner (GOR) relation,
\begin{equation}
m_\pi^2 f_\pi^2 = - 2 m_q \langle \bar qq\rangle \ , 
\label{gor}
\end{equation}
which combines the effects of explicit chiral symmetry breaking,
$m_\pi^2$$\propto$$m_q$, and SB$\chi$S with the pion decay constant as
order parameter.

\begin{figure}[!t]
\begin{center}
\begin{minipage}{0.45\linewidth}
\epsfig{file=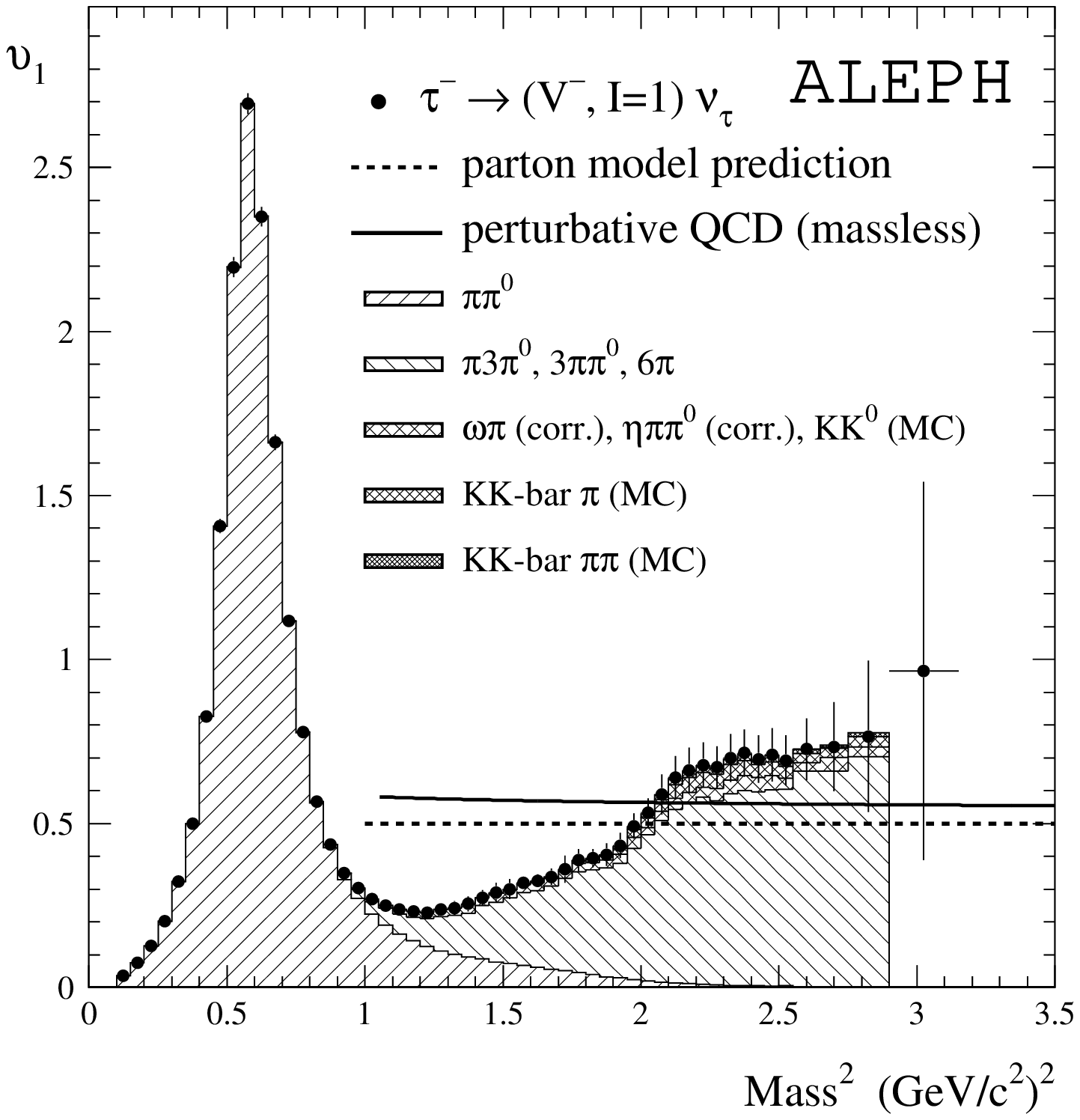,width=\textwidth}
\end{minipage}
\hspace{0.3cm}
\begin{minipage}{0.45\linewidth}
\epsfig{file=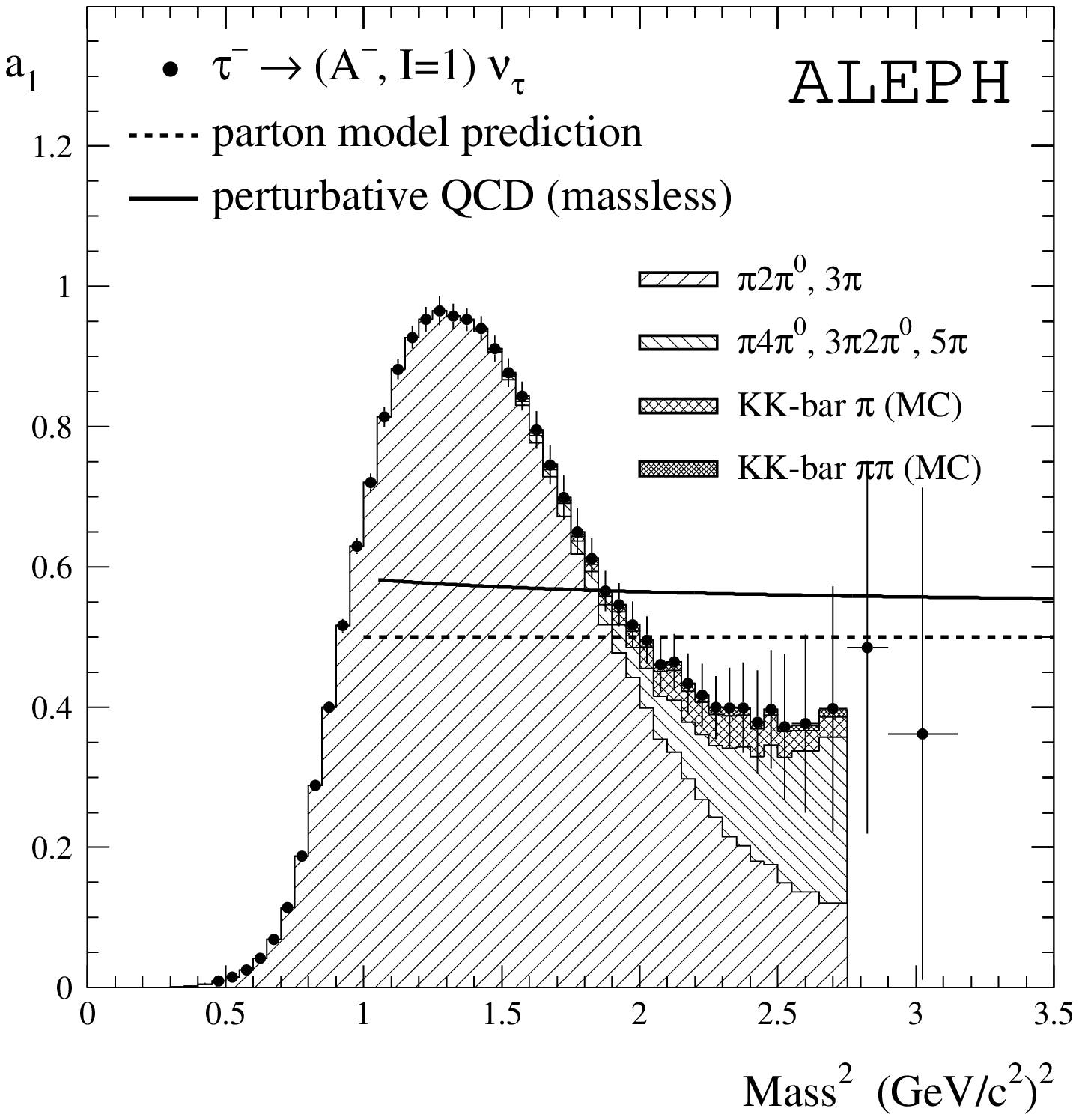,width=\textwidth}
\end{minipage}
\end{center}
\caption{\it Experimental data~\cite{Barate:1998uf} of the
  isovector-vector ($IJ^P$=$11^-$, left panel) and isovector-axialvector
  ($IJ^P$=$11^+$, right panel) spectral functions from hadronic decays of
  $\tau$ leptons (produced in $\sqrt{s}$=91\,GeV $e^+e^-$ annihilation at
  LEP) into even and odd numbers of pions/kaons, respectively. The lines
  indicate theoretical calculations using pQCD.}
\label{fig_aleph}
\end{figure}
One of the best direct empirical evidences for the spontaneous breaking
of chiral symmetry is found in the vector channel, more specifically the
isovector-vector channel ($IJ^P$=$11^+$) and its chiral partner, the
isovector-axialvector one ($IJ^P$=$11^-$), precisely the Noether currents
in Eqs.~(\ref{jVA}). The pertinent spectral
functions have been measured with excellent precision (and a detailed
decomposition of the hadronic final states) at the Large
Electron-Positron collider (LEP) in hadronic $\tau$ decays by the
ALEPH~\cite{Barate:1998uf} and OPAL~\cite{Ackerstaff:1998yj}
collaborations, cf.~Fig.~\ref{fig_aleph}.
In the low-mass region, the strength of each of the two spectral
functions is largely concentrated in a prominent resonance, i.e., the
$\rho$(770) and $a_1$(1260). This very fact indicates that the low-mass 
regime is dominated by nonperturbative effects, while the (large)
difference in mass and width of these resonances signals chiral symmetry
breaking. This connection can be quantified by chiral sum rules
developed by Weinberg~\cite{Weinberg:1967} and others~\cite{Das:1967ek}
in the late 1960's based on current algebra of chiral symmetry.  
These sum rules relate
moments of the difference between vector and axialvector spectral
functions to chiral order parameters. In the chiral limit ($m_\pi$=0)
one has
\begin{eqnarray}
  f_n = - \int\limits_0^\infty \frac{\dd s}{\pi} \ s^n \
  \left[\im \Pi_V(s) - \im \Pi_A(s) \right]  \ ,
  \qquad \qquad \qquad
\label{csr}
\\
f_{-2} = f_\pi^2  \frac{\langle r_\pi^2 \rangle}{3} - F_A \ , \quad
f_{-1} = f_\pi^2 \ , \quad
f_0   = 0 \ ,  \quad
f_1 = -2\pi \alpha_s \langle {\cal O} \rangle  \
\label{fn}
\end{eqnarray}
($r_\pi$: pion charge radius, $F_A$: coupling constant for the radiative
pion decay, $\pi^\pm\to\mu^\pm \nu_\mu \gamma$, $\langle {\cal O_4}
\rangle$: four-quark condensate).

\subsection{Phase Diagram and Chiral Restoration}
\label{ssec_phase-dia}
\begin{figure}[!tb]
\begin{center}
\epsfig{file=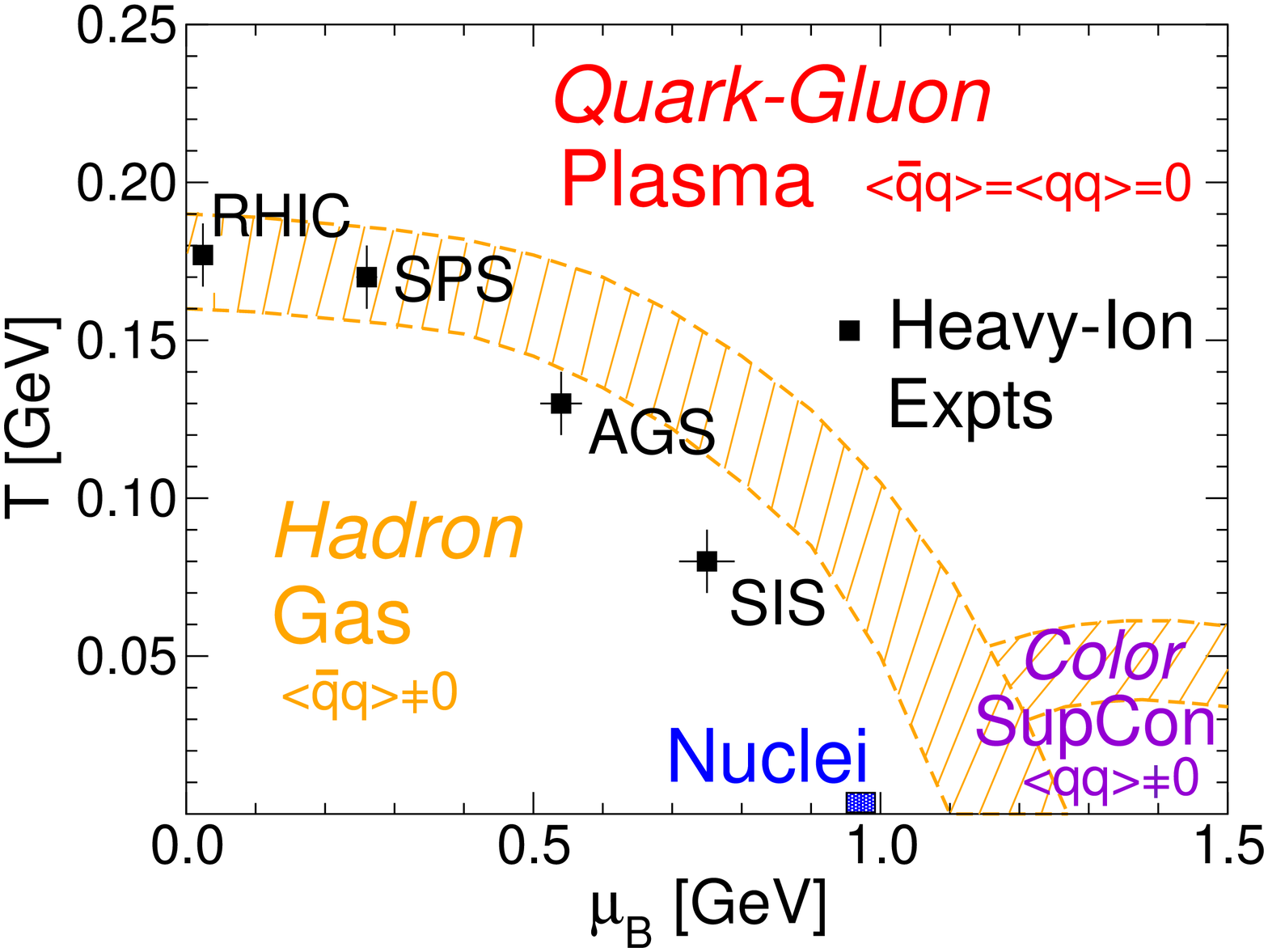,width=0.5\linewidth}
\vspace{-1.0cm}
\end{center}
\caption{\it Schematic QCD phase diagram including empirical extractions
  of $(\mu_B,T)$-values from observed hadron production ratios in
  heavy-ion experiments at different beam
  energies~\cite{BraunMunzinger:2003zd}; the bands indicate lattice-QCD
  and model estimates of the transition regions between HM, QGP and CSC;
  with the HM-QGP transition (along the finite $T$, $\mu_B$=0 axis)
  presumably a cross-over, and the HM-CSC one (along the finite $\mu_B$,
  $T$=0 axis) possibly first order, there is presumably a second order
  endpoint, e.g., around $(\mu_B,T)$$\approx$(400,160)~MeV.}
\label{fig_phase-dia}
\end{figure}

A schematic view of the QCD phase diagram is displayed in 
Fig.~\ref{fig_phase-dia}. It is roughly characterized by
three major regimes (all of which most likely exhibit rich
substructures): hadronic matter (HM) at small and moderate temperature
($T$) and baryon chemical potential ($\mu_B$), Quark-Gluon Plasma (QGP)
at high $T$ and Color Super-Conductors (CSCs) at high $\mu_B$ but low
$T$. The latter may occur in the core of neutron stars, but are unlikely
to be produced in heavy-ion collisions and will not be further discussed
here.  

\begin{figure}[!tb]
\begin{minipage}{0.48\linewidth}
\epsfig{file=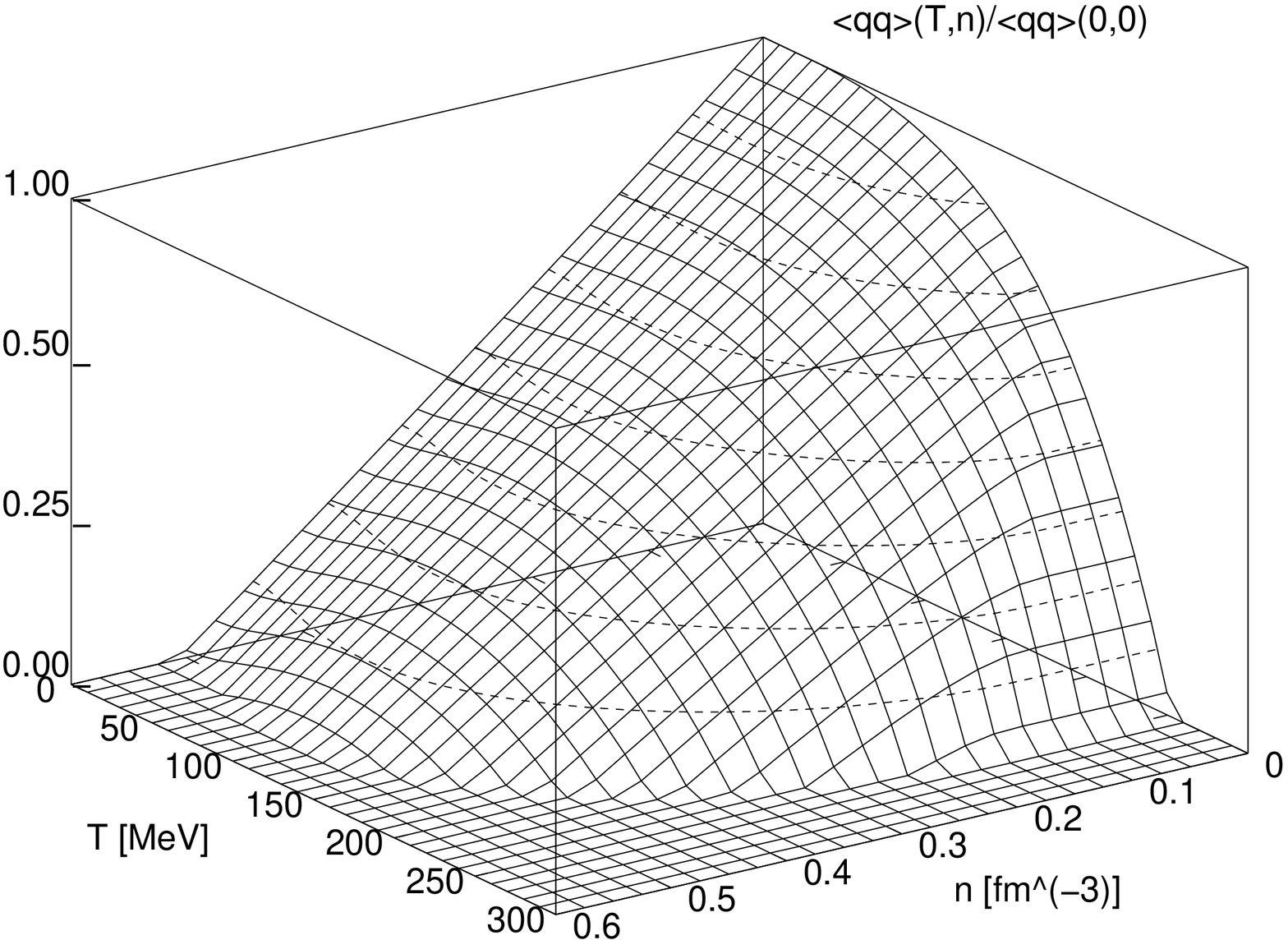,width=0.99 \textwidth,angle=0}
\end{minipage}\hfill
\begin{minipage}{0.48\linewidth}
\epsfig{file=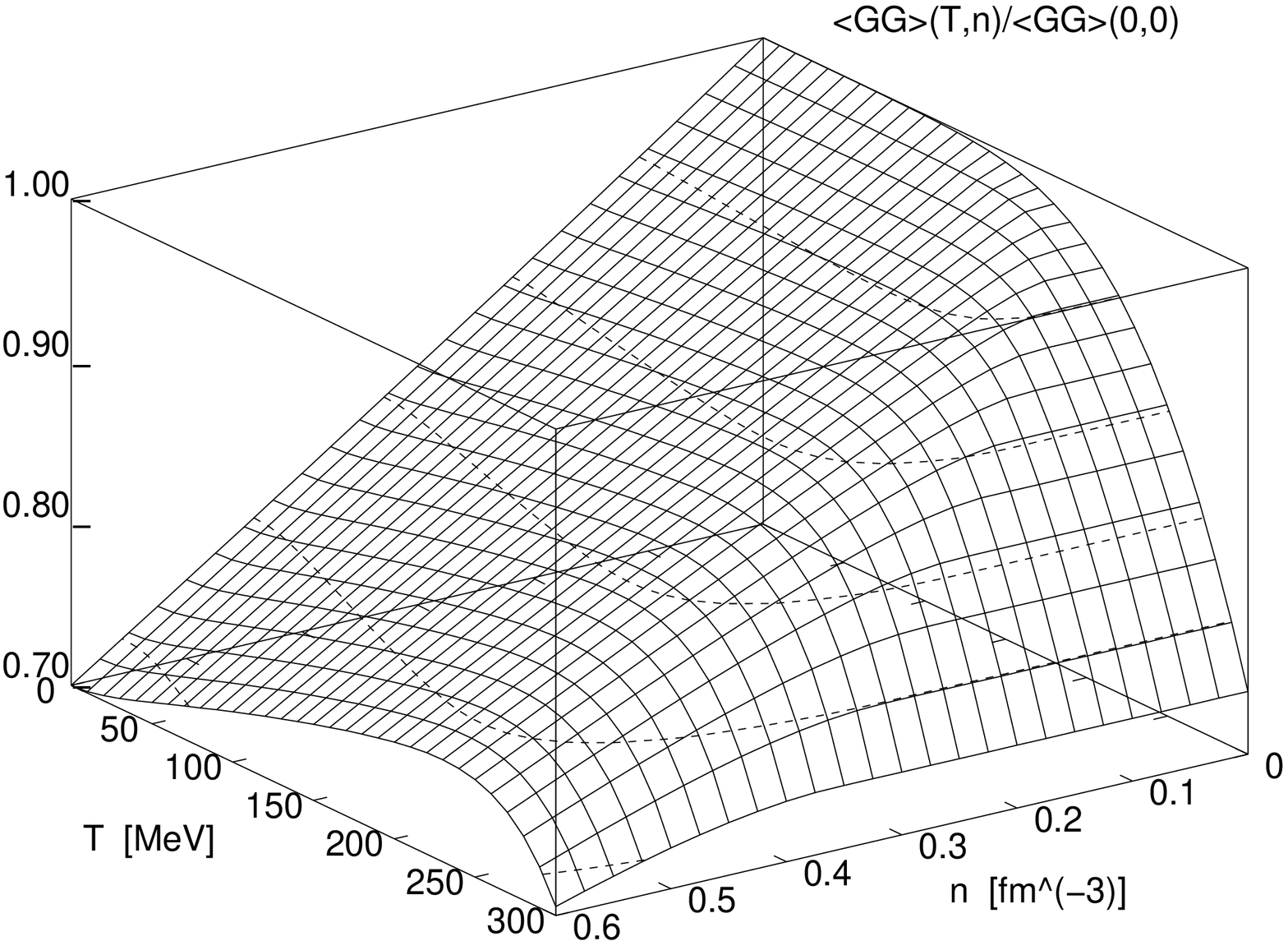,width=0.99 \textwidth,angle=0}
\end{minipage}
\vspace*{3mm}

\begin{minipage}{0.48\linewidth}
\epsfig{file=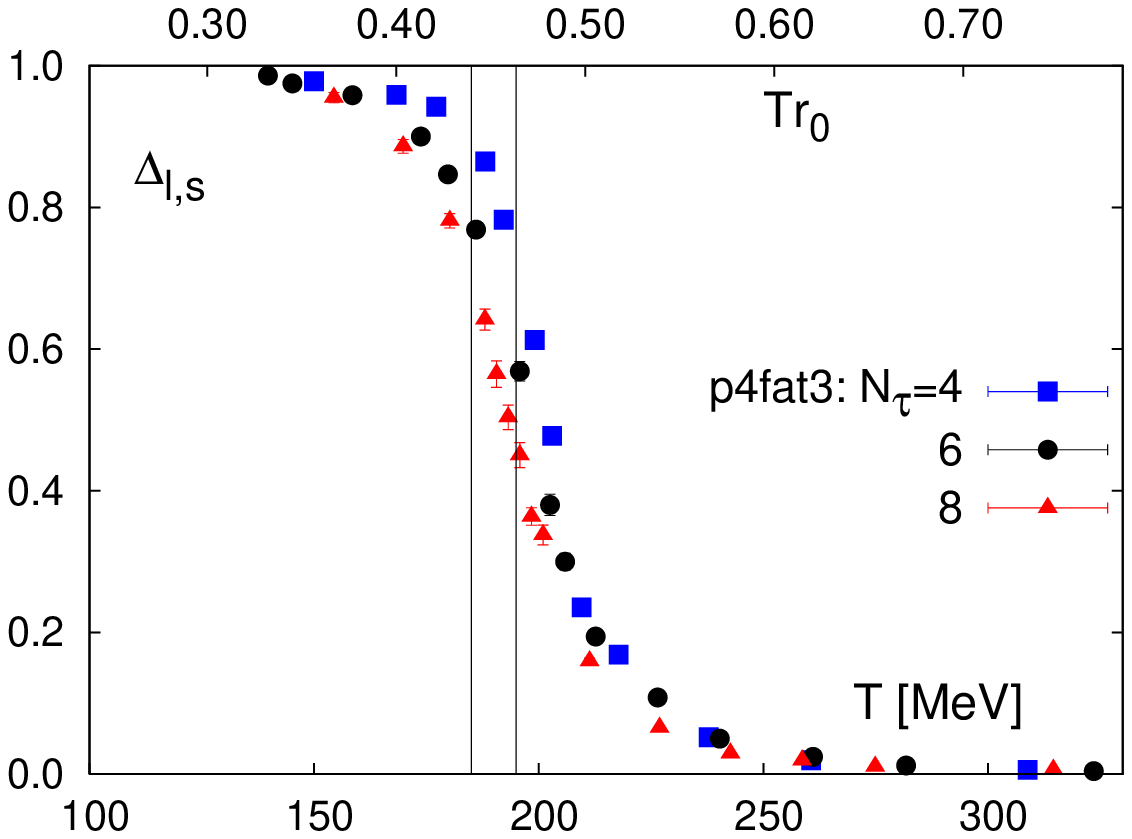,width=1.0\textwidth}
\end{minipage}\hfill
\begin{minipage}{0.48\linewidth}
\vspace{0.5cm}
\epsfig{file=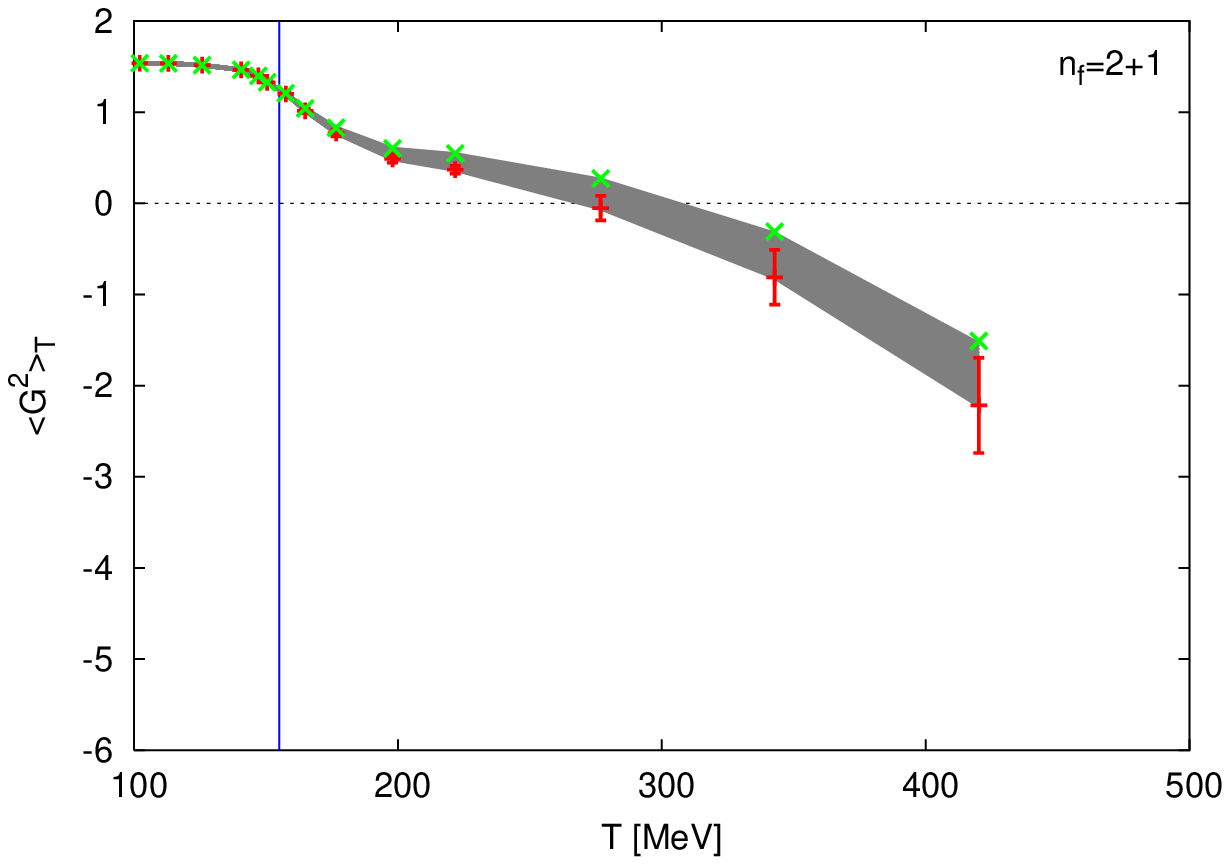,width=1.0\textwidth}
\end{minipage}
\caption{\it Upper panels: density and temperature dependence of the
  chiral (left) and the gluon condensate (right), normalized to their
  vacuum values~\cite{Zschocke:2002ic}, obtained in a low-$T$ and 
  -$\varrho_N$ expansions. Lower panels: $T$ dependence at $\mu_q$=0
  of the (subtracted and renormalized) chiral condensate 
  (left)~\cite{Cheng:2007jq} and of the gluon condensate 
  (right)~\cite{Miller:2006hr} as obtained from $N_f$=2+1 lattice QCD
  computations.}
\label{fig_condens}
\end{figure}

A key issue toward understanding the phase structure of QCD matter
is the temperature and density dependence of its condensates. 
Various condensates serve as order parameters of broken symmetries and
govern the (hadronic) excitation 
spectrum. The latter provides the connections to observables. 
A first estimate of the medium modifications of the
condensates can be obtained in the low-density
limit~\cite{Drukarev:1991fs,Cohen:1991nk}, by approximating the thermal
medium
by non-interacting light hadrons, i.e., pions at finite $T$ an 
nucleons at finite $\varrho_N$. For the quark condensate, this leads to 
a linear density expansion of the type
\begin{equation}
\frac{\tave{\bar qq} (T,\mu_B)}{\langle \bar qq\rangle} \ = \ 
1-\sum\limits_h \frac{\varrho_h^s \Sigma_h}{m_\pi^2 f_\pi^2}  
\ \simeq \ 1 - \frac{T^2}{8f_\pi^2}  - \frac{1}{3} 
\frac{\varrho_N}{\varrho_0} - \cdots \ ,  
\label{qqbar-med}
\end{equation}
where $\rho_h^s$ denotes the scalar density of hadron $h$, and
$\Sigma_h$ denotes its ``$\sigma$''-term ($\Sigma_h/m_q$ may be
interpreted as the number of $\bar qq$ pairs inside hadron $h$ which
diminish the (negative) $\bar qq$ density of the condensate).
In obtaining Eq.~(\ref{qqbar-med}), the GOR relation (\ref{gor}) has 
been used. Alternatively, one can directly use the
definition of the quark condensate in terms of the quark-mass 
derivative of the thermodynamic potential,
\begin{equation}
\tave{\bar qq} = \frac{\partial \Omega}{\partial m_q} \ ,  
\end{equation}
and evaluate the temperature and density-dependent part,
$\bar\Omega(\mu_B,T)\equiv \Omega(\mu_B,T) - \Omega_{\rm vac}$, in the 
free gas approximation. A similar strategy can be adopted for the gluon
condensate, by utilizing its relation, Eq.~(\ref{Tmumu}), to the trace
anomaly,
\begin{eqnarray}
\tave{G^2} =  -(\epsilon-3P) + m_q \tave{\bar qq} \ , 
\end{eqnarray}
and estimating the $\varrho_B$- and $T$-dependent parts of pressure 
and energy density in suitable expansions.
At finite temperature, for a massless pion
gas, one has $\epsilon$=$3P$ and thus no correction to order $T^4$ 
(the system is scale invariant). It turns out that
the lowest-order interaction contribution from (soft) $\pi\pi$
scattering does not contribute either so that the leading temperature
dependence of the gluon condensate arises at order
$T^8$~\cite{Leutwyler:1992ic}. With the leading nuclear-density
dependence as worked out in Refs.~\cite{Drukarev:1991fs,Cohen:1991nk},
one has
 \begin{equation}  
\tave{G^2} = - \langle G^2 \rangle- (m_N - \Sigma_N) \varrho_N - 
\frac{\pi^2}{270}\frac{T^8}{f_\pi^4} 
\left( \ln\frac{\Lambda_p}{T} -\frac{1}{4} \right)
\end{equation}
($\Lambda_p$$\simeq$275\,MeV is a renormalization scale). The above
relations allow for some interesting insights. As already noted in 
Ref.~\cite{Cohen:1991nk}, the linear-density expansions suggest that
the gluon condensate is much less affected then the quark condensate,
cf.~also the upper panels in Fig.~\ref{fig_condens}.
It is not obvious whether recent finite-$T$ lattice computations in QCD
with 2+1 flavors support this picture (see the lower panels of
Fig.~\ref{fig_condens}), especially when approaching the critical
temperature: both $\tave{\bar qq}$ and $\tave{G^2}$ drop significantly
around $T_c$ and reach approximately zero at roughly 1.5\,$T_c$ (the
perturbative interaction contribution to $\epsilon-3P$ renders the gluon
condensate negative at high $T$).  The low-density expansion of the
quark condensate seems to suggest that temperature effects are weaker
than density effects (upper left panel in Fig.~\ref{fig_condens}). This
is, in fact, not the case: as a function of (pion-) {\em density}, the
leading reduction of $\tave{\bar qq}$ in a heat bath is quite
comparable to cold nuclear matter, as determined by the
coefficient in Eq.~(\ref{qqbar-med}) which is in essence given by the
respective $\sigma$ terms, $\Sigma_\pi$$\simeq$70~MeV compared to
$\Sigma_N$=45$\pm$15~MeV.


Another interesting observation can be made when taking the expectation
value of the trace of the energy momentum tensor over a single nucleon
state,
\begin{equation}
\langle N|T^\mu_\mu|N\rangle = - m_N = \langle N|G^2|N\rangle +
 m_q \ \langle N|\bar q q|N\rangle \ . 
\end{equation}
Since the second term (related to the $\sigma$ term) is small (or zero in
the chiral limit), this relation seems to suggest that the major part of
the nucleon mass is generated by the gluon condensate.  This is to be
contrasted with effective quark models (e.g., Nambu Jona-Lasinio) which,
in mean field approximation, attribute the constituent quark mass 
entirely to the quark condensate, 
$m_q^*$=$G_{\rm eff}\langle\bar qq\rangle$.
One should also note that, at least in the QGP phase, the vanishing of
the quark or gluon condensate does not necessarily imply quark
``masses'' to vanish. E.g., in perturbative QCD, partons in the QGP
acquire a thermal mass $m_{q,g}^{\rm th}$$\sim$$gT$. This mass term 
does not break chiral symmetry (its Dirac structure includes a $\gamma_0$
matrix) and presumably persists until close to $T_c$, thus supplanting
the constituent quark mass, $m_q^*$, well before the latter
vanishes. This has, of course, important consequences for the masses of
hadronic states in the vicinity of $T_c$.  In addition, large
binding-energy effects can be present, e.g., for the pion: if the chiral
transition is continuous, the pion's Goldstone-boson nature could very
well imply that it survives as a bound state at temperatures 
above $T_c$~\cite{Hatsuda:1985eb,Brown:2003km,Mannarelli:2005pz}.

As emphasized above, the only known direct way to extract observable
consequences of changes in the QCD condensate structure is to probe
medium modifications in its excitation spectrum. This applies in
particular for the quark condensate which has a rather small impact on
the bulk properties of QCD matter being suppressed by $m_q$ (the
relation of the gluon condensate to the equation of state could, in
principle, be tested via hydrodynamic or transport properties, but this
turned out to be difficult in the context of heavy-ion
collisions~\cite{Teaney:2001av}). The generic model-independent
consequence of $\chi$SR for the in-medium hadronic spectrum is
the degeneracy of the spectral functions within chiral multiplets (i.e.,
for chiral partners) , e.g., $\pi$-``$\sigma$'', $N$-$N^*(1535)$ and
$\rho$(770)-$a_1$(1260). In the ``$\sigma$'' channel (which
asymptotically corresponds to a scalar-isoscalar pion pair), interesting
medium effects have been observed in pion- and photon-induced production
of $S$-wave pion pairs off
nuclei~\cite{Starostin:2000cb,Grion:2005hu,Bloch:2007ka}.  An
accumulation of strength close to the two-pion threshold (which is not
observed in the isotensor $\pi$-$\pi$ channel) has been associated with
an in-medium reduction of the ``$\sigma$''-meson mass as a precursor
effect of $\chi$SR~\cite{Hatsuda:1999kd} (note that the leading-density
approximation, Eq.~(\ref{qqbar-med}), predicts a reduction of the quark
condensate by $\sim$30\% already at normal nuclear matter density).
However, nuclear many-body
effects~\cite{Rapp:1998fx,VicenteVacas:1999xx}, in particular the
renormalization of the pion propagator in the nuclear medium, can 
essentially explain the experimental findings\footnote{Similar results
  are obtained from a transport treatment of pion reinteractions in the
  medium~\cite{Buss:2006yk}.}. This raises an important question: to
what extent do ``conventional'' in-medium effects encode mechanisms of
$\chi$SR? From the point of view of the ``$\sigma$'' spectral function
alone, it is not possible to distinguish whether a softening is caused
by many-body effects or genuine mass changes figuring via medium
modifications of the mass parameter in the underlying effective
Lagrangian.  Thus, a distinction of medium effects into ``conventional''
ones and those associated with an apparent ``direct'' connection to 
$\chi$SR is meaningless. Rather, a careful and exhaustive treatment of
hadronic many-body effects is an inevitable ingredient for evaluating 
mechanisms of $\chi$SR. As already alluded to in the Introduction, a
practical problem of using the $\pi\pi$ decay channel for studying
medium effects are the strong final-state interactions of the individual
pions when exiting the nuclear medium~\cite{Buss:2006yk}. The same
applies to the heavy-ion collision environment, implying that the 
$\pi\pi$ channel can only probe the dilute stages of the produced 
medium. This problem is overcome by dilepton final states, on which we 
will focus in the following.

\section{Vector Mesons in Medium}
\label{sec_vec-mes}

\subsection{Dileptons and Electromagnetic Correlation Function}
\label{ssec_em-corr}
\begin{figure}[!tb]
\begin{minipage}{7.5cm}
\epsfig{file=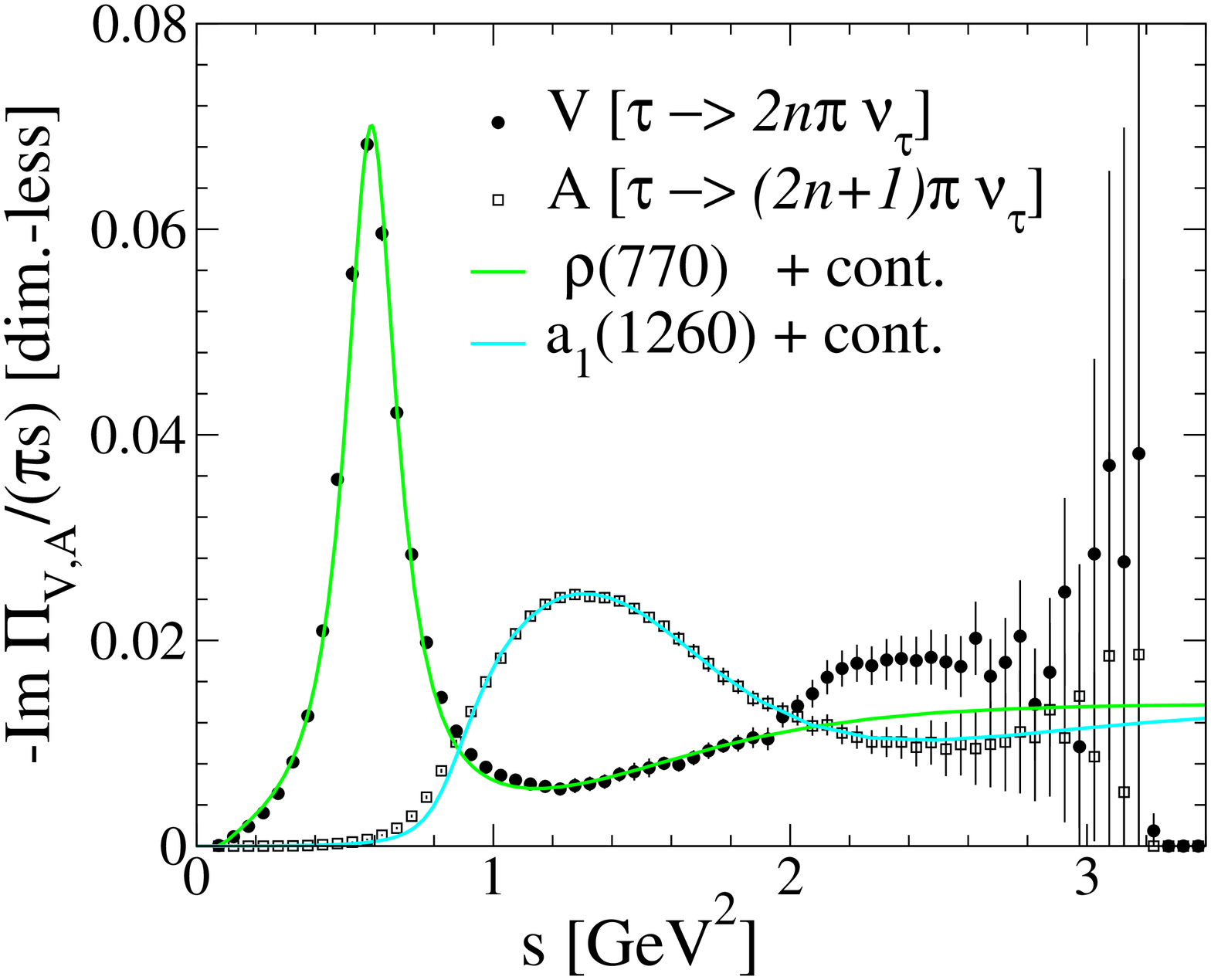,width=7.5cm}
\end{minipage}
\hspace{1cm}
\begin{minipage}{7cm}
\epsfig{file=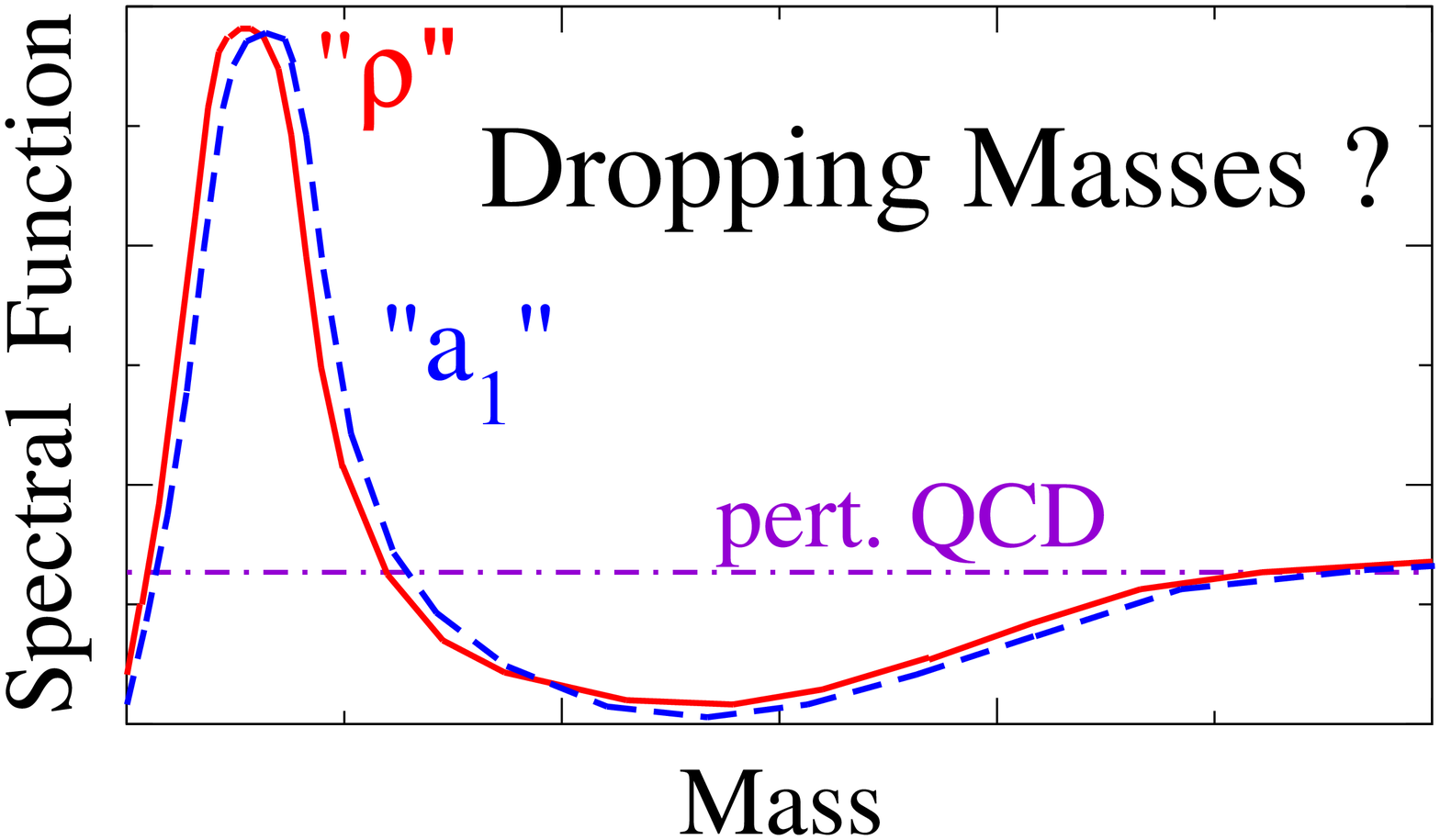,width=6cm}
\epsfig{file=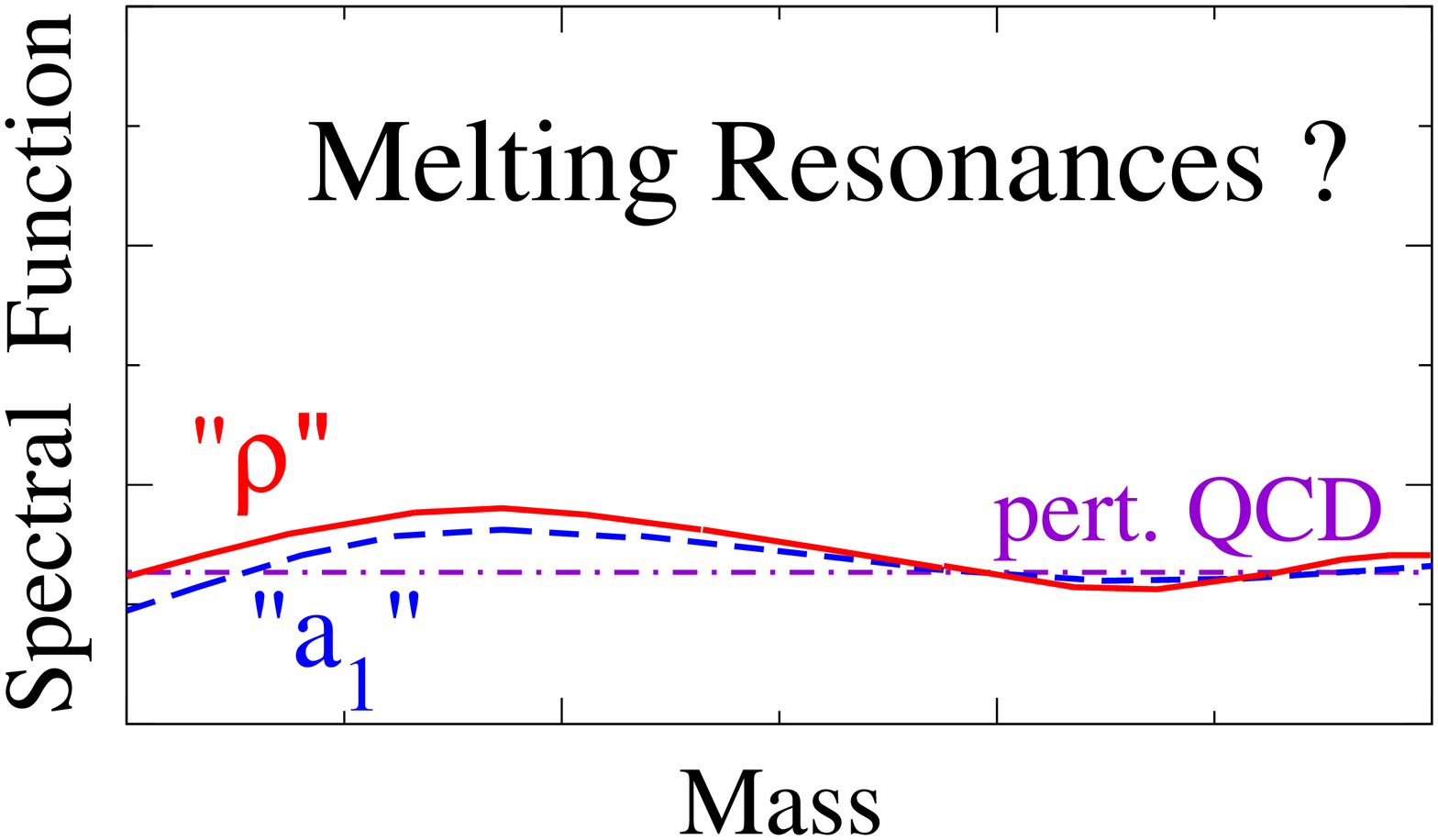,width=6cm}
\end{minipage}
\caption{\it Left panel: vector and axialvector spectral functions as
  measured in hadronic $\tau$ decays~\cite{Barate:1998uf} with model
  fits using vacuum $\rho$ and $a_1$ strength functions supplemented by
  perturbative continua~\cite{Rapp:2002tw}; right panel: scenarios for 
  the effects of chiral symmetry restoration on the in-medium vector- 
  and axial-vector spectral functions.}
\label{fig_VA}
\end{figure}

For a strongly interacting medium in thermal equilibrium the production
rate of dileptons can be cast into the
form~\cite{Feinberg:1976ua,McLerran:1984ay},
\begin{equation}
\frac{dN_{ll}}{d^4xd^4q} = -\frac{\alpha_{\rm em}^2}{\pi^3 M^2} \
       f^B(q_0;T) \  \frac{1}{3} \ g_{\mu\nu} 
\ \im \Pi_{\rm em}^{\mu\nu} (M,q;\mu_B,T) \ .  
\label{Rll}
\end{equation}
This expression is to leading order in the electromagnetic (EM) coupling
constant, $\alpha_{\rm em}$, but exact in the strong interaction. The 
latter is encoded in the EM spectral function, defined via the
retarded correlator of the hadronic EM current, $j_{\rm em}^\mu(x)$,
\begin{equation}
  \Pi_{\rm em}^{\mu \nu}(q_0,q)) = -\ii \int d^4x \ e^{iq\cdot x} \
  \Theta(x_0) \ \tave{[j_{\rm em}^\mu(x), j_{\rm em}^\nu(0)]}   \ . 
\end{equation}
In the vacuum, the spectral strength is directly accessible via the
total cross section for $e^+e^-$ annihilation,
\begin{equation}
\sigma(e^+ e^- \to {\rm hadrons}) = \frac{4\pi \alpha_{\rm em}^2}{s} 
\  \frac{(-12 \pi)}{s} \im\Pi_{\rm em}^{\rm vac}(s) \ ,
\label{sigee}
\end{equation}
recall Fig.~\ref{fig_Pi-em-vac} (the first factor is simply
$\sigma(e^+e^-$$\to$$\mu^+\mu^-)$=$4\pi\alpha_{\rm em}^2/s$). As a function 
of invariant dilepton mass, $M^2$=$q_0^2-\vec q^2$, the spectrum 
basically decomposes into two regimes. In the low-mass region (LMR,
$M$$\le$1\,GeV), the strength is absorbed in the three vector 
mesons $\rho$(770), $\omega$(782) and $\phi$(1020) representing the 
lowest resonances in the two-pion, three-pion and
kaon-antikaon channels, respectively. 
Thus, the EM current is well described within
the vector dominance model (VDM)~\cite{Sakurai:1969} as given by the
field current identity,
\begin{equation}
  j^\mu_{\rm em}(M\le 1~{\rm GeV})= \frac{m_\rho^2}{g_\rho} \rho^\mu + 
  \frac{m_\omega^2}{g_\omega} \omega^\mu +\frac{m_\phi^2}{g_\phi} \phi^\mu
  \, . 
\label{j-meson}
\end{equation}
In the intermediate mass region (IMR, 1\,GeV$<$$M$$\le$3\,GeV), the strength
is reasonably well accounted for by a partonic description,
\begin{equation}
j^\mu_{\rm em}(M>1.5~{\rm GeV}) 
= \sum\limits_{q=u,d,s} e_q \ \bar{q} \gamma^\mu q \ , 
\label{j-quark}
\end{equation}
where $e_q$ denotes the electric quark charge in units of the electron
charge, $e$. The connection between the two representations can be
exhibited by rearranging the charge-flavor content of the quark basis
into hadronic isospin quantum numbers,
\begin{equation}
j_{\rm em}^\mu = \frac{1}{\sqrt{2}} \ \bar \psi \gamma^\mu \psi \
\left[ \frac{\bar uu - \bar dd}{\sqrt{2}} +
\frac{1}{3} \frac{\bar uu + \bar dd}{\sqrt{2}} - 
\frac{\sqrt{2}}{3} \bar ss \right] \ ,
\end{equation}
reflecting the quark content of the (normalized) $\rho$ (isospin
$I$=1), $\omega$ ($I$=0) and $\phi$ ($I$=0) wave functions,
respectively.  Converting the isospin coefficients into numerical
weights in the EM spectral function, one obtains
\begin{equation}
\im\Pi_{\rm em} \sim \left[ \im D_\rho  +
\frac{1}{9}\im D_\omega + \frac{2}{9} \im D_\phi \right] \ ,
\label{ImPi_em-vdm}
\end{equation}
which identifies the isovector ($\rho$) channel as the dominant source
(experimentally it is even larger as given by the electromagnetic decay
widths, $\Gamma_{\rho\to ee}/\Gamma_{\omega\to ee}$$\simeq$~11).
Explicitly evaluating the EM correlators using the currents
(\ref{j-meson}) and (\ref{j-quark}) yields
\begin{equation}
\im \Pi_{\rm em}^{\rm vac}(M) = \left\{
\begin{array}{ll}
 \sum\limits_{V=\rho,\omega,\phi} \left(\frac{m_V^2}{g_V}\right)^2 \
\im D_V^{\rm vac}(M) & , \ M < M_{\rm dual}^{\rm vac} ,
\vspace{0.3cm}
\\
-\frac{M^2}{12\pi} \ (1+\frac{\alpha_s(M)}{\pi} +\dots)  \ N_c
\sum\limits_{q=u,d,s} (e_q)^2  & , \ M > M_{\rm dual}^{\rm vac} \ 
\end{array}  
\right.
\label{ImPi_em}
\end{equation}
($M_{\rm dual}^{\rm vac}$$\simeq$~1.5~GeV, $N_c$=3: number of quark
colors, $D_V=1/[M^2-m_V^2-\Sigma_V]$: vector-meson propagators).  The
associated processes in the thermal dilepton {\em production} rates are,
of course, the inverse of $e^+e^-$ annihilation, i.e., two-pion,
three-pion and $K\bar K$ annihilation (channeled through the $\rho$,
$\omega$ and $\phi$) in a hadronic phase\footnote{Note that the
  dominance of the isovector channel is naturally associated with the
  annihilation of the two lightest constituents in a hadronic medium.}
and $q\bar q$ annihilation in a QGP.  But what about hadronic emission in
the IMR and QGP emission in the LMR? The former follows from
time-reversal invariance of strong interactions: to the extent that the
hadronic final state in $e^+e^-$ annihilation can be represented by a
statistical (thermal) distribution (which is empirically approximately
satisfied), hadron-gas emission in the IMR corresponds to multi-hadron
annihilation ($4\pi, 6\pi \to e^+e^-$, etc., which may 
be built from 2$\rho$, $\pi a_1$, $\pi \omega$, etc.), with a total
strength given by the
partonic continuum. QGP emission in the LMR is, of course, closely
related to a central question of this review: How does the dilute 
hadronic resonance gas rate evolve into the
chirally restored, deconfined QGP rate? At sufficiently low temperatures
and/or baryon densities virial expansions in a hadronic basis can
provide initial insights. With increasing $T$ and $\varrho_B$
resummations become necessary for which many-body
approaches are a suitable tool. It is currently an open question how far
up in $\varrho_B$ and $T$ these calculations are reliable.
Selfconsistent schemes are, in principle, capable of describing
phase-transition dynamics, which, ideally, could be constrained by
unquenched lattice-QCD calculations of the dilepton rate below $T_c$
(energy sum rules turn out to be particularly useful to connect spectral
functions to order parameters). Eventually, in the high-temperature
limit, the LMR rate should recover perturbative $q\bar q$ annihilation, 
where a systematic evaluation of corrections becomes feasible again.  
The remainder of this section is devoted to a discussion of these 
approaches. With hindsight to Sec.~\ref{sec_spectra}, we will focus on 
models for which quantitative applications to dilepton observables have 
been made, with the isovector ($\rho$) channel playing the leading role. 
In the following, for brevity, we refer to the $IJ^P$=$11^\pm$ chiral 
partner channels as vector ($V$) and axialvector ($A$) ones. In the 
vacuum, both can be well represented by a low-lying resonance pole 
($\rho$ and $a_1$) and a continuum above, see left panel of 
Fig.~\ref{fig_VA}. Two schematic scenarios for the degeneration of 
vector and axialvector channels at
chiral restoration (``dropping mass'' and ``resonance melting'') are
sketched in the right panel of Fig.~\ref{fig_VA}.

\subsection{Medium Effects I: Model Independent}
\label{ssec_mod-indep}
In principle, model-independent assessments of medium effects do not 
involve free parameters. These can be realized by virial expansion 
schemes based on experimental input for vacuum spectral functions 
(valid for dilute hadronic matter), perturbative QCD 
calculations (valid in the high-$T$ limit) or first-principle lattice 
QCD computations.

\subsubsection{Chiral Reduction and Mixing}
\label{sssec_mix}
The leading temperature dependence of vector and axialvector
correlators, $\Pi_{V,A}^{\mu\nu}$, i.e., their modification in a 
dilute pion gas, can be inferred from chiral reduction and current
algebra. They allow to simplify 1-pion matrix elements of any
operator according to
\begin{equation}
\langle 0 |{\cal O}|\pi^a\rangle = -\frac{i}{f_\pi} 
\langle 0|[Q_A^a,{\cal O}]|0\rangle \quad , \qquad  
[Q_A^a,j_{V,A}^{\mu,b}] = i\epsilon^{abc} j_{A,V}^{\mu,c}  \ ,  
\end{equation} 
where $\{a,b,c\}$ are isospin indices.  Evaluating the Fourier
transforms of the thermal expectation values in the chiral and soft pion
limit (i.e., $m_\pi$=0 and neglecting any momentum transfer $k$ from 
thermal pions in the heat bath), one obtains the ``mixing'' 
theorem~\cite{Dey:1990ba}
\begin{equation}
\Pi_{V,A}(q) = (1-\varepsilon)~\Pi_{V,A}^{0}(q) +
\varepsilon~\Pi_{A,V}^{0}(q) \
\label{chi-mix}
\end{equation}
with the mixing parameter $\varepsilon$=$T^2/6f_\pi^2$ (the Lorentz
structure remains as in the vacuum). The leading-$T$ effect on the $V$
and $A$ correlators is a mere admixture of the chiral partner with
a corresponding reduction of its original strength, via processes of the
type $\pi+V\leftrightarrow A$ and $\pi+A\leftrightarrow V$; width and
mass of the vacuum correlators are unaffected. For dilepton production, 
this implies a reduced $\rho$ pole strength as well as an enhancement of 
the ``dip'' region, $M$$\simeq$1-1.5\,GeV, where the $a_1$ resonance 
provides a ``maximal feeding''.

When naively extrapolating the mixing expression, Eq.~(\ref{chi-mix}), 
to chiral restoration ($\varepsilon$=1/2), one finds $T_c$=$\sqrt{3}
f_\pi$=160\,MeV. This is, however, misleading for several reasons. First,
this estimate does not coincide with a similar extrapolation for the
vanishing of the chiral condensate, cf.~Eq.~(\ref{qqbar-med}). Second,
even a moderate amendment in terms of a finite pion mass in the scalar 
density shifts the estimate to $T_c$$\simeq$225\,MeV. Both facts underline 
the inadequacy of the extrapolation of a lowest-order result. Third, 
the chiral and soft-pion limits are kinematically not a good
approximation (e.g., at $T$=150\,MeV, thermal pions typically bring in an
energy of $\sim$300-400\,MeV). In cold nuclear matter, a similar mixing
is operative via the coupling of the pion cloud of $\rho$ and $a_1$ to
the nuclear medium~\cite{Krippa:1997ss,Chanfray:1999me}.

A much more elaborate treatment of the chiral reduction formalism has
been conducted in Refs.~\cite{Steele:1996su,Steele:1997tv,Steele:1999hf}.  
These calculations are based on realistic fits to vacuum correlators, do 
not invoke kinematic approximations (chiral or soft-pion limits) and 
include both pion and nucleon ensembles. The leading-density part has
been subjected to constraints from nuclear photo-absorption
including the first and second resonance region via $\Delta$(1232) and
$N$(1520) excitations. This allows for meaningful applications to
dilepton spectra which have been carried out and will be discussed in 
Sec.~\ref{sec_spectra}. Note that these calculations do not explicitly 
invoke the notion of VDM, but the fact that the vacuum correlators are 
constructed with $\rho$ and
$a_1$ pole dominance, which is not upset in the linear density scheme,
implies that VDM is still present upon inclusion of medium effects.

\subsubsection{Lattice QCD and Susceptibilities} 
\label{sssec_lattice}
\begin{figure}[!t]
\hspace{-0.8cm}
\begin{minipage}{0.5\textwidth}
\epsfig{file=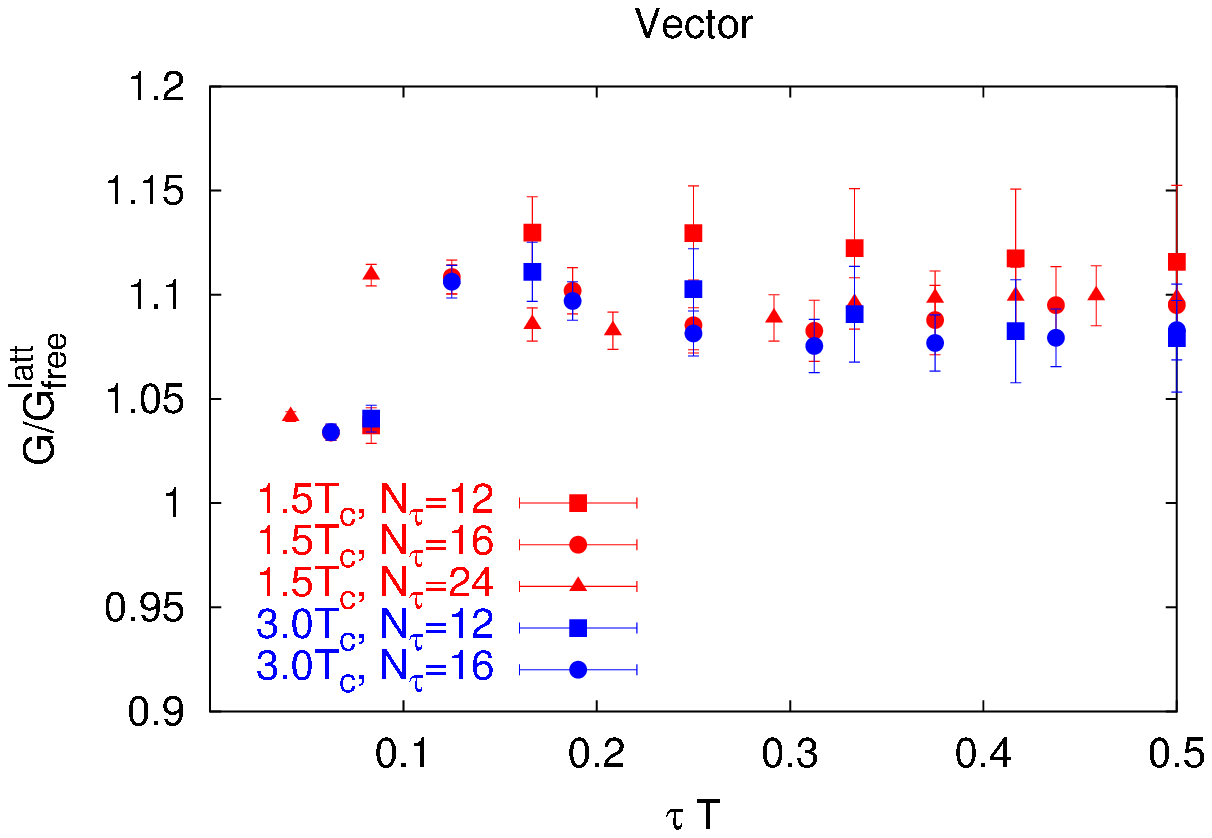,width=1.10\linewidth,height=0.23\textheight}
\end{minipage}
\hspace{0.5cm}
\begin{minipage}{0.5\textwidth}
\epsfig{file=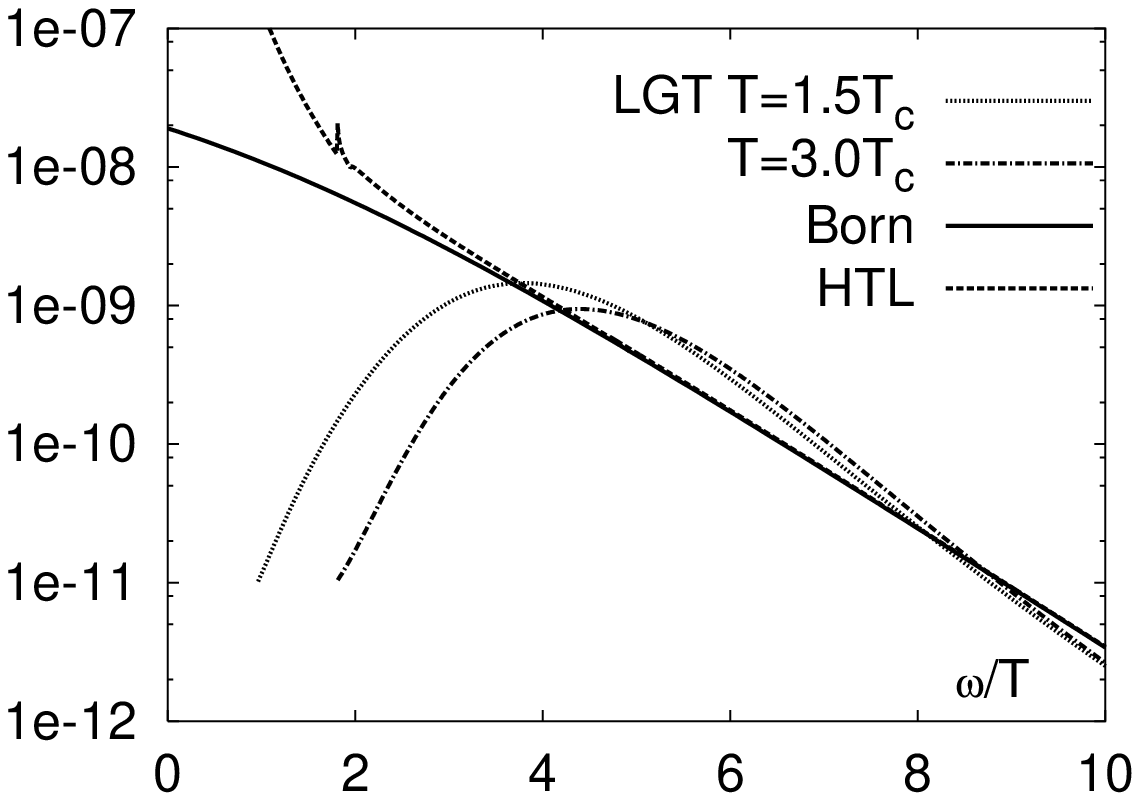,width=1.0\linewidth,angle=0}
\end{minipage}
\caption{\it Left panel: vector correlation function as a
  function of Euclidean time as evaluated in quenched lattice QCD for a
  gluon plasma at temperatures above $T_c$~\cite{Karsch:2001uw}.
  The in-medium correlators are normalized to the free one using the
  integration Kernel at the same temperature (the so-called
  ``reconstructed'' correlator).  Right panel: thermal dilepton rates,
  $dN/(d^4qd^4x)$, in quenched lattice QCD as extracted from the
  correlation functions shown in the left panel using the maximum 
  entropy method. The lattice results are compared to calculations in
  perturbation theory, either to leading order (${\cal O}(\alpha_s^0)$)
  $q\bar q$ annihilation (solid line) or within the hard-thermal-loop
  (HTL) framework~\cite{Braaten:1990wp} (dashed line).  All rates are
  calculated at a total pair 3-momentum of $q$=0, i.e., the dilepton
  energy, $\omega$=$q_0$, equals its invariant mass, $M$.  }
\label{fig_dilep-lat}
\end{figure}

First-principle computations of light-hadron correlation functions in
medium are based on a lattice discretized form of the finite-$T$ QCD 
partition function. Besides a finite lattice spacing, additional 
approximations currently involve the restriction to finite volumes as 
well as the use of unphysically large up- and down-quark masses in the 
simulations. Furthermore, the implementation of chiral symmetry is not 
trivial in the lattice formulation. The numerical evaluation of the 
QCD path integral,
is facilitated by transforming the action to imaginary (Euclidean) time,
which converts the oscillatory behavior of the integrand in the partition
function into an exponential damping. The pertinent Euclidean correlation
function, $\Pi(\tau)$, is related to the physical spectral function,
$\rho=-2 \im\Pi$, via
\begin{equation}
\Pi(\tau,q;T)=
\int\limits_0^\infty \frac{dq_0}{2\pi} \ \rho(q_0,q;T) \
\frac{\cosh[(q_0(\tau-1/2T)]}{\sinh[q_0/2T]} \ . 
\label{G-tau}
\end{equation}
The resulting Euclidean vector correlators in ``quenched" 
QCD\footnote{In the "quenched" approximation the fermionic part of the 
QCD action is neglected in the evaluation of the Euclidean path integral. 
This amounts to neglecting fermion loops.} 
above $T_c$ shows a moderate enhancement over the free correlator, 
cf.~left panel of Fig.~\ref{fig_dilep-lat}~\cite{Karsch:2001uw}.
The extraction of the spectral function requires an inverse integral
transform over a finite number of $\tau$ points\footnote{The (anti-)
periodicity of the boson (fermion) fields at finite $T$ restricts the
Euclidean time direction to the interval $[0,\beta]$ where $\beta=1/T$.}
which can only be achieved with a probabilistic treatment based on the 
``Maximum Entropy Method"~\cite{Asakawa:2000tr}. The resulting 
strength function has been
inserted into the dilepton rate and is compared to perturbative QCD
(pQCD) rates in the right panel of Fig.~\ref{fig_dilep-lat}. The
leading-order pQCD corresponds to the $q\bar q$ strength distribution 
in Eq.~(\ref{ImPi_em}), lower line, while the hard-thermal-loop (HTL) 
improved rate is from Ref.~\cite{Braaten:1990wp}. The latter shows the 
expected divergence for $M\to 0$ which is caused by the Bose factor and 
photon propagator which overcome the $\rho\propto q_0$ dependence of a
retarded correlation function (cf.~also Ref.~\cite{Moore:2006qn}).  This
feature is not shared by the lattice result which might be an artifact
of, e.g., the finite-volume restriction (it would also suggest a small
or even vanishing photon production rate). On the other hand, the
enhancement in the Euclidean correlator translates into an enhanced
dilepton rate at energies of a few times the temperature. Whether this
reflects a broad resonance structure is not clear at present.

\begin{figure}[!tb]
\begin{center}
\begin{minipage}{0.46\linewidth}
\epsfig{file=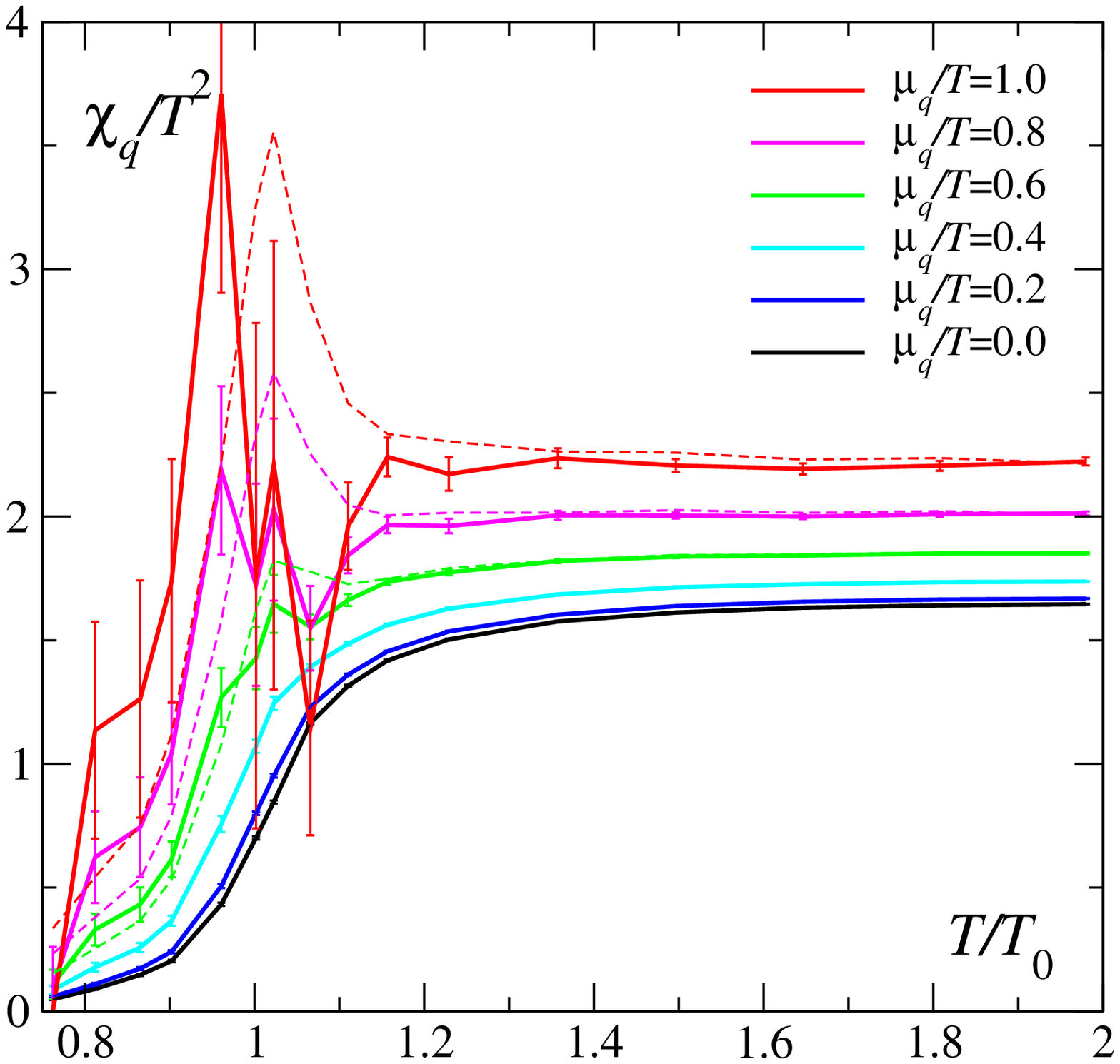,width=0.95\textwidth}
\end{minipage}
\hfill
\begin{minipage}{0.46\textwidth}
\epsfig{file=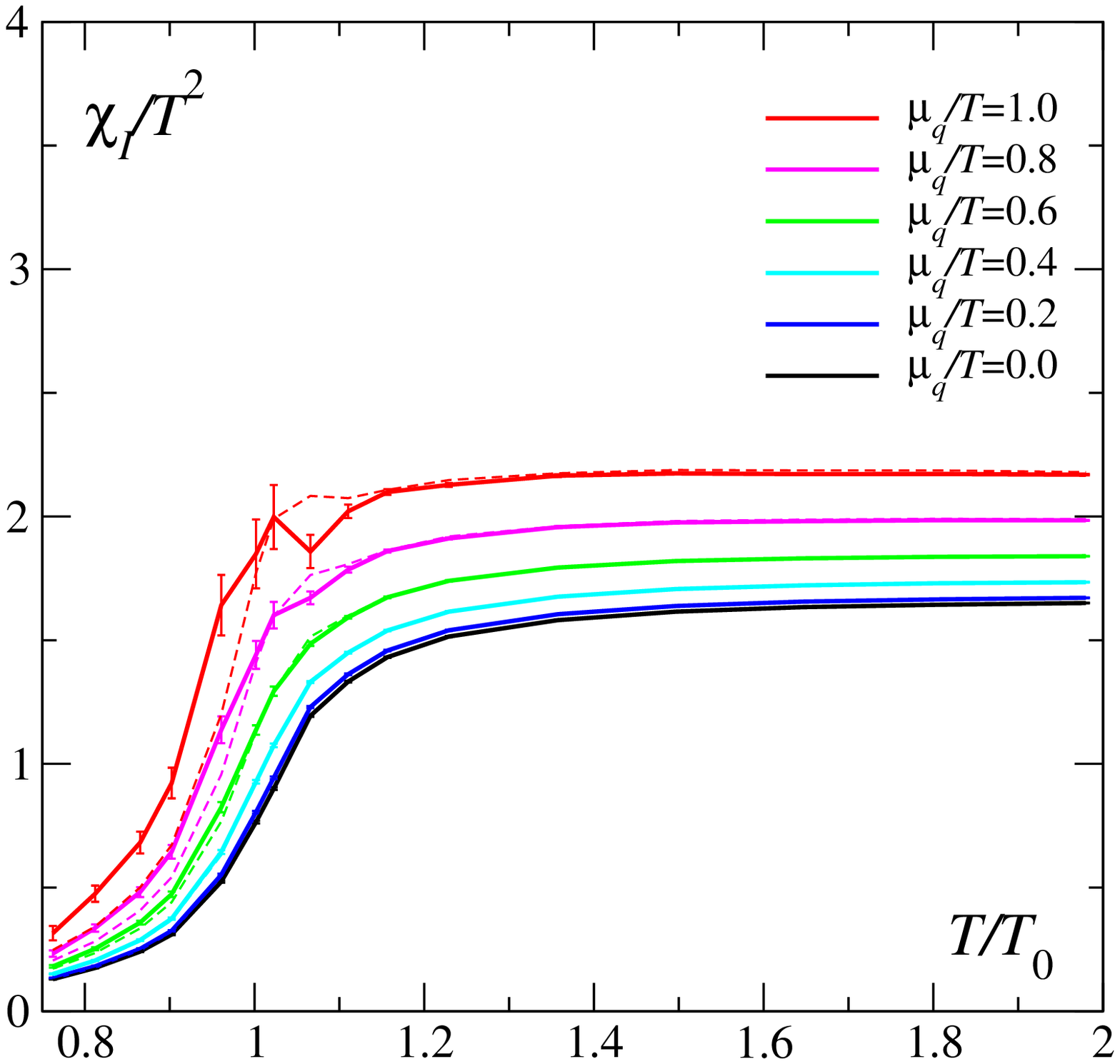,width=0.95\textwidth}
\end{minipage}
\end{center}
\caption{\it Isoscalar (left) and isovector (right) quark-number
  susceptibility for various quark chemical potentials, $\mu_q=\mu_B/3$,
  as computed in unquenched lattice QCD~\cite{Allton:2005gk}.}
\label{fig_chi-lat}
\end{figure}

Additional constraints from lattice QCD are provided by susceptibilities
which are defined as second-order derivatives of the thermodynamic
potential. In our context, the quark-number susceptibilities are of
special interest,
\begin{equation}
\chi_\alpha \sim \frac{\partial^2 \Omega}{\partial \mu_\alpha^2} \sim
\Pi_\alpha(q_0=0,q\to 0) \ ,
\end{equation}
which can be decomposed in isoscalar ($\mu_q=(\mu_u+\mu_d)/2$) and
isovector ($\mu_I=(\mu_u-\mu_d)/2$) channels carrying the quantum
numbers of the $\omega$ and $\rho$, respectively. The spacelike limits
of the correlators basically represent the screening masses in the
respective channels. Lattice QCD computations of the quark-number
susceptibilities indicate that both $\rho$ and $\omega$ channels behave
smoothly with temperature for small chemical potentials, see
Fig.~\ref{fig_chi-lat}.
However, as $\mu_q$ increases, $\chi_q$ develops a peak whereas $\chi_I$
remains smooth. The former indicates an increase in the (local)
baryon-number fluctuations and may be a precursor of the baryon-number
discontinuity between hadronic and QGP phase as one is approaching a
first-order line. Remarkably, this is not seen for the isospin
fluctuations.


\subsection{Sum Rules and Order Parameters}
\label{ssec_sum-rules}
Sum rules are currently the most promising tool to connect the
nonperturbative physics encoded in spectral functions to the
condensate structure of the QCD vacuum. In particular, the Weinberg sum
rules directly relate order parameters of $\chi$SR to the axial-/vector
spectral functions, which, in the medium, have not been exploited much 
to date. 

\subsubsection{Chiral Sum Rules}
The Weinberg and DMO sum rules~\cite{Weinberg:1967,Das:1967ek},
Eqs.~(\ref{csr}), directly relate moments of the ``vector minus
axialvector'' spectral functions to chiral order parameters.
This is a rather fortunate situation in view of the dominant role that
the isovector-vector ($\rho$) channel plays in dilepton production, recall
Eq.~(\ref{ImPi_em-vdm}). For $N_f$=2, the $\omega$ is a chiral singlet,
while in the strangeness sector ($\phi$), i.e., for $N_f$=3, chiral
symmetry becomes much less accurate (e.g., $\tave{\bar ss}$ persists
much farther into the QGP).

As has been shown in Ref.~\cite{Kapusta:1993hq}, the Weinberg sum rules
remain valid at finite temperature, albeit with two important
modifications induced by the breaking of Lorentz invariance caused
by the heat bath which defines a preferred rest frame: (i) each energy
sum rule applies for a fixed three-momentum, and (ii) at finite
three-momentum, the vector and axialvector spectral functions split 
into longitudinal and transverse modes,
\begin{equation}
\Pi_{V}^{\mu\nu} =  \Pi_{V,A}^{T} P_T^{\mu\nu} +
\Pi_{V,A}^{L} P_L^{\mu\nu} \ , 
\label{Pi-pol}
\end{equation}
with individual sum rules for each of them. The
explicit form is as follows:
\begin{eqnarray}
-\int\limits_0^\infty \frac {\dd q_0^2}{\pi (q_0^2-q^2)}
 \left[\im\Pi_V^L(q_0,q) - \im\Pi_A^L(q_0,q) \right]
&=& 0 ,
\label{wsr1med}
\\
-\int\limits_0^\infty \frac{\dd q_0^2}{\pi}
 \left[\im\Pi_V^{L,T}(q_0,q) - \im\Pi_A^{L,T}(q_0,q) \right]
&=& 0 ,
\label{wsr2med}
\\
-\int\limits_0^\infty q_0^2 \frac{\dd q_0^2}{\pi}
 \left[\im\Pi_V^{L,T}(q_0,q) - \im\Pi_A^{L,T}(q_0,q) \right]
&=& -2\pi \alpha_s \langle\langle {\cal O}_4 \rangle\rangle  \ .
\label{wsr3med}
\end{eqnarray}
In writing Eqs.~(\ref{wsr1med})-(\ref{wsr3med}) the pionic piece of
the (longitudinal) axialvector correlator has been absorbed into the
definition of the in-medium spectral function, $\im \Pi_A^{L}(q_0,q)$;
in the vacuum and in the chiral limit it is represented by a sharp 
state, $\im \Pi_\pi^{\mu\nu}$=$f_\pi^2 M^2 \delta(M^2)
P_L^{\mu\nu}$. In this form it only contributes to the first sum rule,
Eq.~(\ref{wsr1med}). However, in matter (and for $m_\pi$$>$0) this 
is no longer true since the pion is expected 
to undergo substantial medium effects.

The in-medium chiral sum rules constitute a rich source of constraints
on both energy and three-momentum dependence of in-medium spectral
functions.  The energy moments demonstrate that chiral restoration
requires degeneracy of the entire spectral functions. Combining lQCD
computations of order parameters with effective model calculations 
thus provides a promising synergy for deducing
chiral restoration from experiment~\cite{David:2006sr}.

\subsubsection{QCD Sum Rules}
\begin{figure}[!t]
\center
\epsfig{file=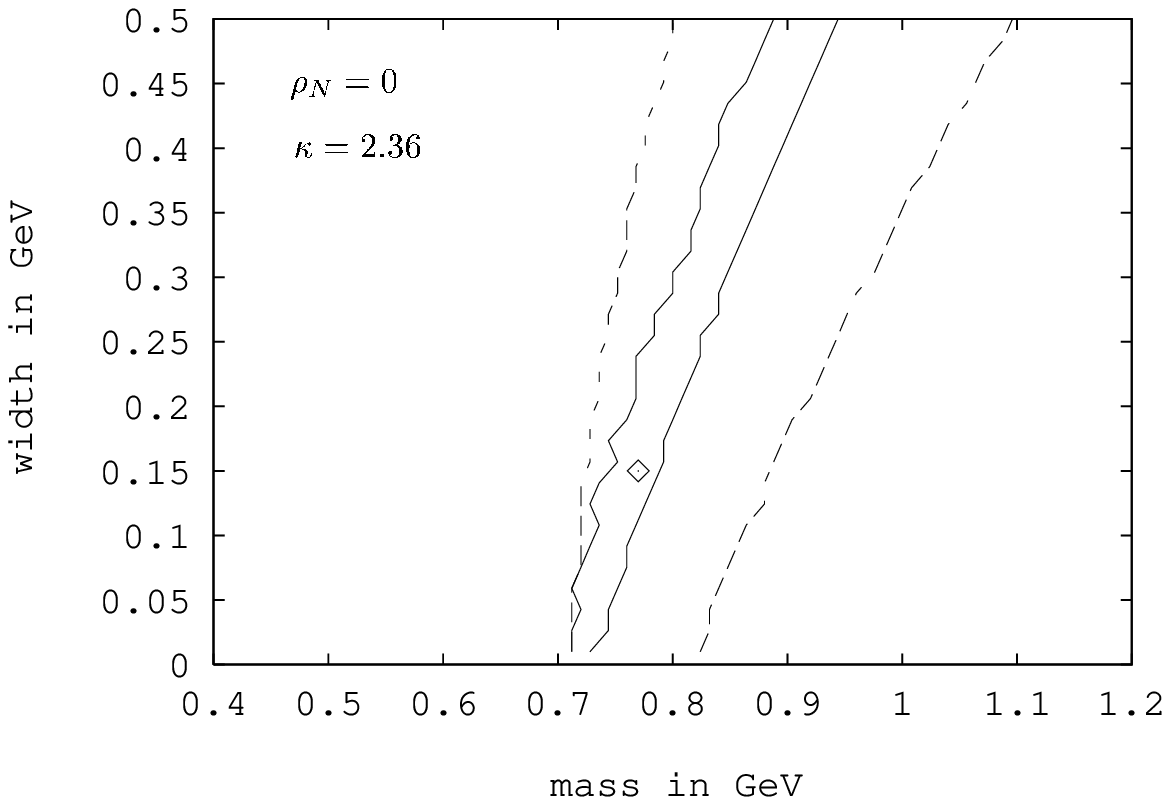,width=0.45\linewidth,angle=0}
\hspace{0.3cm}
\epsfig{file=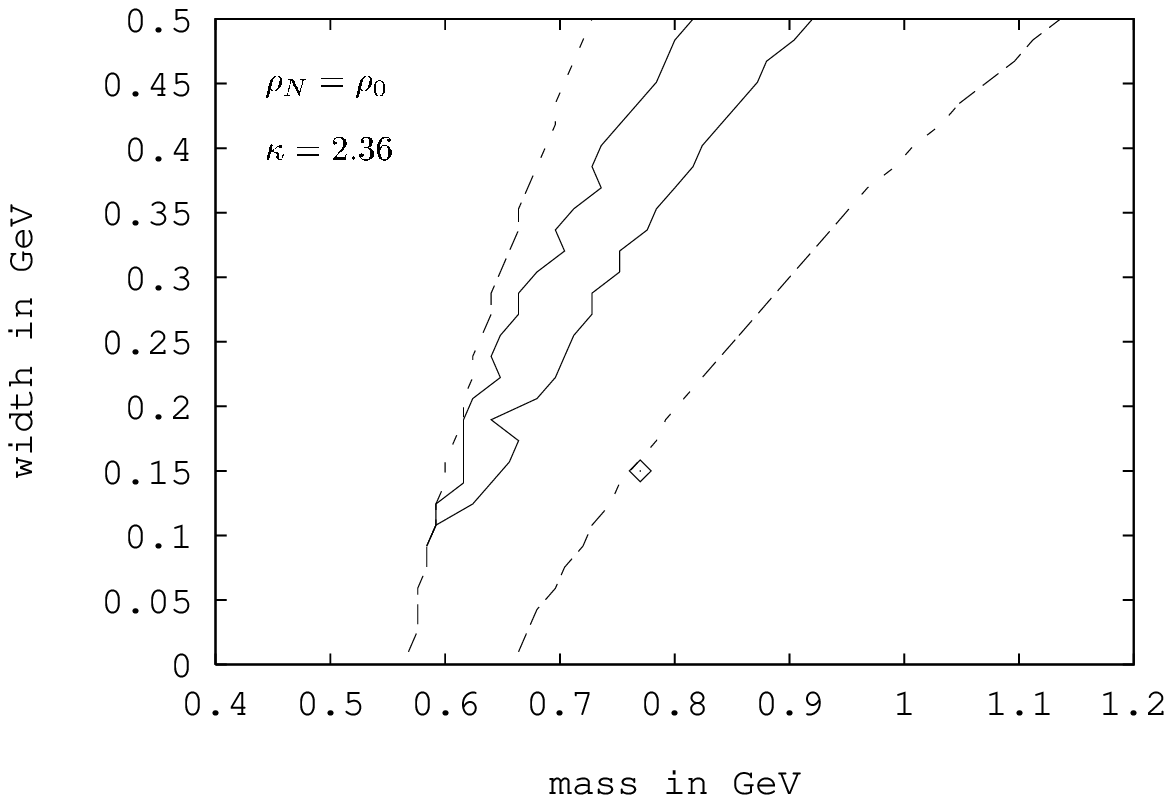,width=0.45\linewidth,angle=0}
\caption{\it QCD sum rule constraints on $\rho$ meson mass and width as
  inferred from Breit-Wigner parameterizations of its spectral
  function~\cite{Leupold:1997dg} (left panel: vacuum, right panel: cold
  nuclear matter at saturation density). ``Allowed'' regions of mass and
  width are indicated by the bands between solid and dashed curves,
  corresponding to maximal deviations between the l.h.s and r.h.s. of
  the SR of 0.2\% and 1\%, respectively.  The diamond depicts the
  corresponding vacuum parameters.} 
\label{fig_qcdsr}
\end{figure}

QCD sum rules are based on a (subtracted) dispersion relation for a
correlation function in a given hadronic channel $\alpha$, formulated 
for spacelike momenta $q^2$=$-Q^2$$<$0~\cite{Shifman:1978bx},
\begin{equation}
\Pi_\alpha(Q^2) = \Pi_\alpha(0) +\Pi_\alpha'(0)~Q^2
+ Q^4 \int \frac{\dd s}{\pi s^2} \frac{\im \Pi_\alpha(s)}{s+Q^2} \ .
\label{qcdsr}
\end{equation}
The right-hand-side ({\it rhs}) contains the spectral function which is
usually related to observables or evaluated in model calculations. On
the left-hand-side ({\it lhs}), the correlation function is expanded 
into a power series of $1/Q^2$ (operator-product expansion = OPE) where 
the (Wilson) coefficients contain perturbative contributions as well as
vacuum-expectation values of quark and gluon operators (the
nonperturbative condensates; for practical purposes the 
convergence of the OPE is improved by means of a so-called Borel
transformation which we do not discuss here). The explicit form of the
OPE for vector and axialvector correlators reads (truncating higher
order terms in $m_q$, $\alpha_s$, etc.)
\begin{eqnarray}
\frac{\Pi_V^{\rm vac}}{Q^2} = -\frac{1+\frac{\alpha_s}{\pi}}{8\pi^2}  
\ln\frac{Q^2}{\mu^2} + \frac{m_q \langle\bar qq\rangle} {Q^4} + 
\frac{1}{24Q^4} \ \langle\frac{\alpha_s}{\pi} {G_{\mu\nu}^a}^2 \rangle
- \frac{112\pi\alpha_s}{81Q^6} \ \kappa \ \langle\bar qq\rangle^2 + \cdots 
\label{Pi_V-ope}
\\
\frac{\Pi_A^{\rm vac}}{Q^2} = -\frac{1+\frac{\alpha_s}{\pi}}{8\pi^2}
\ln\frac{Q^2}{\mu^2} - \frac{m_q \langle\bar qq\rangle} {Q^4} +
\frac{1}{24Q^4} \ \langle\frac{\alpha_s}{\pi} {G_{\mu\nu}^a}^2\rangle
+ \frac{176\pi\alpha_s}{81Q^6} \ \tilde\kappa \ \langle{\bar qq}\rangle^2   
+ \cdots
\end{eqnarray} 
where the four-quark condensates have been approximated by factorizing
them into the squared two-quark condensate with parameters $\kappa$,
$\tilde\kappa$ which simulate intermediate states other than the ground
state (the scale $\mu$ is typically chosen around 1~GeV).  Note that
SB$\chi$S is nicely reflected by the opposite signs of the 
quark-condensate terms in $\Pi_V$ and $\Pi_A$, while the ``flavor-blind"
gluon condensate enters with the same sign. Qualitatively, the 
(positive) gluon condensate actually induces a softening of the spectral
function (i.e., a larger weight at small $s$ in the dispersion 
integral)~\cite{Hatsuda:1991ez}. On the other hand, for the vector 
channel, the negative contributions from the quark condensates on the 
{\it lhs} of the sum rule push spectral-function strength to larger $s$, 
relative to the axialvector channel (this may seem surprising in view 
of the masses of the pertinent resonances, $m_\rho$=0.77\,GeV vs.
$m_{a_1}$=1.23\,GeV; recall, however, that the (longitudinal) axialvector
channel contains a contribution from the axialvector current of the
pion). Inserting numerical values, $\alpha_s$=0.35,
$m_q$=0.005~GeV, $\langle{\bar qq}\rangle$=(-0.25\,GeV)$^3$ and
$\langle\frac{\alpha_s}{\pi} {G_{\mu\nu}^a}^2\rangle$=0.012\,GeV$^4$,
leads to 
\begin{eqnarray}
\frac{\Pi_V^{\rm vac}}{Q^2} = \frac{1}{8\pi^2} \left( 
-1.11 \ln\frac{Q^2}{\mu^2} - \frac{0.0062~{\rm GeV}^4}{Q^4} + 
\frac{0.039~{\rm GeV}^4}{Q^4}
- \frac{0.029~{\rm GeV}^6~\kappa}{Q^6}  \right) \ ,  
\end{eqnarray}
illustrating that the leading contributions arise from the gluon and
four-quark condensates (especially for typical values of
$\kappa$$\simeq$2.5), while the impact of the quark condensate is rather
moderate. For the vector channel, and in vacuum, there is a large
cancellation between the gluon and 4-quark condensate terms. However, in 
the medium this is presumably lifted, especially at low $T$ and $\rho_B$ 
where quark and gluon condensates change rather differently. The stronger 
reduction of the ``repulsive'' 4-quark condensate relative to the 
``attractive'' gluon condensate induces a softening of the spectral 
function in the dispersion integral. The softening can be satisfied by 
both broadening and/or a downward mass
shift~\cite{AK93,Klingl:1997kf,Leupold:1997dg,Ruppert:2005id,Kwon:2008vq}.
Quantitative studies (which also include effects of non-scalar
condensates induced by hadron structure of the heat-bath particles)
based on Breit-Wigner model spectral functions are displayed in
Fig.~\ref{fig_qcdsr}.
For the axialvector channel, the reduction in both condensates suggests
a substantial loss of soft-mode strength which points at the dissolution
of the pion mode (whose polestrength is given by $f_\pi$) as a
consequence of (the approach toward) $\chi$SR.

Finally, it is instructive to compare $\omega$ and $\rho$ mesons: while
their OPE side is rather similar (governed by 4-quark condensates),
the subtraction constant, $\Pi_V(0)=\varrho_N/4M_N$ to leading order in
$\varrho_N$, makes a difference. It is given by the Thompson limit of
the $VN$ scattering amplitude and turns out to be identical in the
$\rho$ and $\omega$ sum rule. However, since Im~$\Pi_\rho$ is larger
than Im~$\Pi_\omega$ by an isospin factor of
$(g_\omega/g_\rho)^2$$\simeq$9 (recall Eq.~(\ref{ImPi_em-vdm})), the
finite-$\varrho_N$ subtraction actually stabilizes the $\omega$ sum rule,
implying stronger medium effects (softening) on the
$\rho$ than on the $\omega$ (it amounts to a ``repulsive'' contribution
on the OPE side counterbalancing the reduction in the 4-quark
condensate).

\subsection{Medium Effects II: Chiral Effective Models}
\label{ssec_eff-mod}
Model-independent and/or low-density approaches as discussed above
provide valuable constraints on the vector and axialvector correlators 
and their connections to QCD vacuum structure. However, 
quantitative calculations suitable for comparison with experiment
require the construction of effective models. As indicated in the
Introduction, in the low-mass region most of the  thermal dilepton yield
in heavy-ion collisions is expected to emanate from the hot/dense
hadronic phase (even at collider energies), especially from the $\rho$
channel.  Hadronic chiral Lagrangians are therefore a suitable
starting point, extended by the implementation of the low-lying vector
mesons. This is usually done by a local gauging procedure of the chiral
pion Lagrangian, thus realizing the gauge principle at the composite
(hadronic) level. The most common approaches are based on non-linear
realizations of chiral symmetry (i.e., without explicit $\sigma$
meson) within the Hidden Local Symmetry (HLS)~\cite{Bando:1984ej} or
Massive Yang Mills (MYM)~\cite{Gomm:1984at} schemes.  Rather than
reviewing these in a comprehensive form, we here focus on recent
developments with relevance for dilepton production, i.e., the ``vector
manifestation'' (VM) scenario of SB$\chi$S within
HLS~\cite{Harada:2003jx} (Sec.~\ref{sssec_vm}), as well as hadronic
many-body theory within MYM (Sec.~\ref{sssec_hmbt}).

\subsubsection{Hidden Local Symmetry and Vector Manifestation}
\label{sssec_vm}
\begin{figure}[!t]
\begin{minipage}{0.5\textwidth}
\epsfig{file=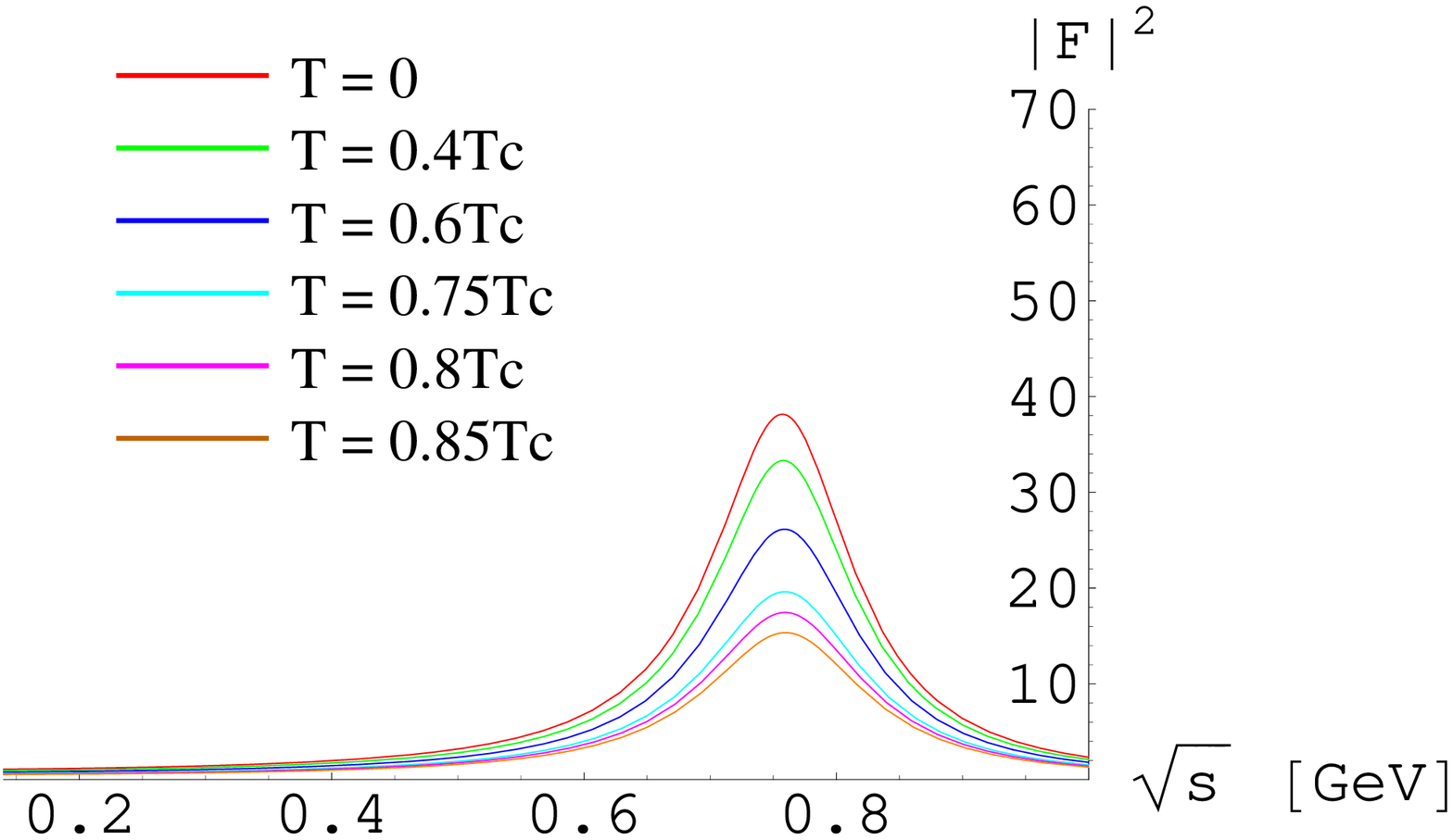,width=0.98\textwidth}
\end{minipage}
\begin{minipage}{0.5\textwidth}
\epsfig{file=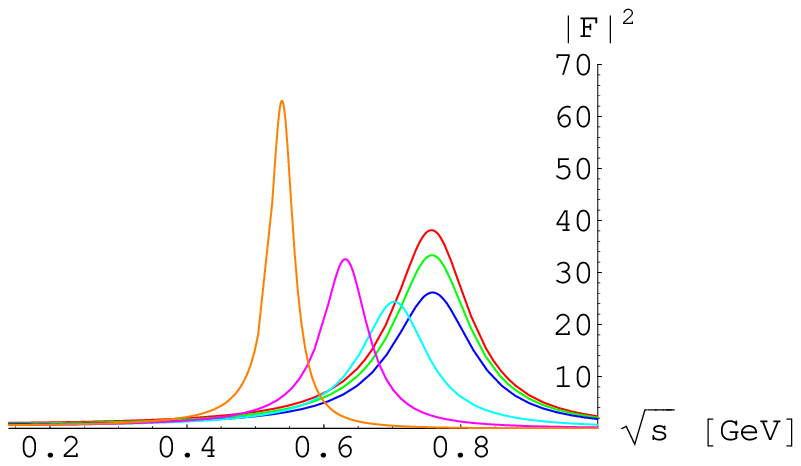,width=0.98\textwidth}
\end{minipage}
\caption{\it Pion EM formfactor at finite temperature in the HLS vector
manifestation framework~\cite{Harada:2006hu}; left panel: with finite-$T$ 
loop effects; right panel: additionally including a $T$-dependence of
the bare $\rho$ mass above $T$=0.7\,$T_c$.}
\label{fig_Piem-vm}
\end{figure}

Within the HLS framework, an alternative realization of chiral symmetry
in the meson spectrum has been suggested in 
Ref.~\cite{Harada:2003jx}, by identifying the chiral partner of the 
pion with the (longitudinal) $\rho$ (rather than with the $\sigma$).
This ``vector manifestation" of chiral symmetry has been shown to give 
a satisfactory phenomenology of hadronic and EM decay branchings in 
the vacuum.  When applied within a finite-$T$ loop expansion, the 
$\rho$-meson mass was found to be affected at order $T^4$ (consistent
with chiral symmetry), showing a slightly repulsive shift. However, when 
matching the hadronic axial-/vector correlators to pQCD in the spacelike 
regime (using an OPE), a reduction of the bare $\rho$ mass has been 
inferred, consistent with ``Brown-Rho" scaling~\cite{Brown:1991kk}.
In addition, vector dominance was found to be violated in the medium,
leading to a gradual decoupling of the $\rho$ from the EM current
toward the critical temperature. However, the finite-$T$ EM formfactor, 
which determines the dilepton production rate, clearly shows the downward 
moving $\rho$ peak~\cite{Harada:2006hu}, see Fig.~\ref{fig_Piem-vm}.
An interesting question is how these features develop in the presence
of finite baryon density.

\subsubsection{Massive Yang-Mills and Hadronic Many-Body Theory}
\label{sssec_hmbt}
As in HLS, the basic building block of the MYM Lagrangian is the chiral
pion Lagrangian based on the unitary pion field,
\begin{equation}
U=\exp(i\sqrt{2}\phi/f_\pi) \ ,\qquad
\phi\equiv \phi_a\frac{\tau_a}{\sqrt{2}} \ .
\label{ufield}
\end{equation}
Hadronic gauge fields, $A^\mu_{L,R}$ are introduced via the covariant
derivative,
\begin{equation}
D^\mu U=\partial^\mu-ig (A_L^\mu U - U A_R^\mu)
\end{equation}
and supplemented with kinetic and mass terms (with bare mass $m_0$).
One has
\begin{eqnarray}
{\cal L}_{\rm mym} &=& \frac{1}{4} f_\pi^2 \
\tr \left[ D_\mu U D^\mu U^\dagger\right]
-\frac{1}{2} \tr \left[ (F_L^{\mu\nu})^2+(F_R^{\mu\nu})^2\right]
+m_0^2 \ \tr \left[(A_L^\mu)^2+(A_R^\mu)^2\right]
\nonumber\\
 && -i\xi \ \tr \left[D_\mu U D^\mu U^\dagger F_L^{\mu\nu} +
D_\mu U D^\mu U^\dagger F_R^{\mu\nu} \right] +
\sigma \ \tr \left[F_L^{\mu\nu} U F_{R \mu\nu} U^\dagger \right] \ , 
\label{Lmym}
\end{eqnarray}
where the last two (non-minimal) terms are necessary to achieve a
satisfactory phenomenology in the vacuum.  After the identifications
$\rho^\mu\equiv V^\mu$=$A_R^\mu +A^\mu_L$, $A^\mu$=$A_R^\mu -A^\mu_L$ 
(and a field redefinition of the axialvector field to remove a
$\partial^\mu\vec\pi A^\mu$ term), the leading terms of the MYM
Lagrangian take the form
\begin{eqnarray}
{\cal L}_{\rm mym}&=& \frac{1}{2} m_\rho^2 \vec{\rho}_\mu^2
+\frac{1}{2}\left[m_\rho^2+g^2 f_\pi^2\right] \vec{a_1}^2_\mu +
g^2 f_\pi \vec{\pi} \times \vec{\rho}^\mu \cdot \vec{a_1}_\mu +
\nonumber\\
 & & g_{\rho\pi\pi}^2 \left[ \vec{\rho}_\mu^2 \vec{\pi}^2 
-\vec{\rho}^\mu \cdot \vec{\pi} \ \vec{\rho}_\mu \cdot \vec{\pi} \right] 
+ g_{\rho\pi\pi} \vec\rho_\mu \cdot (\vec\pi\times\partial^\mu\vec\pi) 
+ \dots 
\label{Lmym2}
\end{eqnarray}
($g_{\rho\pi\pi}^2$=$\frac{1}{2}g^2$). Note that the
Higgs mechanism induces the {\em splitting} of $\rho$ and $a_1$ masses,
\begin{equation}
m_{a_1}^2 = m_\rho^2 + g^2 f_\pi^2 \ , \quad m_\rho^2=m_0^2 \ ,  
\end{equation}
which is entirely due to SB$\chi$S (via $f_\pi$). The bare $\rho$ mass 
itself is an {\em external} parameter, which is different from the
HLS scheme discussed in the previous section.  Electromagnetism is
readily included into the MYM Lagrangian by adding the vector dominance
coupling~\cite{Gomm:1984at}
\begin{equation}
{\cal L}_{\rho\gamma} = \frac{em_\rho^2}{g_{\rho\pi\pi}} \ B_\mu 
\ \rho_3^\mu \ ,  
\end{equation}
where $B_\mu$ denotes the photon field. In this scheme, VDM remains
valid in the medium, and the task of computing the low-mass isovector
axial/-vector correlators amounts to assessing the medium modifications
of $\rho$ and $a_1$ mesons.

The $\rho$-meson propagator in hot and dense hadronic
matter can be written as
\begin{equation}
D_\rho^{L,T}(q_0,q;\mu_B,T) = 
\frac{1}{M^2-m_V^2-\Sigma_{\rho\pi\pi}^{L,T}-\Sigma_{\rho M}^{L,T}
-\Sigma_{\rho B}^{L,T}}
 \ ,  
\end{equation}
with transverse and longitudinal modes as defined in Eq.~(\ref{Pi-pol}).
The key quantities are the in-medium selfenergies, $\Sigma_\rho^{L,T}$,
which may be classified as follows (cf.~Fig.~\ref{fig_self-dia}):
\begin{figure}[!t]
\begin{center}
\epsfig{file=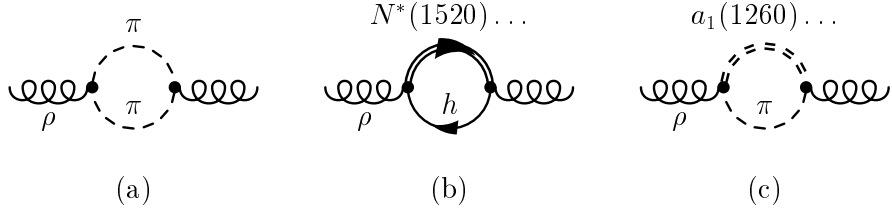,width=0.8\textwidth}
\end{center}
\vspace{-0.5cm}
\caption{\it Graphical representation of self-energy diagrams
  characterizing the interactions of the $\rho$ meson in hot and dense
  hadronic matter: (a) renormalization of its pion cloud due to modified
  pion propagators, and direct interactions of the $\rho$ meson with (b)
  baryons and (c) mesons, typically approximated by baryon- and
  meson-resonance excitations~\cite{Urban:1999im,Rapp:1999us}.  }
\label{fig_self-dia}
\end{figure}
$\Sigma_{\rho\pi\pi}$ accounts for the pion cloud of the $\rho$, which 
in the vacuum gives rise to its finite width via $\rho\to\pi\pi$. 
Direct interactions of the $\rho$ with mesons ($M$=$\pi$, $K$, $\rho$, 
$\dots$) and baryons ($B$=$N$, $\Lambda$, $\Delta$, $\dots$) from the 
heat bath are represented by $\Sigma_{\rho M}$ and $\Sigma_{\rho B}$, 
respectively; they vanish in the vacuum.
In terms of underlying scattering processes, the latter are typically
resonance excitations (e.g., $\rho\pi\to a_1$ or $\rho N\to N(1520)$)
while medium modifications of pions (e.g., $\pi N\to\Delta$) in
$\Sigma_{\rho\pi\pi}$ correspond to, e.g., $t$-channel $\pi$
exchange processes ($\rho N\to\pi\Delta$). When evaluating interactions
which are not directly constrained by chiral (or gauge) symmetry
(especially those involving higher resonances), phenomenological
information is essential for a reliable determination of the
coupling constants (and cutoff parameters in the hadronic formfactors
to account for the finite size of the hadrons). The simplest form of 
such constraints are hadronic decay widths of resonances 
(e.g., $a_1\to\rho\pi$), supplemented by
radiative decays (e.g., $a_1\to\gamma\pi$). However, especially for
``subthreshold'' states (e.g., $\omega\to\rho\pi$ or $N(1520)\to\rho N$),
where the coupling is realized via the low-energy ($\pi\pi$ decay) tail
of the $\rho$ spectral function, empirical information can be rather
uncertain. In this case, comprehensive constraints inferred from
scattering data become invaluable. Unfortunately, in practice this is
only possible for $\rho N$ interactions (e.g., via $\pi N\to \rho N$ or
$\gamma N$ scattering), but, as it turns out, the modifications of the
$\rho$ due to interactions with nucleons are generally stronger than 
with pions. In addition, by using nuclear targets, one has the 
possibility to constrain (or test) the modifications in nuclear 
{\em matter}, rather than on a single nucleon (which corresponds 
to the leading-order density effect).
     
Let us start by discussing finite-$T$ effects. Calculations of the
$\rho$ propagator in a hot pion gas based on the MYM
scheme~\cite{Song:1995ga} have shown small medium effects. An extended
analysis~\cite{Rapp:1999qu} of the $\rho$ in hot meson matter, including 
resonance excitations ($\rho\pi\to a_1, \omega,h_1, \pi', a_2(1320)$, 
$\rho K\to K^*, K_1$, $\rho\rho\to f_1(1285)$) and pion Bose 
enhancement in $\Sigma_{\rho\pi\pi}$, leads to total broadening
of $\sim$80\,MeV at $T$=150\,MeV (corresponding to a pion density
$\varrho_\pi$=0.12\,fm$^{-3}$$\simeq$0.75\,$\varrho_0$), with little mass
shift. Approximately $\sim$20\,MeV of the broadening is due to the
$\pi\pi$ Bose factor (cf.~also Ref.~\cite{Dobado:2002xf}) and
$\sim$50-60\,MeV due to meson resonances. The latter is comparable to
Refs.~\cite{Haglin:1994xu,Eletsky:2001bb} which are directly based on
$\rho\pi$ and $\rho K$ scattering amplitudes.

\begin{figure}[!t]
\begin{minipage}{0.57\linewidth}
\epsfig{file=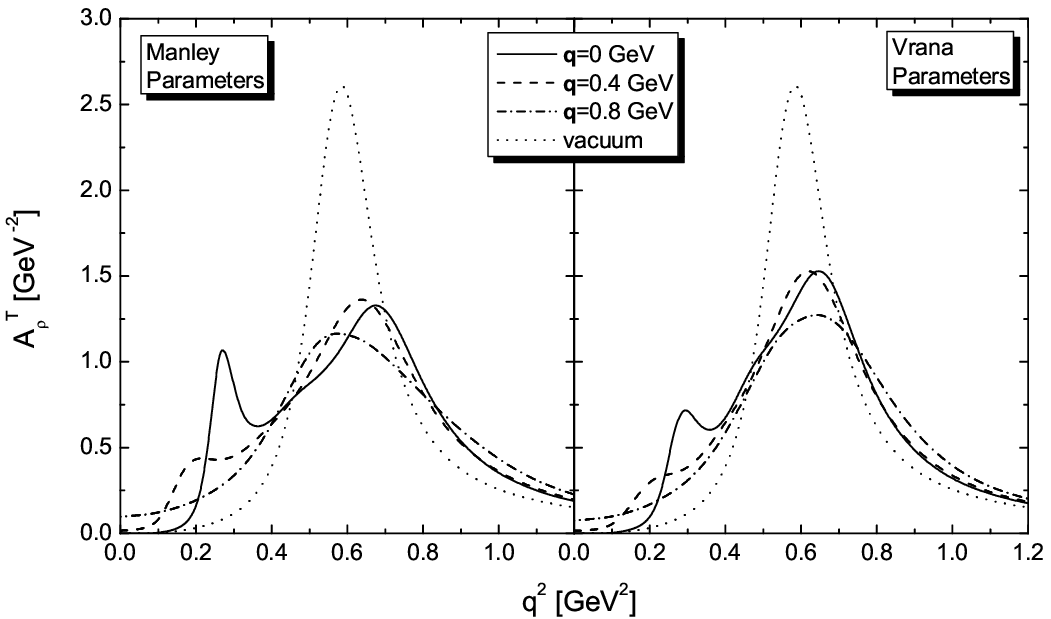,width=\textwidth,angle=0}
\end{minipage}
\hfill
\begin{minipage}{0.4\textwidth}
\vspace{-0.4cm}
\epsfig{file=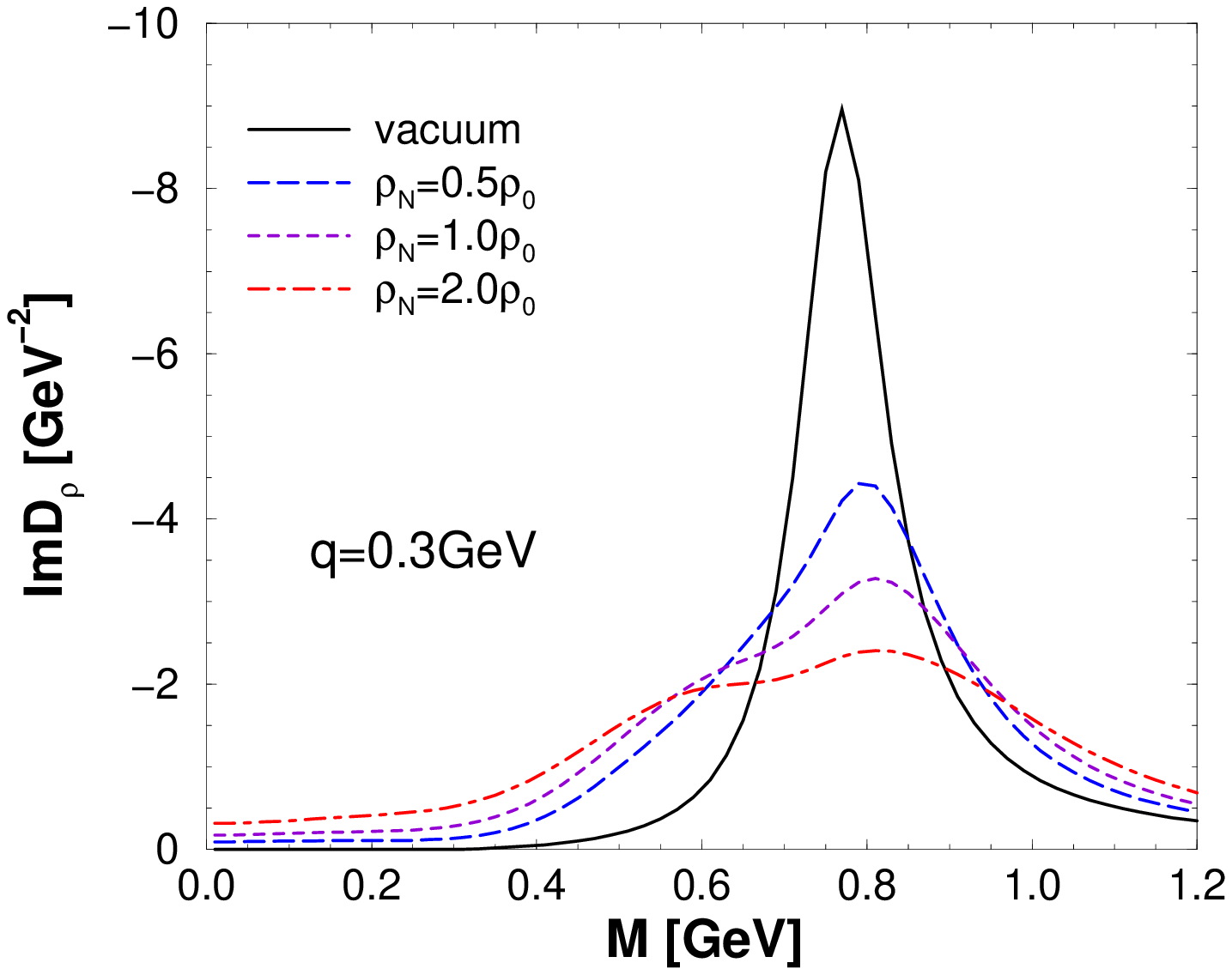,width=\textwidth,angle=0}
\end{minipage}
\caption{\it Comparison of $\rho$-meson spectral functions in cold
  nuclear matter within the hadronic many-body approaches of
  Refs.~\cite{Post:2003hu} (left panels, based on two different phase
  shift analysis of $\pi N$
  scattering~\cite{Manley:1984jz,Vrana:1999nt}) and
  \cite{Urban:1999im,Rapp:1999us} (right panel).  }
\label{fig_Arho-cold}
\end{figure}
\begin{figure}[!b]
\vspace{0.3cm}
\center
\epsfig{file=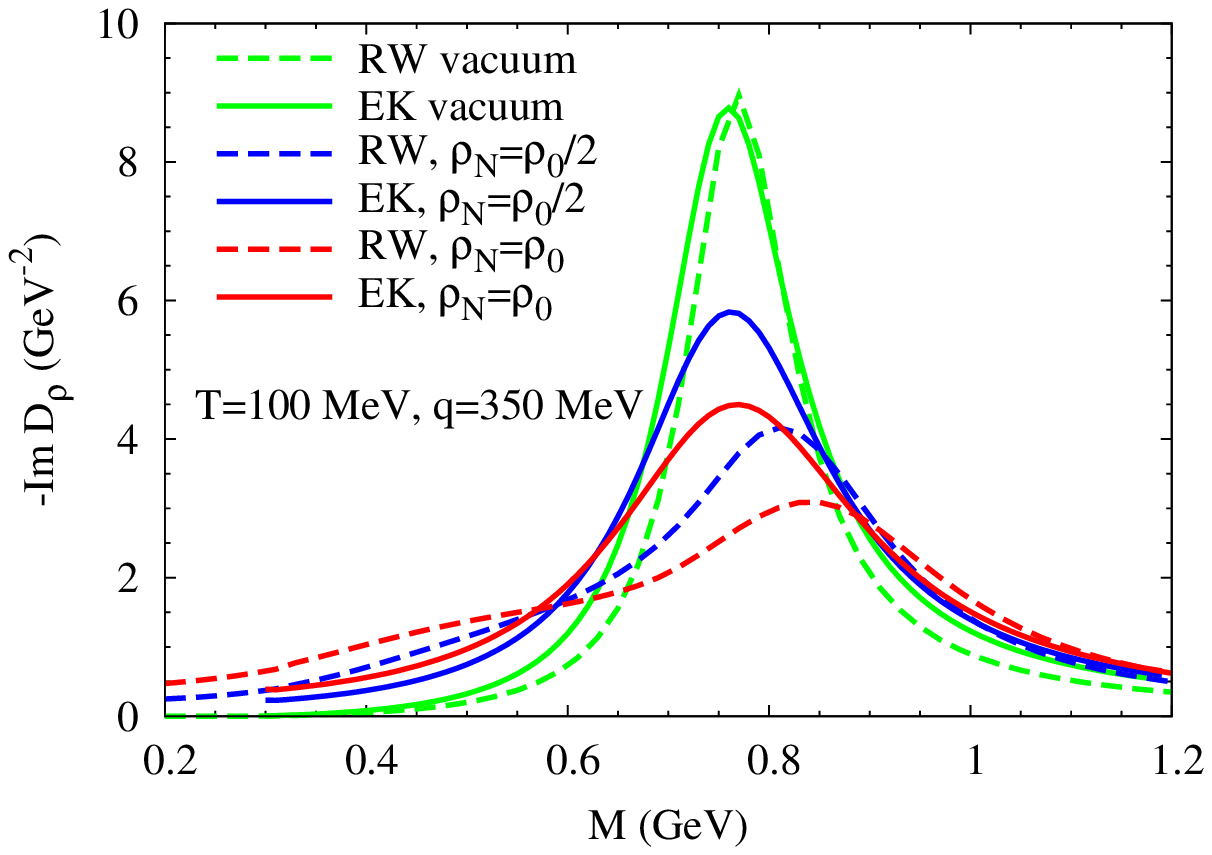,width=0.48\linewidth,angle=0}
\hspace{0.2cm}
\epsfig{file=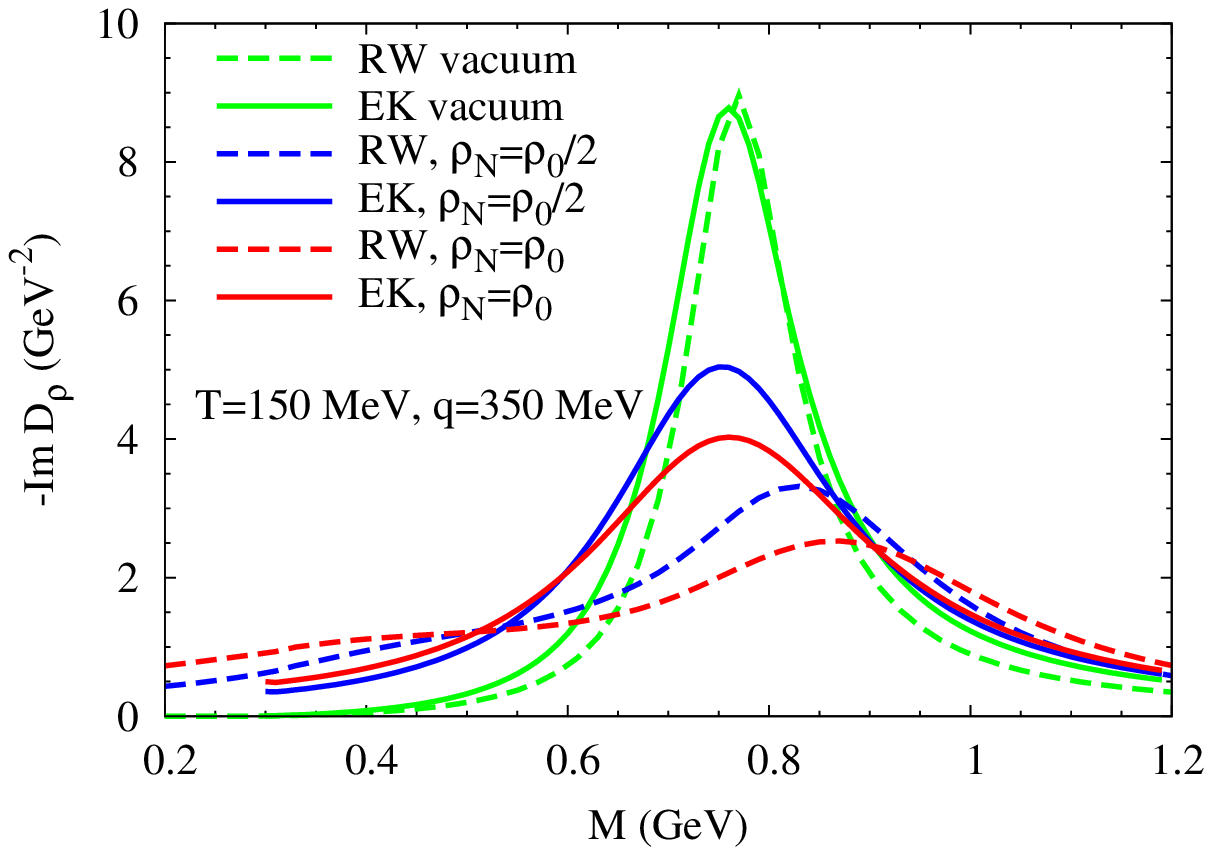,width=0.48\linewidth,angle=0}
\caption{\it Comparison of $\rho$-meson spectral functions in hot and
  dense hadronic matter from Refs.~\cite{Rapp:1999us} (dashed lines) and
  \cite{Eletsky:2001bb} (solid lines) at temperatures of $T$=100\,MeV
  (left panel) and $T$=150\,MeV (right panel) for {\em nucleon} densities
  of $\varrho_N$=0.5,1.0~\,$\varrho_0$ (the corresponding nucleon chemical
  potentials are $\mu_N$=673,745\,MeV for $T$=100\,MeV and
  $\mu_N$=436,542\,MeV for $T$=150\,MeV, respectively).}
\label{fig_Arho-had}
\end{figure}
Next, we turn to modifications in cold nuclear matter.
Fig.~\ref{fig_Arho-cold} shows two calculations in which the underlying
$\rho$ self-energies have been rather thoroughly constrained. In
Ref.~\cite{Post:2003hu} (left panels), a $\rho N$ resonance model
(corresponding to $\Sigma_{\rho N}$) has been constructed utilizing a
detailed analysis of empirical $\pi N\to\rho N$ phase shifts and
inelasticities~\cite{Manley:1984jz,Vrana:1999nt}. The resulting $\rho$
spectral functions are displayed at normal nuclear density (taken as
$\varrho_N$=0.15\,fm$^{-3}$) for various three-momenta and two distinct
data sets for constraints. A substantial broadening of close to
$\sim$200\,MeV is found, with a slight upward peak shift of a few tens of
MeV; the three-momentum dependence is relatively 
weak. In Refs.~\cite{Urban:1999im,Rapp:1999us}, $\Sigma_{\rho N}$ and 
an in-medium pion cloud, $\Sigma_{\rho\pi\pi}$ (incorporating $P$-wave 
``pisobar'' nucleon- and $\Delta$-hole excitations and
associated vertex corrections), have been calculated and constrained by
total photoabsorption data on the nucleon and nuclei~\cite{Rapp:1997ei},
as well as total $\pi N\to\rho N$ cross sections. The resulting spectral
functions are quite similar to the ones of Ref.~\cite{Post:2003hu}, with
a somewhat stronger broadening of $\sim$300\,MeV at 
$\varrho_N$=0.16\,fm$^{-3}$ and a comparable mass shift of $\sim$40\,MeV. 
It is quite remarkable that the predicted in-medium mass and width of
$\sim$(810,450)\,MeV are in good agreement with the QCD sum rule
constraints derived in Ref.~\cite{Leupold:1997dg}, cf.~right panel in 
Fig.~\ref{fig_qcdsr}. Both broadening and mass shift decrease at higher 
three-momentum, e.g., 
($\Delta m_\rho,\Delta\Gamma_\rho$)$\simeq$(30,150)\,MeV at $q$=1\,GeV.  
Both calculations~\cite{Post:2003hu,Urban:1999im} include a rather
strong coupling to $\rho N(1520)N^{-1}$ excitations (appearing as 
a low-mass peak or shoulder in the $\rho$ spectral function).
This has been questioned in Ref.~\cite{Lutz:2001mi} based on a coupled 
channel analysis of $S$-wave $\rho N$ and $\omega N$ scattering, where
all nucleon resonances but the $\Delta$(1232) are generated dynamically
via four-point interactions. The (generated) $N(1520)$ is deduced to 
primarily couple
to $\omega N$ rather than $\rho N$, entailing an in-medium $\rho$
with significantly less broadening.

\begin{figure}[!t]
\begin{minipage}{0.5\textwidth}
\epsfig{file=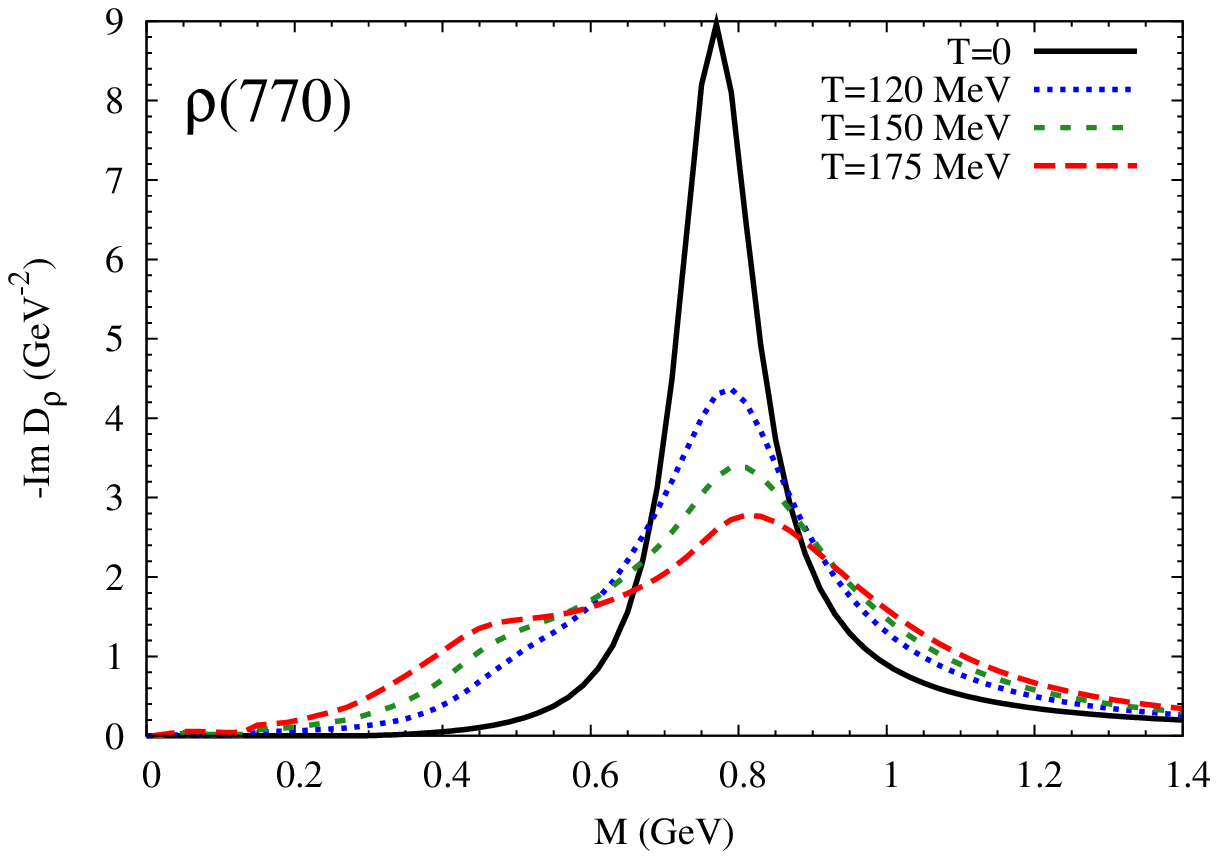,width=0.98\linewidth,angle=0}
\end{minipage}
\hspace{0.1cm}
\begin{minipage}{0.5\textwidth}
\epsfig{file=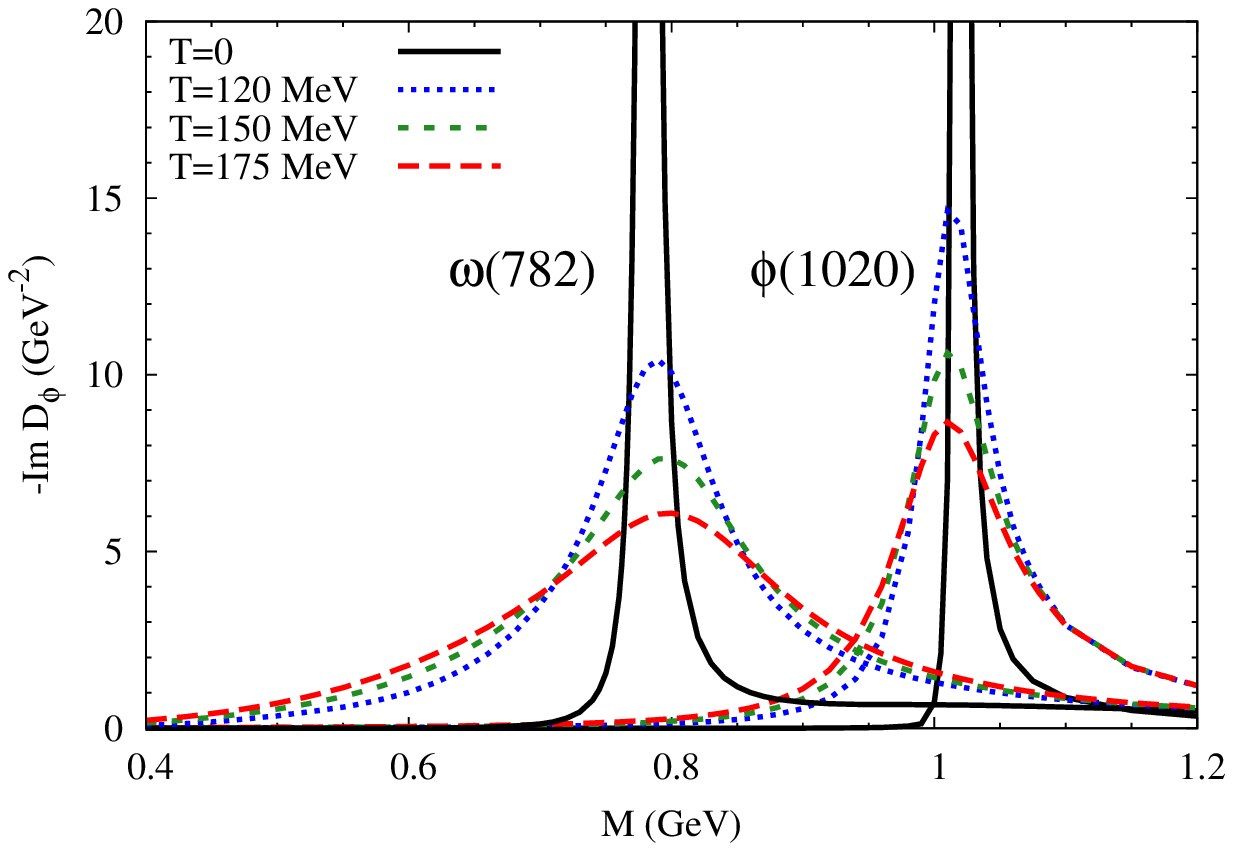,width=0.98\linewidth,angle=0}
\end{minipage}
\vspace{0.3cm}
\begin{minipage}{0.5\textwidth}
\vspace{0.7cm}
\epsfig{file=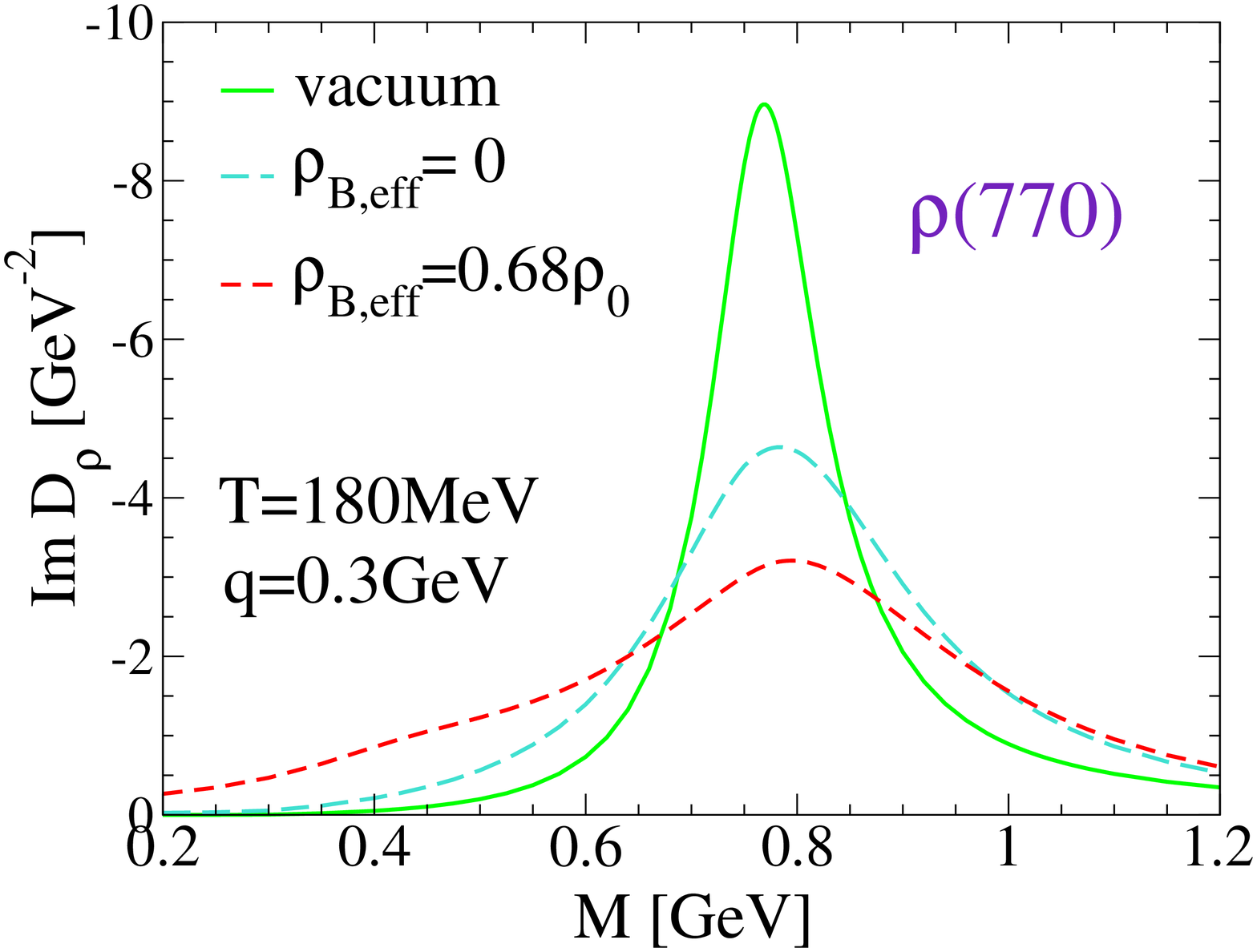,width=0.98\linewidth,angle=0}
\end{minipage}
\hspace{0.1cm}
\begin{minipage}{0.5\textwidth}
\vspace{0.7cm}
\epsfig{file=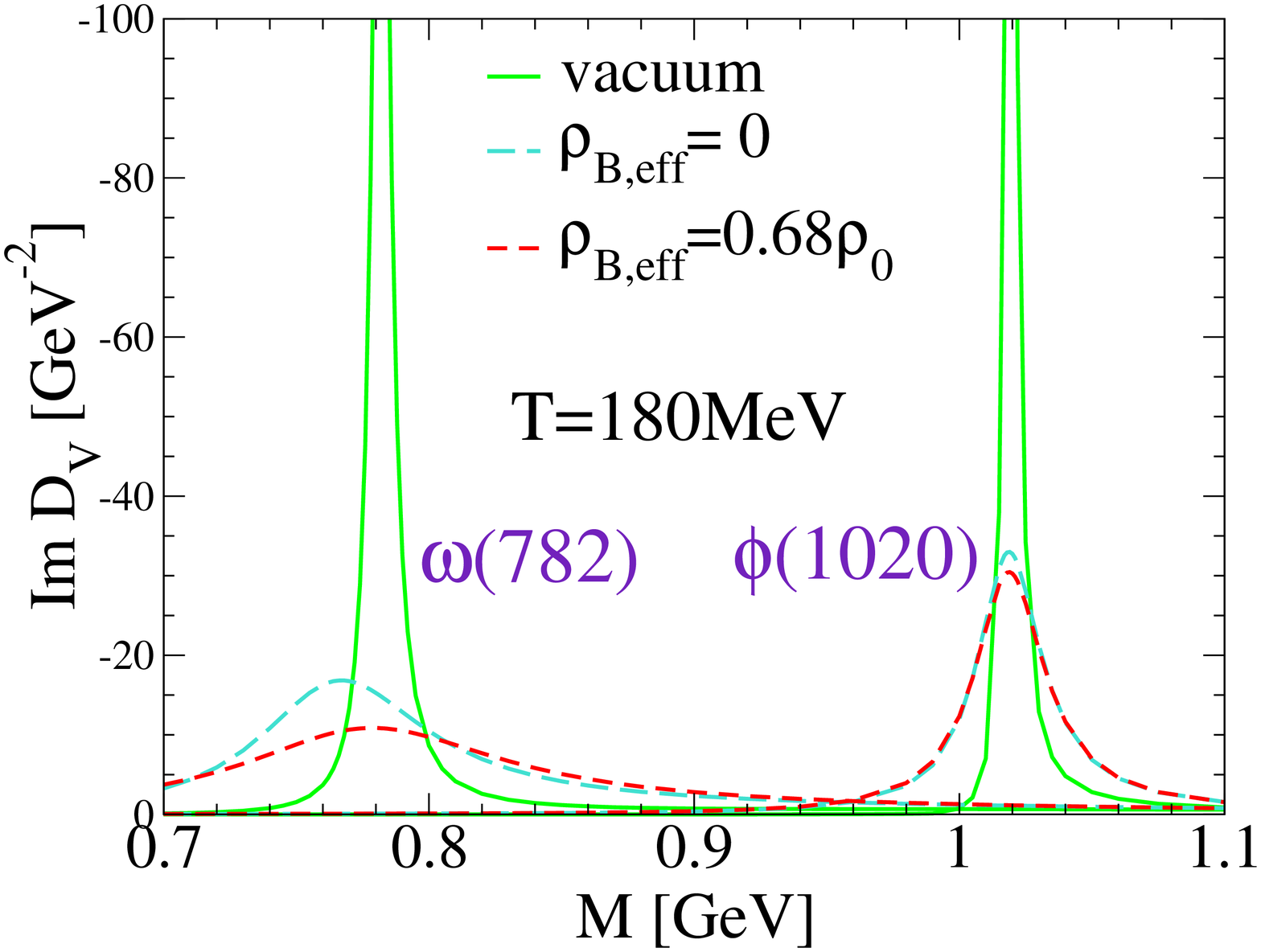,width=0.98\linewidth,angle=0}
\end{minipage}
\caption{\it In-medium spectral functions of light vector mesons in hot
  and dense hadronic matter based on the hadronic many-body approach of
  Refs.~\cite{Rapp:1997ei,Rapp:1999us,Rapp:2000pe}; upper panel: under
  conditions resembling heavy-ion collisions at SPS (i.e., along an
  isentropic trajectory in the phase diagram which preserves the
  measured hadron ratios determined at $(\mu_B^{\rm chem},T_{\rm chem})
  \simeq (230,175)$~MeV)~\cite{vanHees:2007th}; lower panels: at
  $(\mu_B,T)\simeq (25,180)$~MeV (resembling chemical freeze-out at
  RHIC) with (short-dashed lines) and without (long-dashed lines) medium
  effects induced by anti-/baryons~\cite{Rapp:2004zh}.}
\label{fig_Arho69}
\end{figure}
Finally, we turn to a hot and dense hadronic medium as
expected to be formed in high-energy heavy-ion collisions. In
Fig.~\ref{fig_Arho-had} the $\rho$ spectral functions of the hadronic
many-body calculations~\cite{Urban:1999im,Rapp:1999us} are compared to
those obtained in the scattering-amplitude approach of
Ref.~\cite{Eletsky:2001bb}. The latter exhibit less broadening and
a small (if any) downward mass shift of the $\rho$ peak,
compared to the upward shift in the many-body approach (mostly
induced by baryonic effects). Consequently, in terms of spectral
strength, the discrepancies between the two calculations are largest
for masses around $M$$\simeq$0.7\,GeV, as well as for very low mass,
$M$$\le$0.4\,GeV, by up to a factor of $\sim$2. This mass region is
much magnified in thermal dilepton production rates due to the
Boltzmann factor and a photon propagator $\propto 1/M^2$. However,
the amplitude approach only accounts for interactions with pions and
nucleons, while the many-body calculations include estimates of $\rho$
interactions with strange baryons and resonances~\cite{Rapp:1999us}.
This difference may account for some of the discrepancy.
 
In preparation for applications to dilepton spectra in URHICs, we
summarize in Fig.~\ref{fig_Arho69} in-medium $V$-meson spectral
functions in the many-body approach under conditions relevant for SPS
(upper panels)~\cite{Rapp:1999us,vanHees:2007th} and RHIC (lower
panels)~\cite{Rapp:2000pe}. The $\rho$ meson (left
panels) ``melts'' when extrapolated to temperatures close to
the expected phase boundary. Baryons play an essential role in the
melting, even at RHIC (where the net baryon density is small), since the
relevant quantity is the {\em sum} of baryon and antibaryon
densities. The effects due to baryons and antibaryons are most 
prominent as an enhancement in the mass region below
$M$$\simeq$0.5\,GeV. The $\omega$ and especially $\phi$ spectral
functions (right panels) appear to be more robust. One should also point
out that at $T$=120,150\,MeV in the upper panels appreciable pion 
and kaon chemical potentials are present which sustain
larger hadron densities and thus support stronger medium effects than in
chemical equilibrium.

\subsection{Thermal Dilepton and Photon Rates}
\label{ssec_rates}
\begin{figure}[!t]
\begin{center}
\begin{minipage}[]{0.48\linewidth}
\epsfig{file=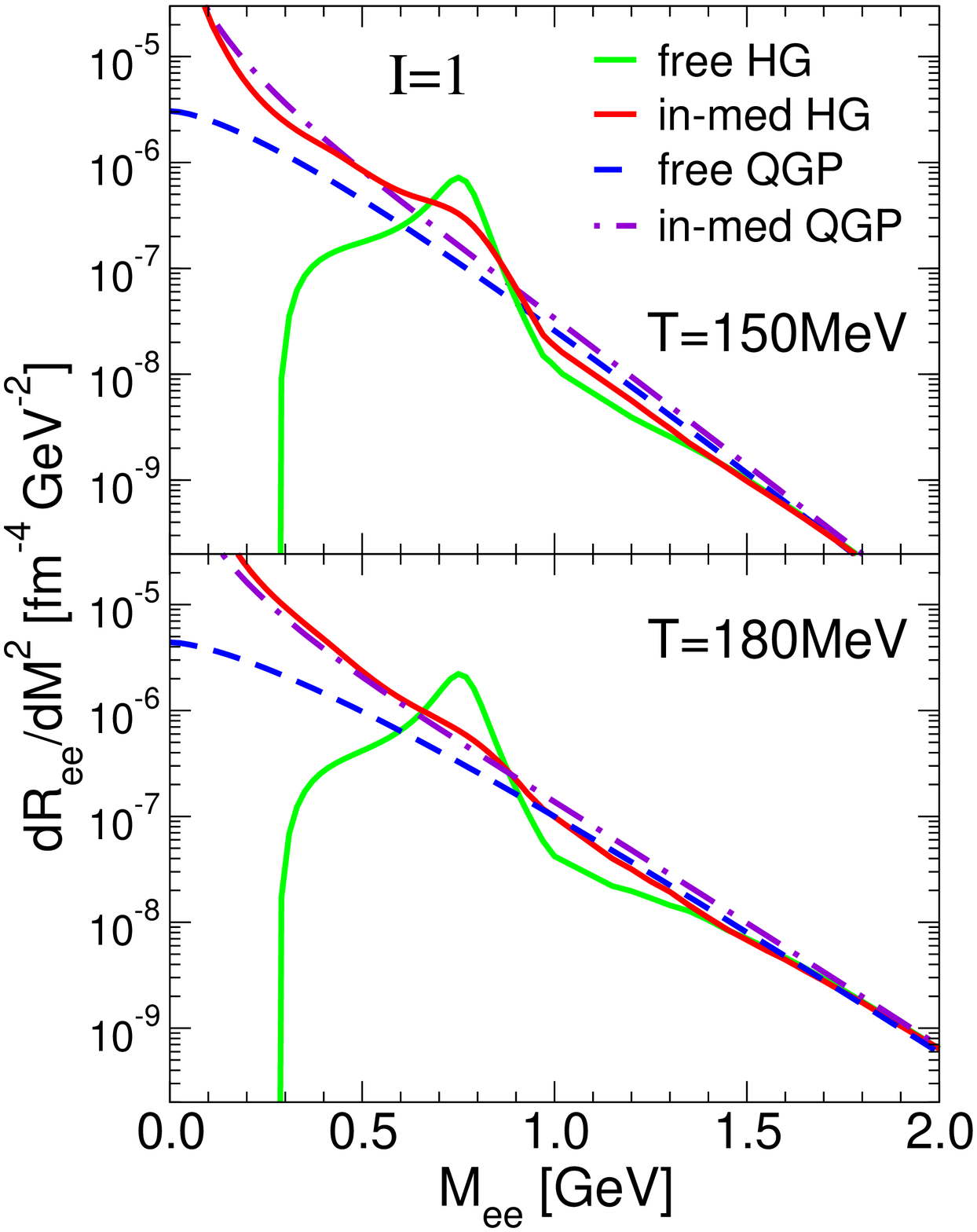,width=1.0\textwidth}
\end{minipage}
\hfill
\begin{minipage}[]{0.48\linewidth}
\epsfig{file=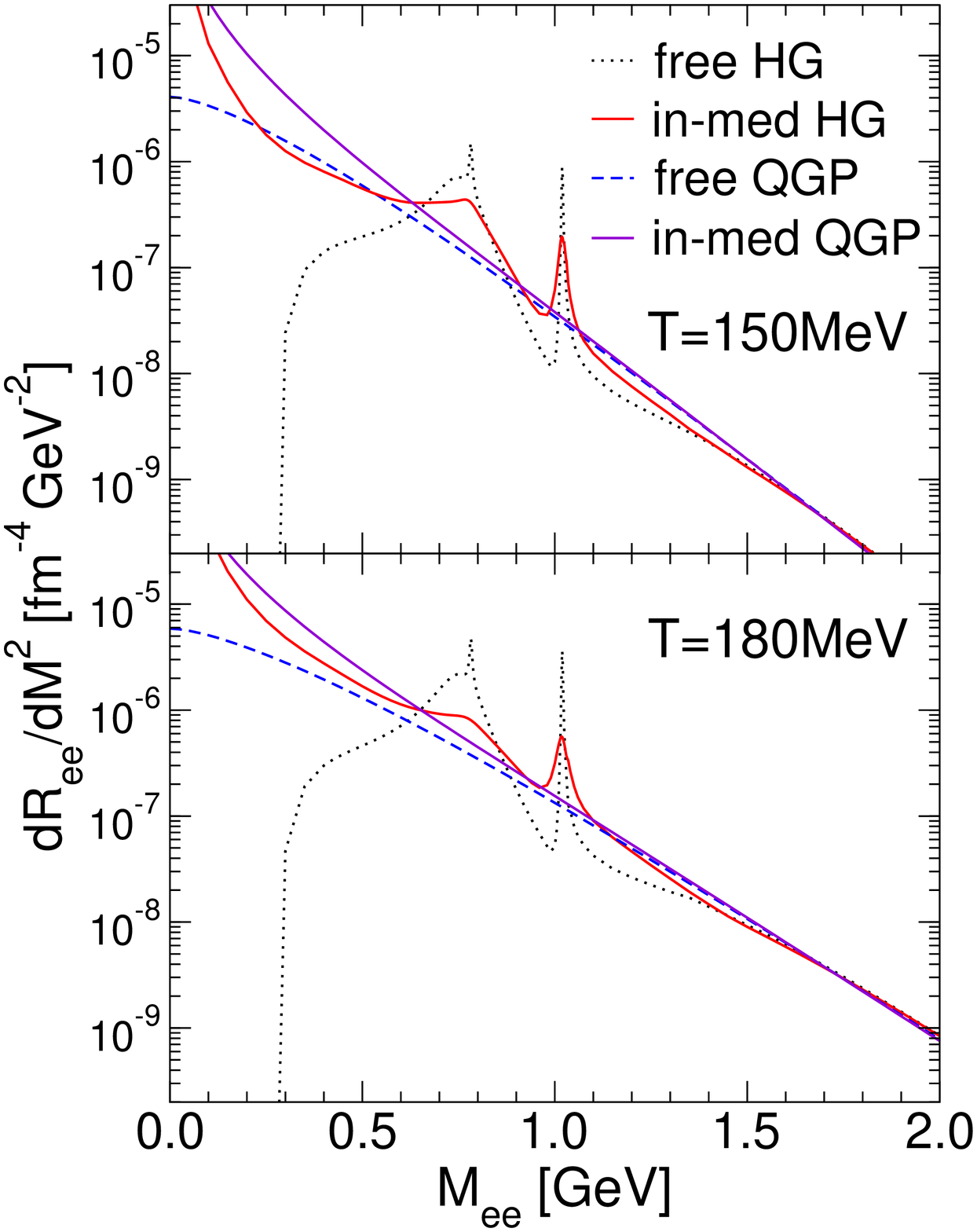,width=1.0\textwidth,angle=0}
\end{minipage}
\end{center}
\caption{ \it Three-momentum integrated thermal dilepton rates at fixed
  temperature~\cite{Rapp:2002tw,Rapp:2002mm} for the vacuum $\rho$  
  (dotted lines), the hadronic   
  many-body approach of Refs.~\cite{Rapp:1997ei,Rapp:1999us} (solid 
  lines) and for the QGP using either free $q\bar q$ annihilation 
  (dashed line) or hard-thermal loop improved 
  rates~\cite{Braaten:1990wp} (dash-dotted line). The left panel 
  refers to the isovector ($\rho$) channel, under conditions 
  resembling heavy-ion collisions at the SPS (fixed $\mu_B$=270\,MeV).
  The right panel additionally includes isoscalar ($\omega$ and $\phi$) 
  channels and corresponds to small $\mu_B$$\simeq$25\,MeV appropriate 
  for the conditions at collider energies.}
\label{fig_dlrates}
\end{figure}

The in-medium vector-meson propagators discussed in the preceding
section are converted to thermal dilepton rates via Eqs.~(\ref{ImPi_em}) 
(upper line) and (\ref{Rll}). This is based on the assumption that
VDM for the EM correlator remains valid in the medium\footnote{Strictly 
speaking, the EM correlator of Refs.~\cite{Urban:1999im,Rapp:1999us} 
includes corrections to VDM in the baryon sector as determined via 
photoabsorption spectra on the nucleon and nuclei~\cite{Rapp:1997ei}; 
the assumption is that 
this modified version of VDM is not affected at higher densities and 
at finite temperature.}. The resulting three-momentum integrated thermal 
dilepton rates are summarized in Fig.~\ref{fig_dlrates}. 
The left panel, which displays the isovector channel, reiterates that 
the $\rho$ resonance signal disappears from the mass spectrum as one 
approaches the putative phase boundary. The hadronic rates also include
an estimate of the leading-$T$ chiral mixing effect, Eq.~(\ref{chi-mix}),
in the mass region $M$=1-1.5\,GeV.  The comparison to perturbative 
$q\bar q$ annihilation reveals that the the top-down extrapolated QGP 
rate closely coincide with the bottom-up extrapolated in-medium 
hadronic one, especially in case of the HTL-improved $q\bar q$ rate. 
This feature suggests that the hadronic rate has indeed approached 
$\chi$SR (since the QGP rates are chirally symmetric at any finite 
order in perturbation theory)~\cite{Rapp:1999if,Rapp:1999us}.
The ``matching" of QGP and hadronic rates occurs directly in the 
timelike regime without the need for in-medium changes of the bare
parameters in the effective Lagrangian. Medium effects due to
baryons play an important role in this mechanism; the situation is
similar for small $\mu_B$ and close to $T_c$ where the sum
of baryon and antibaryon densities is appreciable, see right panel of
Fig.~\ref{fig_dlrates}. $\omega$ and especially $\phi$ mesons appear to
be more robust, possibly surviving above $T_c$.  
The dilepton rates in the vector manifestation of HLS~\cite{Harada:2006hu}
look rather different; based on the pertinent pion EM formfactor, 
Fig.~\ref{fig_Piem-vm}, a distinct $\rho$ peak survives in the rate up 
to temperatures of at least $T$=0.85\,$T_c$$\simeq$155\,MeV (assuming 
$T_c$$\simeq$180\,MeV). 

\begin{figure}[]
\begin{minipage}[]{0.43\linewidth}
\epsfig{file=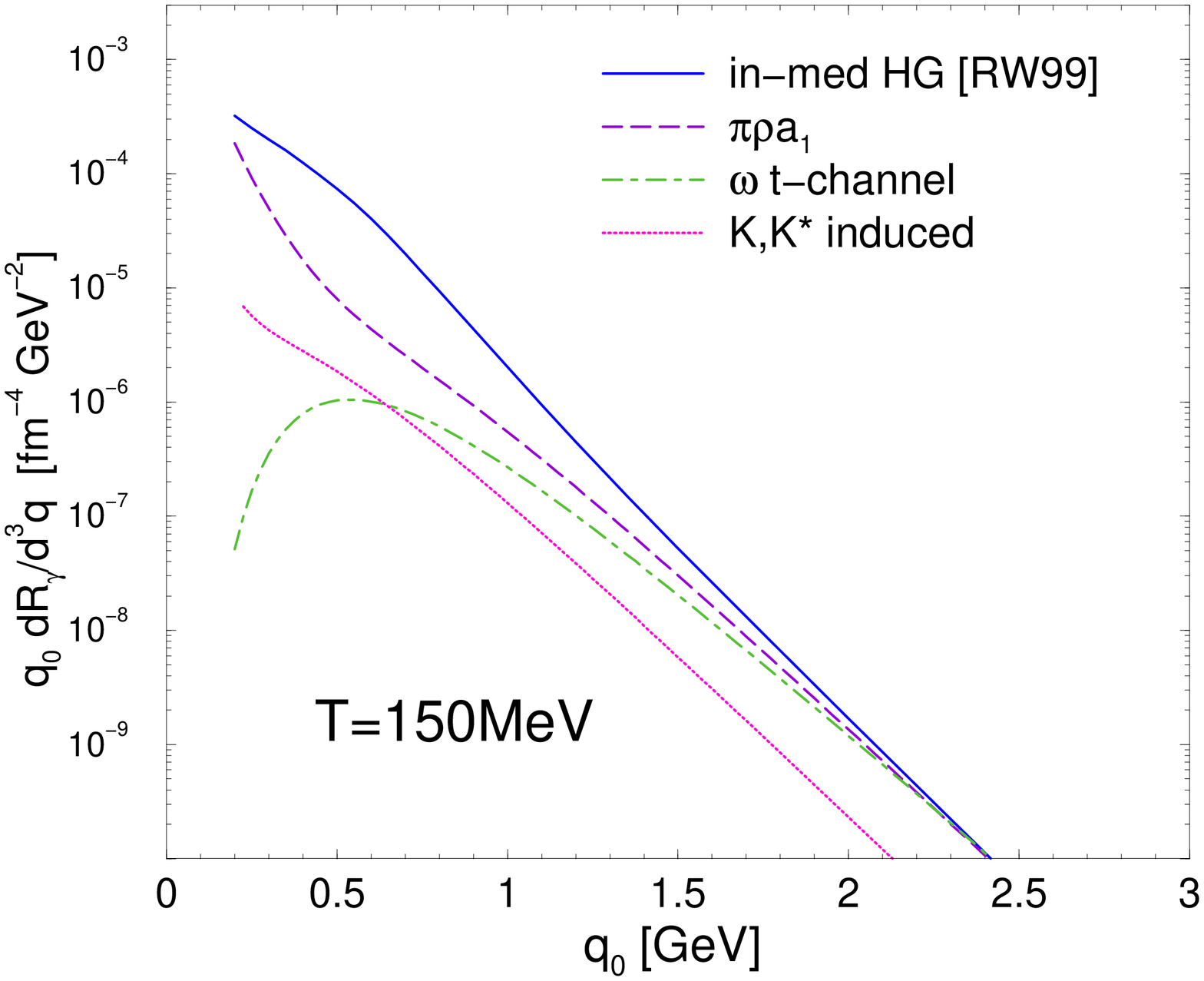,width=0.95\textwidth,angle=-90}
\end{minipage}
\hfill
\begin{minipage}[]{0.48\linewidth}
\epsfig{file=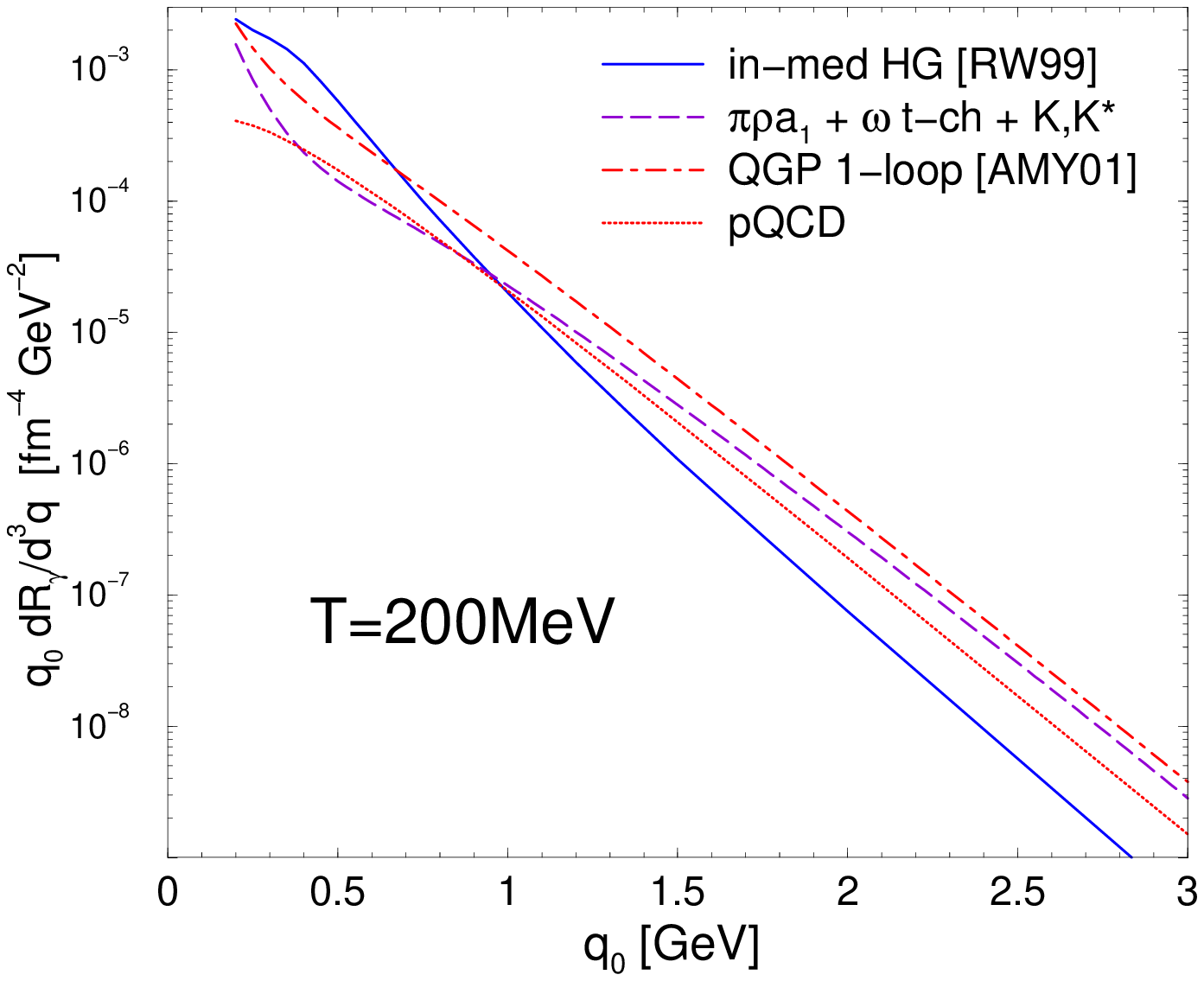,width=0.95\textwidth}
\end{minipage}
\caption{ \it Thermal photon production rates in hot hadronic
  matter~\cite{Turbide:2003si} (left panel), and compared to 
  1-loop~\cite{Kapusta:1991qp} and full leading-order~\cite{Arnold:2001ms} 
  QGP emission (right panel).}
\label{fig_phrate}
\end{figure}

Emission rates of dileptons are closely related to those of real photons
which are determined by the lightlike limit ($q_0=\mid\vec q\mid$) of 
the EM spectral function,
\begin{equation}
q_0\frac{dN_{\gamma}}{d^4xd^3q} = -\frac{\alpha_{\rm em}}{\pi^2} \
       f^B(q_0;T) \  \im \Pi_{\rm em}(M=0,q;\mu_B,T) \ .
\label{Rgamma}
\end{equation}
In Ref.~\cite{Turbide:2003si} the in-medium $\rho$ of 
Refs.~\cite{Rapp:1997ei,Rapp:1999us} has been found to constitute the 
dominant hadronic source of thermal photons for momenta up to 
$q$$\simeq$1\,GeV; above, $t$-channel meson exchange reactions not 
included in the spectral function (most notably $\pi$ and $\omega$
exchange in $\pi\rho\to\pi\gamma$) take over, cf.~left panel of 
Fig.~\ref{fig_phrate}. 
Similar to the dilepton case, at temperatures of 150-200\,MeV, the 
strength of the combined thermal rate for hadronic photon production 
turns out be very comparable to perturbative QGP emission, especially
for the complete leading-order result~\cite{Arnold:2001ms}.   

\section{Interpretation of Dilepton Spectra}
\label{sec_spectra}
In this section we will scrutinize experimental results for dilepton 
spectra in light of the theoretical developments elaborated above.  
A brief discussion of production experiments off nuclei, representing
cold nuclear matter up to saturation density, will be followed by a 
more extensive study of invariant-mass and momentum spectra in heavy-ion
reactions involving hot and dense matter possibly probing the
transition regime to the QGP.

\subsection{Medium Effects in Nuclei}
\label{ssec_nuclei}
Dilepton production experiments off nuclei have the advantage over 
heavy-ion collisions that the medium is well-defined. Medium- to heavy 
ground-state nuclei resemble in their interior infinitely extended 
nuclear matter at vanishing temperature.  Therefore the experiments 
probe to a large extent the properties of ``cold nuclear matter".
However, a good knowledge of the production process is required, 
and medium effects are typically rather moderate, further reduced by 
surface effects and decays outside the nucleus especially at large 
three-momenta (which, in turn, are needed in the production process). 

\begin{figure}[!tb]
\begin{minipage}{0.6\linewidth}
\epsfig{file=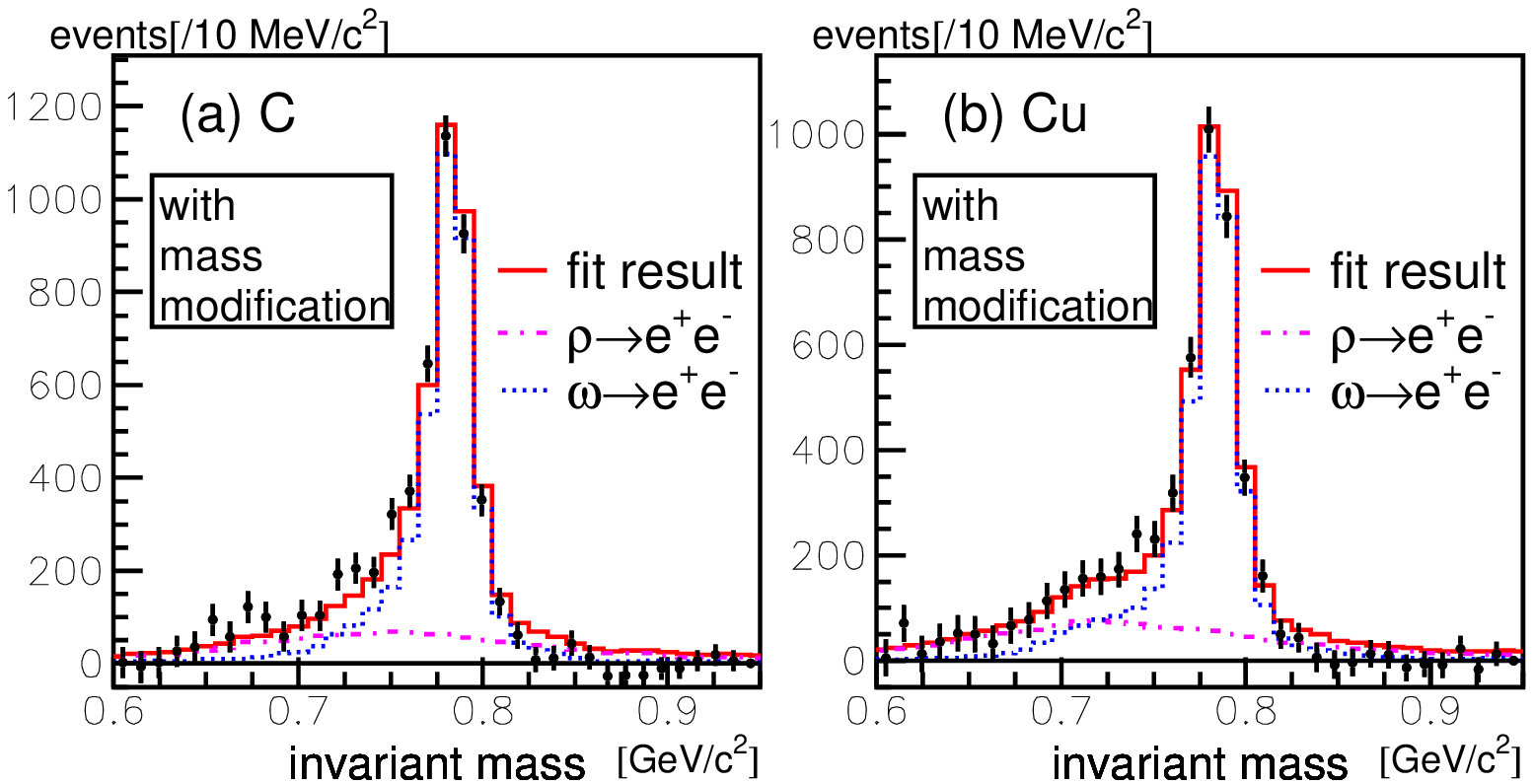,width=1.0\textwidth}
\end{minipage}
\hspace{0.5cm}
\begin{minipage}{0.35\linewidth}
\vspace{0.2cm}
\epsfig{file=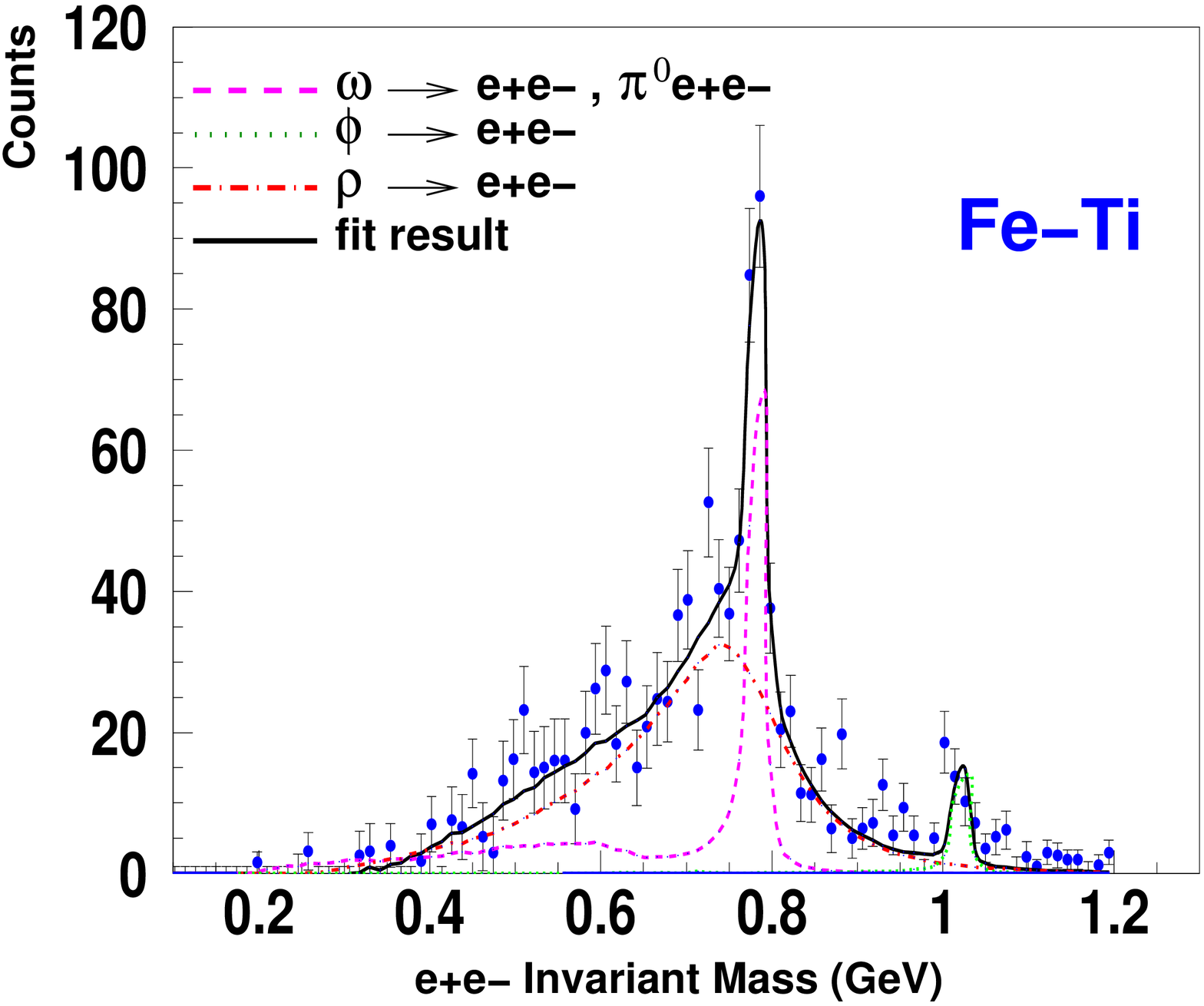,width=0.97\textwidth}
\end{minipage}
\caption{\it Invariant-mass spectra of $e^+e^-$ in proton- 
  (left)~\cite{Naruki:2005kd} and photon-induced 
  (right)~\cite{Clas:2007mga} reactions off nuclear targets. 
  }
\label{fig_vmes-nuc}
\end{figure}

The E325 experiment at KEK~\cite{Naruki:2005kd} used 12\,GeV proton 
projectiles and
found significant differences in the spectra between $C$ and $Cu$ 
targets, see left panel of Fig.~\ref{fig_vmes-nuc}.  
After subtraction of combinatorial background as well as $\eta$ and 
$\omega$ Dalitz decays, the best fit to the excess spectra using 
$\omega$ and $\rho$ Breit-Wigner distributions was obtained with a mass 
shift of ca. 9\% at nuclear matter density, and a $\rho/\omega$ ratio 
of about $\sim$0.45.  

The CLAS experiment~\cite{Clas:2007mga} used a photon beam at 
Jefferson Lab with incident 
energies $E_\gamma$=0.6-3.8\,GeV, directed on various nuclear targets.
After subtraction of the combinatorial background, determined
with absolute normalization, the invariant-mass spectra are best 
reproduced with a $\rho$ spectral distribution with a mass and width of 
($m_\rho$,$\Gamma_\rho$)=(775$\pm$5,220$\pm$15)\,MeV, cf.~ right 
panel of Fig.~\ref{fig_vmes-nuc}. These values are well reproduced 
by Boltzmann transport calculations~\cite{Wood:2008ee},
and are consistent with the predictions of 
Refs.~\cite{Urban:1999im,Rapp:1999us} at $\varrho_N$=0.5\,$\varrho_0$ and 
$q$=1\,GeV, where ($\Delta m_\rho,\Delta\Gamma_\rho$)$\simeq$(15,75)\,MeV,
see Ref.~\cite{Riek:2008ct} for a recent calculation.
An apparent difference between the E325 and CLAS spectra is that the
background subtraction in the former removes any excess for 
$M$$\simeq$0.85-1\,GeV; this suppresses (and possibly shifts down) 
the $\rho$ contribution in the E325 fit. 

Photoproduction experiments ($E_\gamma$=0.8-1.12\,GeV) of $P$-wave 
$\pi^+\pi^-$ pairs off $^2$H, $^3$He and $^{12}$C were conducted by 
the TAGX collaboration~\cite{Huber:2003pu}. The  
spectra for the $^{12}$C target support medium 
effects in line with hadronic many-body $\rho$ spectral 
functions~\cite{Urban:1999im,Rapp:1999us}.   

\subsection{Heavy-Ion Collisions}
\label{ssec_hics}
In contrast to production experiments off nuclei, the (energy-)
density of the medium created in heavy-ion collisions undergoes
a rapid evolution after initial nuclear impact until break-up.
Even under the simplifying assumption of local thermal equilibrium,
a good knowledge of the temperature and baryon-density evolution
is necessary to convert the dilepton rates discussed above into a
space-time integrated spectrum  In addition, sources other than thermal
radiation have to be considered, especially toward higher mass
{\em or} $q_t$ where the assumption of equilibrium becomes increasingly
questionable.  These issues are addressed in Sec.~\ref{sssec_sources}.
Phenomenological analyses of
dilepton spectra focus on recent SPS data from NA60 and CERES/NA45 in
Secs.~\ref{sssec_na60} and \ref{sssec_na45-wa98}, respectively, with a
short digression to direct photons (W98) and a brief outlook to future
experiments in Sec.~\ref{sssec_future}.

\subsubsection{Thermal Evolution and Dilepton Sources}
\label{sssec_sources}

As discussed in Sec.~\ref{ssec_urhics}, hadronic observables in URHICs
point at a reasonable degree of thermalization of the bulk medium produced
in these reactions. Therefore, we here focus on hydrodynamic and 
expanding fireball approaches utilizing the assumption of local thermal
equilibrium.\footnote{Comparisons of dilepton spectra
computed in hydrodynamic/fireball and transport calculations (based on
similar in-medium spectral functions) have shown rather good agreement, 
see e.g. Refs.~\cite{Cassing:1997jz,Rapp:1999us},
Ref.~\cite{Huovinen:2002im}, or 
Refs.~\cite{vanHees:2007th,Bratkovskaya:2008bf}.}  

Thermal emission spectra follow from the convolution of the production 
rate, Eq.~(\ref{Rll}), over the space-time evolution of the medium,
\begin{equation}
\frac{\dd N_{ll}}{\dd M \dd y} 
= \frac{M}{\Delta y} \int\limits_0^{\tau_{\mathrm{fo}}}
\dd\tau \int\limits_{V_{\mathrm{FB}}} \dd^3x \int \frac{\dd^3 q}{q_0} \ 
\frac{\dd N_{ll}}{\dd^4 x~\dd^4 q}(M,q;T,\mu_N,\mu_s,\mu_i) \  
\mathrm{Acc}(M,q_t,y),
\label{th-rad}
\end{equation}
where $\mathrm{Acc}(M,q_t,y)$ accounts for the detector acceptance and
$\Delta y$ denotes the corresponding rapidity interval. 
The temperature and chemical potentials in general depend 
on space-time, ($\tau$,$\vec x$). Note that while $\mu_N$ and $\mu_s$ 
correspond to exact conservation of baryon number and strangeness, 
effective chemical potentials $\mu_i=\mu_{\pi,K,\eta,...}$ are needed 
to preserve the experimentally observed hadron ratios in the evolution 
of the hadronic phase between chemical ($T_{\rm ch}$=155-175\,MeV) and 
kinetic freezeout ($T_{\rm fo}$=100-140\,MeV).     

An overview of several key input parameters of three thermal 
approaches~\cite{vanHees:2006ng,vanHees:2007th,Dusling:2006yv,
Dusling:2007kh,Renk:2006qr,Ruppert:2007cr}, which have been used to 
compute dimuon spectra in comparison to NA60 data, is compiled in 
Tab.~\ref{tab_fireball} (see also Ref.~\cite{Skokov:2005ut}). 
The overall range of the underlying parameters
and assumptions is rather similar. This is not a coincidence 
but a consequence of constraints from measured hadron spectra at SPS
energies~\cite{Bearden:1996dd,Appelshauser:1997rr,Antinori:2001yi,Adamova:2002wi} 
which all of the three models have been subjected to.  
There are, however, noticeable differences. E.g., all approaches operate
with a for SPS energies ``canonical" formation time of $\tau_0$=1\,fm/$c$,
but the initial peak temperature in Ref.~\cite{Ruppert:2007cr} is 
about 15\% larger than in Ref.~\cite{Dusling:2006yv} (e.g., due to 
differences in the underlying QGP EoS). Averaging over the initial 
spatial density profile typically leads to a 15\% smaller {\em average} 
temperature~\cite{Huovinen:2001wx}; thus, $\bar T_0$$\simeq$190\,MeV in
Ref.~\cite{vanHees:2006ng} is quite consistent with 
$T_0^{\rm max}$$\simeq$220\,MeV in Ref.~\cite{Dusling:2006yv}. 
\begin{table}
\begin{center}
\begin{tabular}{|p{0.12\linewidth}||p{0.24\linewidth}|p{0.24\linewidth}|p{0.24\linewidth}|}
\hline
\quad & DZ & RR & HR\\
\hline \hline
$T_0$ & 220~MeV (peak) & 250~MeV (peak) & 190~MeV (average)\\
\hline
$T_{c}$ & 170~MeV & 170~MeV & 175(160/190)~MeV
\\
\hline
$T_{\text{fo}}$ &  130~MeV  & 130~MeV & 120(135)~MeV
\\
\hline
spatial & Glauber (initial) & Woods-Saxon & isotropic
\\
\hline
$v_{\text{fo}}^{\mathrm{s}}$ & $\sim$0.5-0.55 & $0.57$ & $0.53$ \\
\hline
$v(r)$ & approx. linear & $\rho_t \propto \sqrt{r}$ & $v_t \propto r$ \\
\hline
$\tau_{\text{FB}}$ & $\sim$~8-9~fm/$c$ & $\sim 7.5 \;
\mathrm{fm}/c$ & $\sim 6.5 \; \mathrm{fm}/c$ \\
\hline
QGP-EoS & massless ($N_f$=3) & quasi-particle model &
massless ($N_f$=2.3) \\
\hline
HG-EoS & lowest SU($3$) multipl. & $m_{B,M }\leq$~2,~1.5~GeV &
$m_{B,M }\leq$~2,~1.7~GeV
\\ \hline
$s/\varrho_B$ & 42 & 26(?) & 27
\\ \hline
$\mu_{\pi}^{{\rm fo}}$ & 0 & ? ($\neq 0$) & 80(35)~MeV
\\ \hline
EM rates & chiral virial & empirical~scatt.~ampl. & hadronic many-body
\\ \hline
\end{tabular}
\end{center}
\caption{\it Fireball parameters employed in the 
  calculations of dilepton spectra in In(158~AGeV)-In collisions
  in Refs.~\cite{Dusling:2006yv,Dusling:2007kh} (DZ), 
  \cite{Renk:2006qr,Ruppert:2007cr} (RR) 
  and~\cite{vanHees:2006ng,vanHees:2007th} (HR).} 
\label{tab_fireball}
\end{table}
The slightly larger expansion velocity in 
Ref.~\cite{Ruppert:2007cr} (surface velocity $v_{\rm fo}^{\rm s}$=0.57 
at thermal freezeout), together with its square-root radial 
profile, imply larger boost factors in the $q_t$ spectra which becomes
significant at high momenta. In this approach preliminary NA60 pion 
spectra in semicentral In-In are saturated by thermal emission 
over the entire measured range up to $p_T$$\simeq$3\,GeV. 
Alternatively, in Ref.~\cite{vanHees:2007th}, based on an analysis of 
pion spectra in Pb-Au and S-Au collisions at SPS, the thermal component
was found to account for the pion yields only up to $p_T$$\simeq$1\,GeV,
requiring the introduction of a ``primordial" component associated 
with initial hard scattering of the incoming nucleons. This 
interpretation is supported by the observation that the pion spectra
for $p_T$$\ge$2\,GeV essentially scale with the number of binary $N$-$N$ 
collisions ($N_{\rm coll}$), indicating that the hard component 
dominates the spectra at these momenta. The preliminary NA60 pion 
spectra are also well predicted in this approach, with a crossing of 
thermal and hard components at $p_T$$\simeq$1.2\,GeV. 
As discussed in Sec.~\ref{ssec_urhics}, a valuable 
indicator of the degree of thermalization is the elliptic flow, 
$v_2(p_T)$. At SPS energies, ideal hydrodynamics overpredicts this 
quantity even at low $p_T$ by about 30-50\% (possibly due to 
neglecting effects of finite viscosity, in connection with initial 
temperatures in the vicinity of $T_c$ where the EoS is presumably 
rather soft). Moreover, the experimental $v_2(p_T)$ in semicentral Pb-Au 
levels off at $p_T$=1.5-2\,GeV~\cite{Alt:2003ab,Wurm:2004gx}, indicative 
for a transition to a kinetic regime, while hydrodynamic results keep 
rising, overpredicting $v_2$($p_T$=2\,GeV) by about a factor of $\sim$2.  

Concerning effective chemical potentials for pions (and other
stable particles) between chemical and thermal
freezeout, their main effect is a faster cooling in the evolution
of $T(\tau)$ (the equation of state, $P(\epsilon)$, is largely 
unaffected)~\cite{Rapp:2002fc,Teaney:2002aj}. E.g., for 
$\mu_\pi$=80\,MeV at $T$=120\,MeV~\cite{vanHees:2006ng}, the pion 
density, $\varrho_\pi(T,\mu_\pi)$, is enhanced by a factor of 
$\sim$e$^{80/120}$$\sim$2 relative to $\mu_\pi$=0, and as large
as $\varrho_\pi$(T=150\,MeV,0). Therefore, thermal freezeout at 
($T_{\rm fo}$,$\mu_\pi^{\rm fo}$)=(130,0)\,MeV~\cite{Dusling:2006yv} 
corresponds to a smaller pion density (and thus larger volume) than at 
($T_{\rm fo}$,$\mu_\pi^{\rm fo}$)=(120,80)\,MeV~\cite{vanHees:2006ng}, 
consistent with the longer lifetime in Ref.~\cite{Dusling:2006yv}.  

Implications of varying critical and chemical-freezeout temperatures for 
dilepton spectra have been studied in Ref.~\cite{vanHees:2007th}. 
The value of $T_c$ affects the {\em relative} partition of QGP and hadronic 
emission, especially at masses $M$$\ge$1\,GeV where the Boltzmann factor 
augments the sensitivity to earlier phases and the hadronic 
rates are not enhanced by resonances. However, if hadronic and QGP rates 
are ``dual" around $T_c$, this distinction is largely academic.    
Smaller $T_{\rm ch}$'s lead to smaller $\mu_i$'s in the subsequent hadronic
evolution, and thus higher kinetic freezeout temperatures, e.g.,
($T_{\rm fo}$,$\mu_\pi^{\rm fo}$)=(135,35)~MeV for $T_{\rm ch}$=160~MeV.

Experimentally measured dilepton spectra contain sources other than
thermal radiation represented by Eq.~(\ref{th-rad}). A systematic
evaluation of these sources has recently been conducted in
Ref.~\cite{vanHees:2007th}, in terms of (i) final-state decays
and (ii) primordial sources.  

Dilepton decays of long-lived hadrons (mostly $\eta$, $\eta'$, $\omega$ 
and $\phi$ mesons) after thermal freezeout, commonly referred to as 
``hadron decay cocktail", are usually based on chemical freezeout for 
their abundance and thermal freezeout for their $p_T$ spectra. The 
situation is more involved for the $\rho$-meson, since its continuous 
regeneration implies relative chemical equilibrium with pions until 
thermal freezeout (to a certain extent this may also apply to $\omega$ 
and $\phi$) . In addition, its short lifetime is not well separated 
from the typical duration of the freezeout process. 
In Ref.~\cite{Rapp:1999us}, the final generation of $\rho$ decays has
been approximated by an extra 1~fm/$c$ of fireball lifetime. However,
as has been clarified in Refs.~\cite{Rapp:2007zza,vanHees:2007th}, 
when treating the final generation of $\rho$'s as a cocktail decay,
the time dilation of the moving $\rho$'s generates a hardening of its
$q_t$ spectrum by a factor $\gamma_t$=$M_t/M$ ($M_t^2$=$M^2$+$q_t^2$). 
The resulting spectrum recovers the standard Cooper-Frye~\cite{cf74} 
description for freezeout at a fixed time in the laboratory frame 
(cf.~also Refs.~\cite{Knoll:2008sc,Cassing:2008sv}). This, in turn,
implies that the apparent temperature of the radiation formula 
(\ref{th-rad}) is smaller than the actual temperature figuring into 
the Boltzmann factor (independent of flow effects) by about
$\sim$10\%.\footnote{The
time dilation factor for $\rho$ decays in the thermal radiation 
formula is compensated by the same time dilation in $\rho$ formation, 
as a consequence of detailed balance.} 

In analogy to the pion-$p_T$ spectra discussed above, the $\rho$ spectra
are expected to have a primordial component (emanating from hard $N$-$N$
collisions) which does not equilibrate with the medium. Such a component
has been introduced in Refs.~\cite{Rapp:2007zza,vanHees:2007th} 
including a schematic treatment for Cronin effect and jet-quenching as
inferred from pion spectra in S-Au and Pb-Au collisions at 
SPS~\cite{dEnterria:2005cs} (also note
that, at high $q_t$, this component scales with $N_{\rm coll}$, rather
than $N_{\rm part}$ as for the (low-$p_T$) cocktail).  

Another primordial dilepton source is the well-known Drell-Yan (DY) 
process, i.e., quark-antiquark annihilation in binary $N$-$N$ 
collisions. To leading order $q \bar{q} \rightarrow e^+ e^-$ is 
$\mathcal{O}(\alpha_s^0 \alpha^2)$ and can be reliably
calculated in perturbation theory at sufficiently large masses, 
$M$$\gsim$2~GeV, utilizing parton distribution functions as 
input~\cite{Halzen:1978et}. A finite pair momentum, $q_t$$>$0, can be
generated by intrinsic parton $k_t$ and at next-to-leading order (NLO) 
(the latter is the dominant effect). The extrapolation of DY to small 
masses is problematic, but at a scale of $q_t$$\simeq$2\,GeV its
contribution to dilepton spectra at SPS is potentially sizable.
In Ref.~\cite{vanHees:2007th} it has been suggested to estimate the
spectrum of slightly virtual DY pairs, i.e., for $M^2\ll q_t^2$, by 
an extrapolation of a finite-$q_t$ DY expression to zero mass 
and constrain the resulting photon spectrum by measured photon 
spectra in $p$-$A$ collisions. 
     

Semileptonic final-state decays of correlated of $D$ and $\bar{D}$ mesons 
(i.e., corresponding to an associately produced $c\bar c$ pair) lead to an 
irreducible dilepton signal. The pertinent mass spectrum is, in fact, 
sensitive to reinteractions of the charm quarks and/or hadrons in the 
medium. At SPS energies, the relevance of this effect for correlated
$D\bar D$ decays is currently an open question~\cite{Arnaldi:2008er}. 
Theoretical calculations discussed in the following 
are employing $N_{coll}$-extrapolated spectra from $p$-$p$ collisions 
(based on PYTHIA~\cite{Sjostrand:2000wi} simulations) as 
provided by the NA60 collaboration~\cite{Arnaldi:2006jq}.

\subsubsection{CERN-SPS I: NA60}
\label{sssec_na60}

In section we discuss several calculations of $\mu^+\mu^-$ spectra in 
semicentral In(158\,AGeV)-In collisions as measured by NA60.
The excellent mass resolution and statistics of the data allowed 
for a subtraction of the hadronic cocktail (excluding $\rho$
and $D\bar D$ decays), resulting in the so-called ``excess spectra"
(in more recent, acceptance-corrected, NA60 spectra~\cite{Arnaldi:2008fw} 
correlated $D\bar D$ decays are also subtracted, with some caveat as to 
their medium modifications, as mentioned above).

\begin{figure}[!t]
\begin{minipage}[]{0.5\textwidth}
\epsfig{file=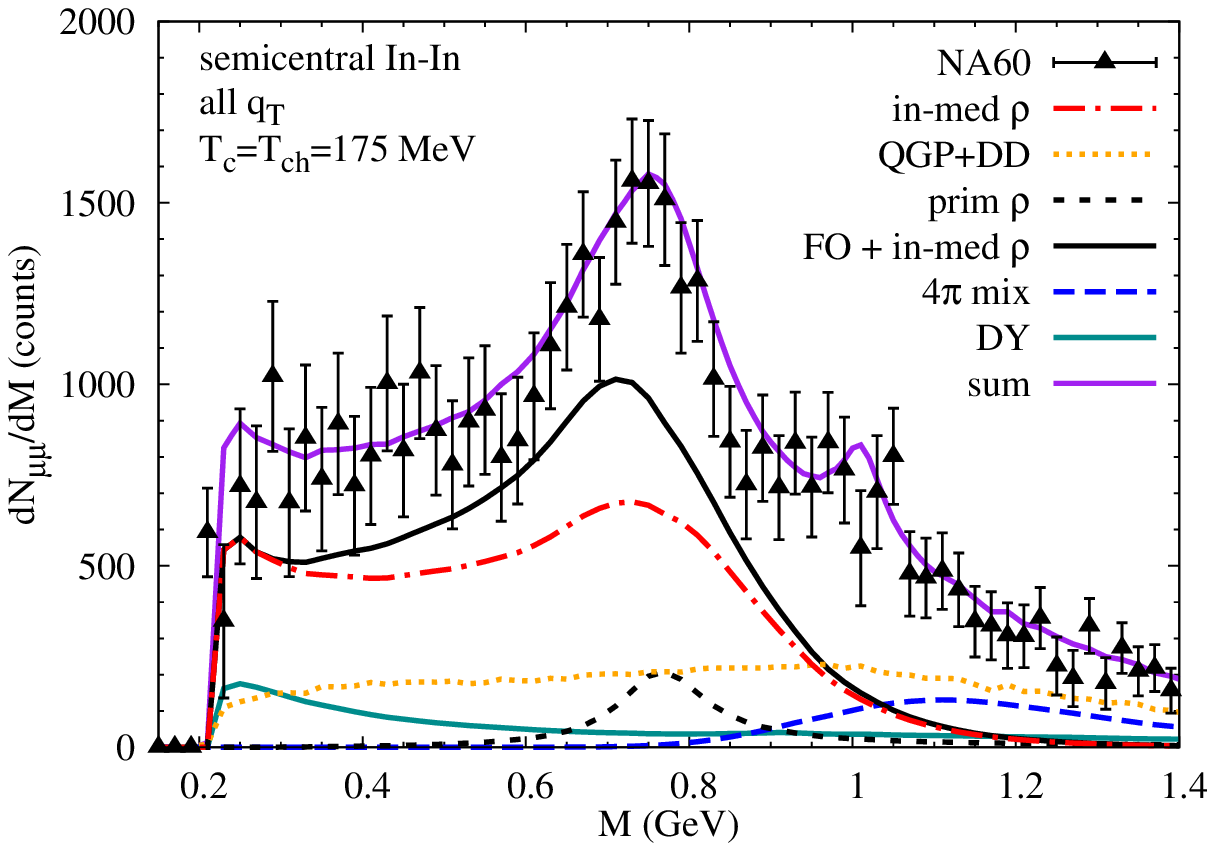,width=0.95\textwidth}
\end{minipage}
\hspace{0.1cm}
\begin{minipage}[]{0.5\textwidth}
\epsfig{file=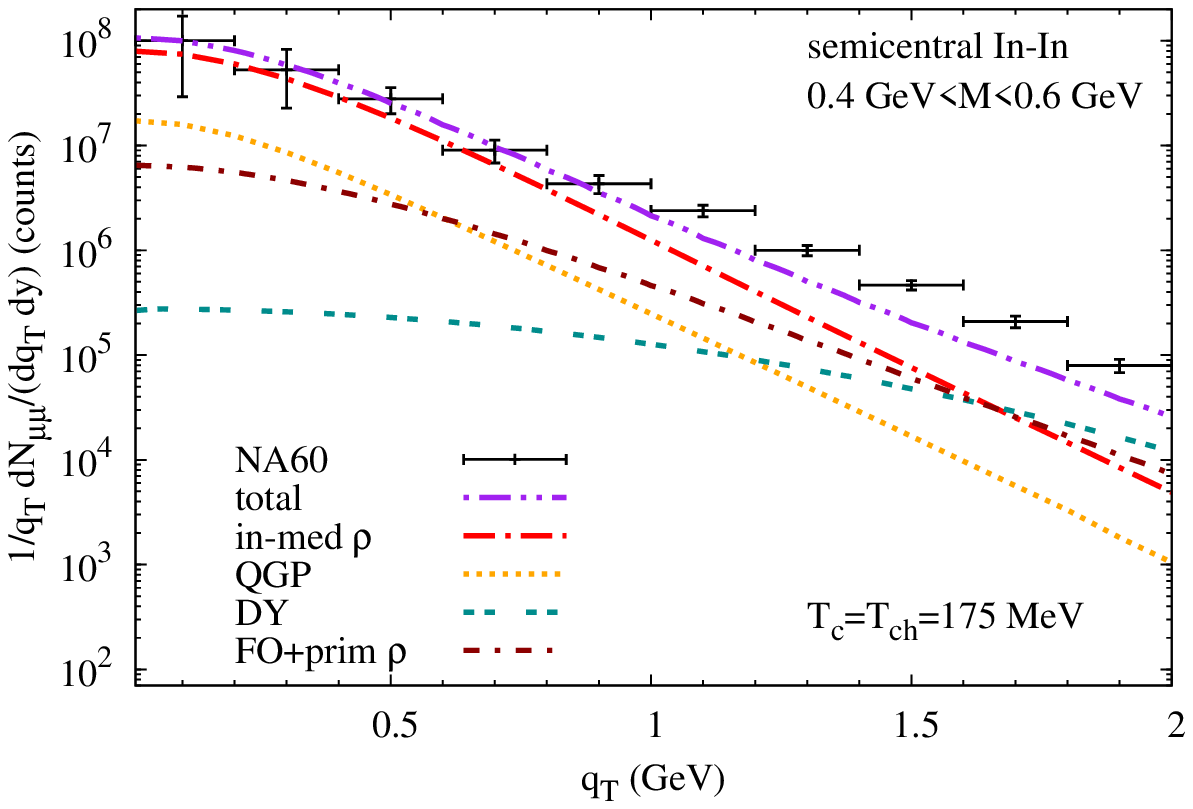,width=0.95\textwidth}
\end{minipage}
\begin{minipage}[]{0.5\textwidth}
\epsfig{file=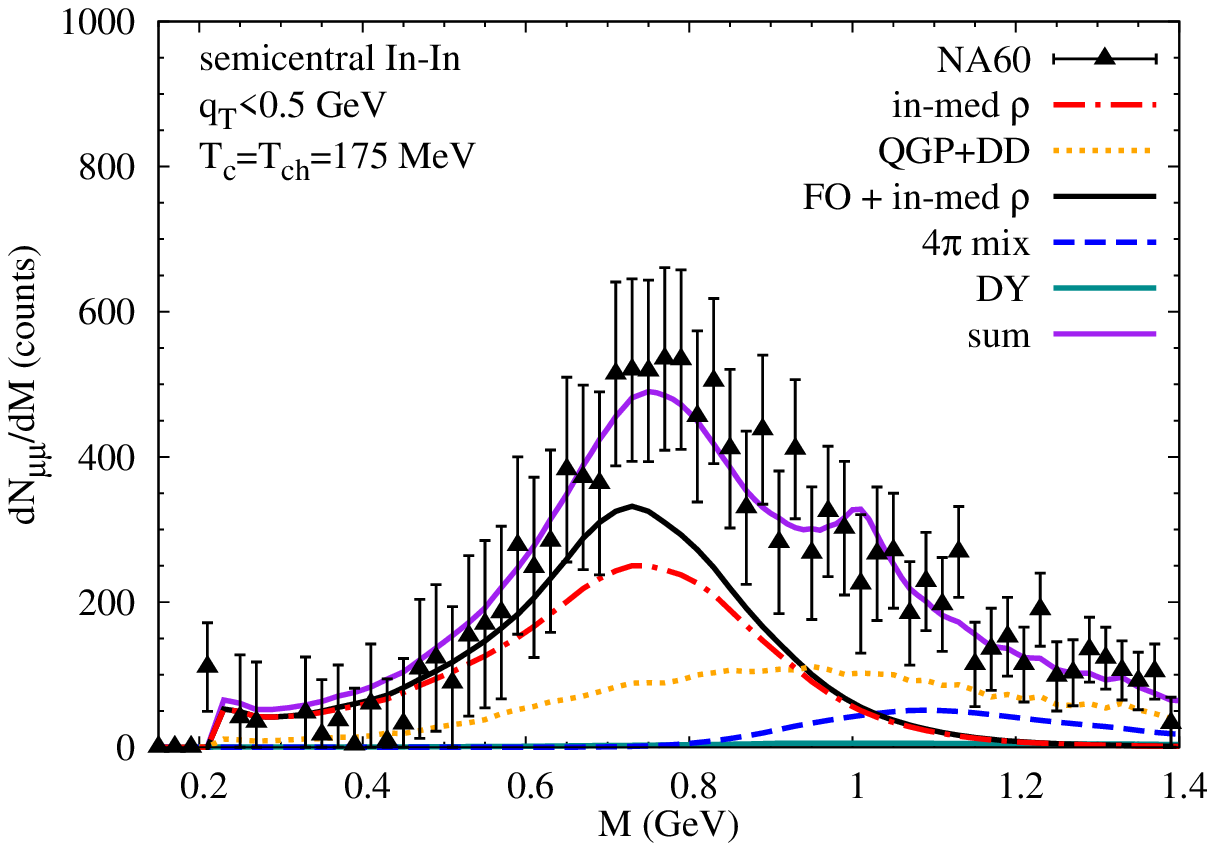,width=0.95\textwidth}
\end{minipage}
\hspace{0.1cm}
\begin{minipage}[]{0.5\textwidth}
\epsfig{file=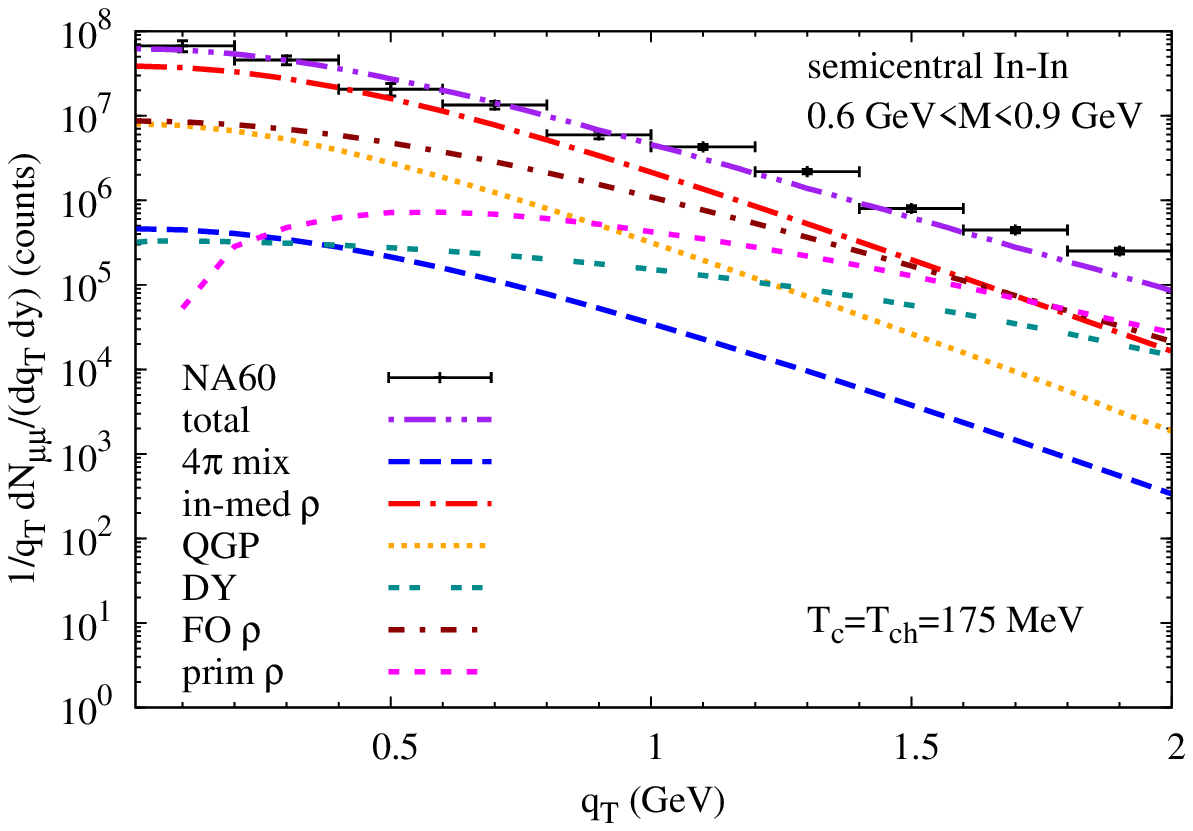,width=0.95\textwidth}
\end{minipage}
\begin{minipage}[]{0.5\textwidth}
\epsfig{file=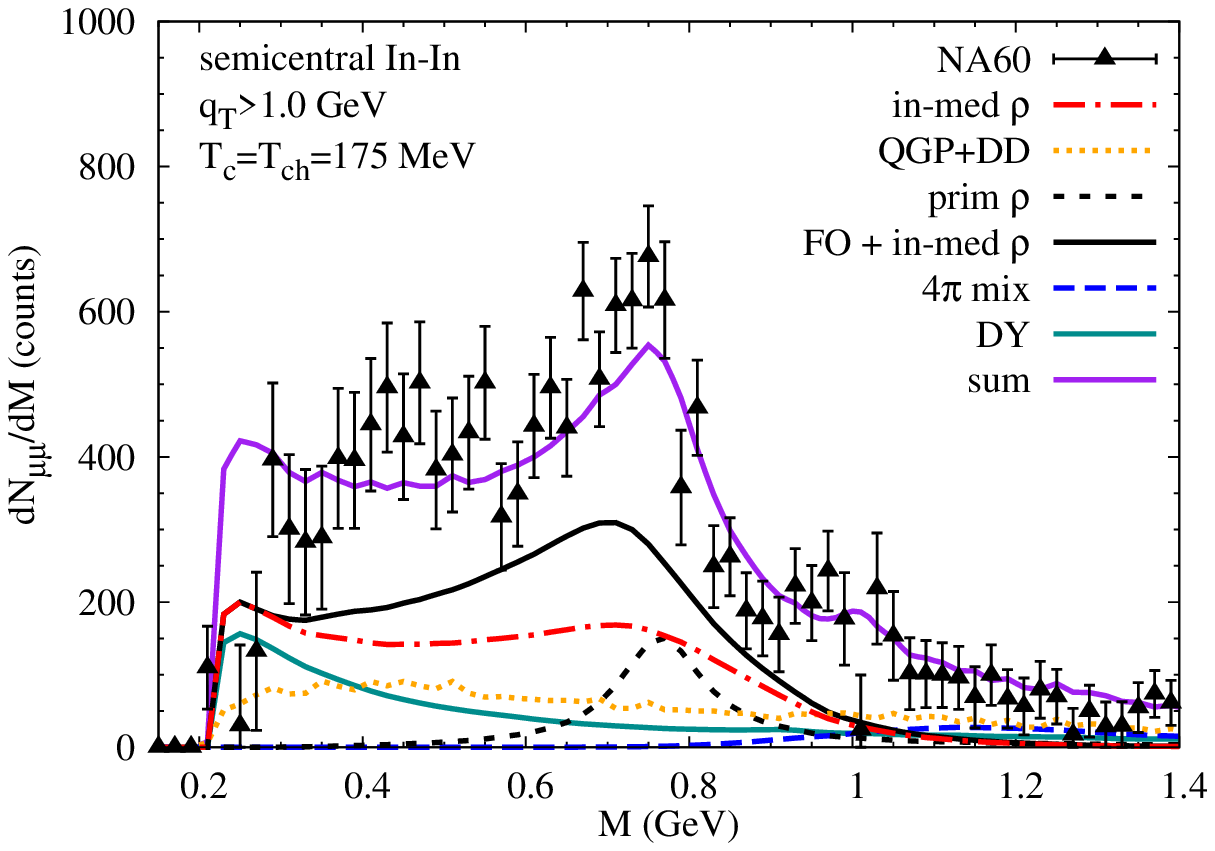,width=0.95\textwidth}
\end{minipage}
\hspace{0.1cm}
\begin{minipage}[]{0.5\textwidth}
\epsfig{file=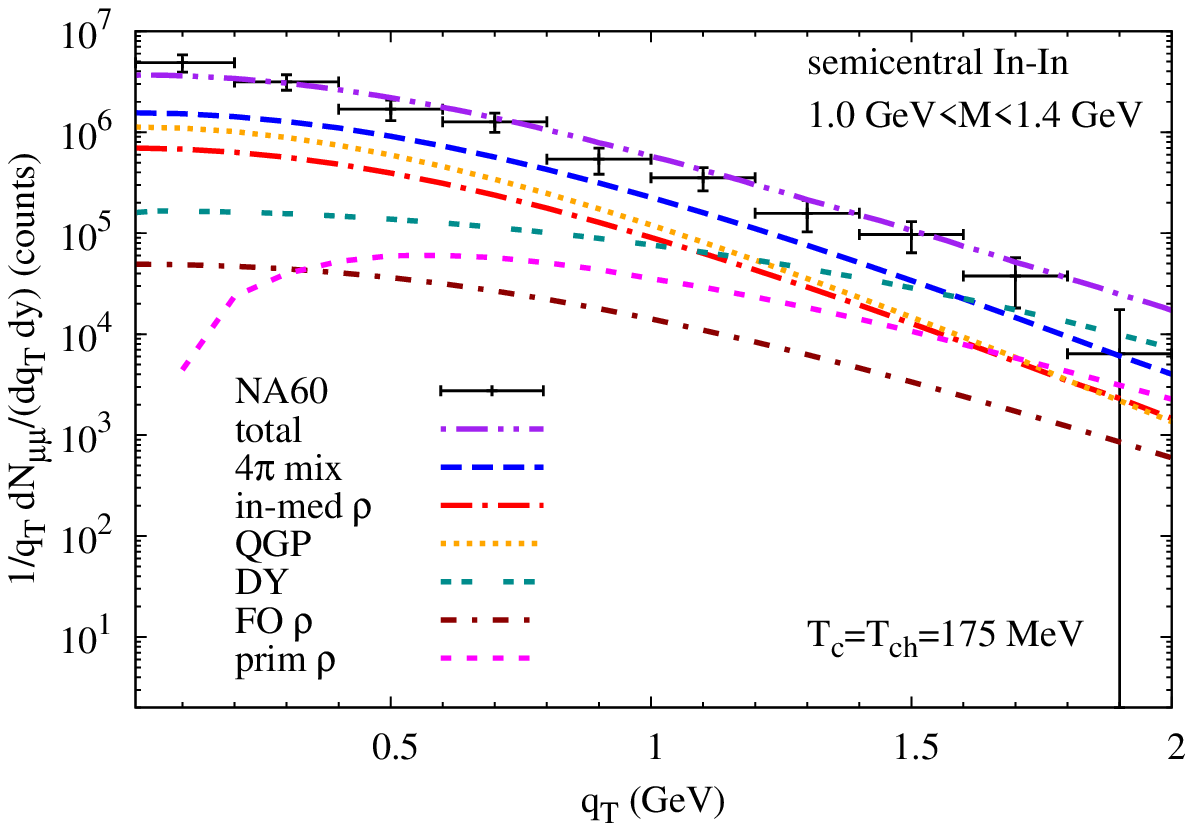,width=0.95\textwidth}
\end{minipage}
\caption{\it Dimuon invariant-mass (left panels) and 
  acceptance-corrected transverse-momentum (right panels) spectra in 
  semi-central {\rm In(158~AGeV)-In} collisions. 
  Calculations~\cite{vanHees:2007th} for thermal emission
  utilizing in-medium $\rho$, $\omega$ and $\phi$ spectral functions based 
  on hadronic-many body theory~\cite{Rapp:1999us,Rapp:2000pe}, $4\pi$ 
  annihilation including chiral mixing~\cite{vanHees:2006ng} and QGP
  emission, supplemented by non-thermal sources (Drell-Yan annihilation,
  primordial and freeze-out $\rho$-meson, open-charm decays), are
  compared to NA60
  data~\cite{Arnaldi:2006jq,Damjanovic:2007qm,Arnaldi:2007ru}.}
\label{fig_na60-HR}
\end{figure}

\begin{figure}[!t]
\begin{minipage}[]{0.48\linewidth}
\epsfig{file=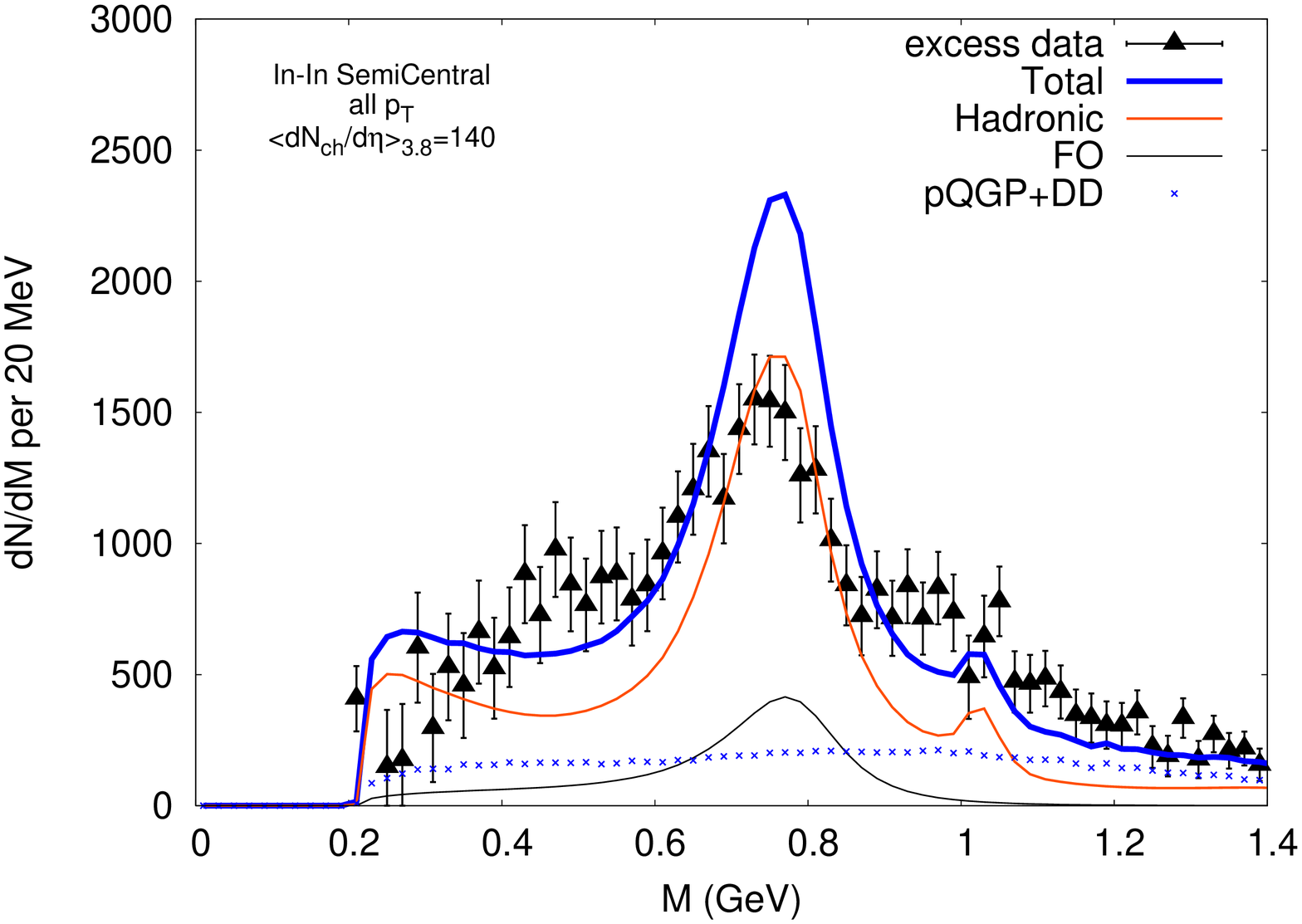,width=\textwidth}
\end{minipage}
\hfill
\begin{minipage}[]{0.48\linewidth}
\epsfig{file=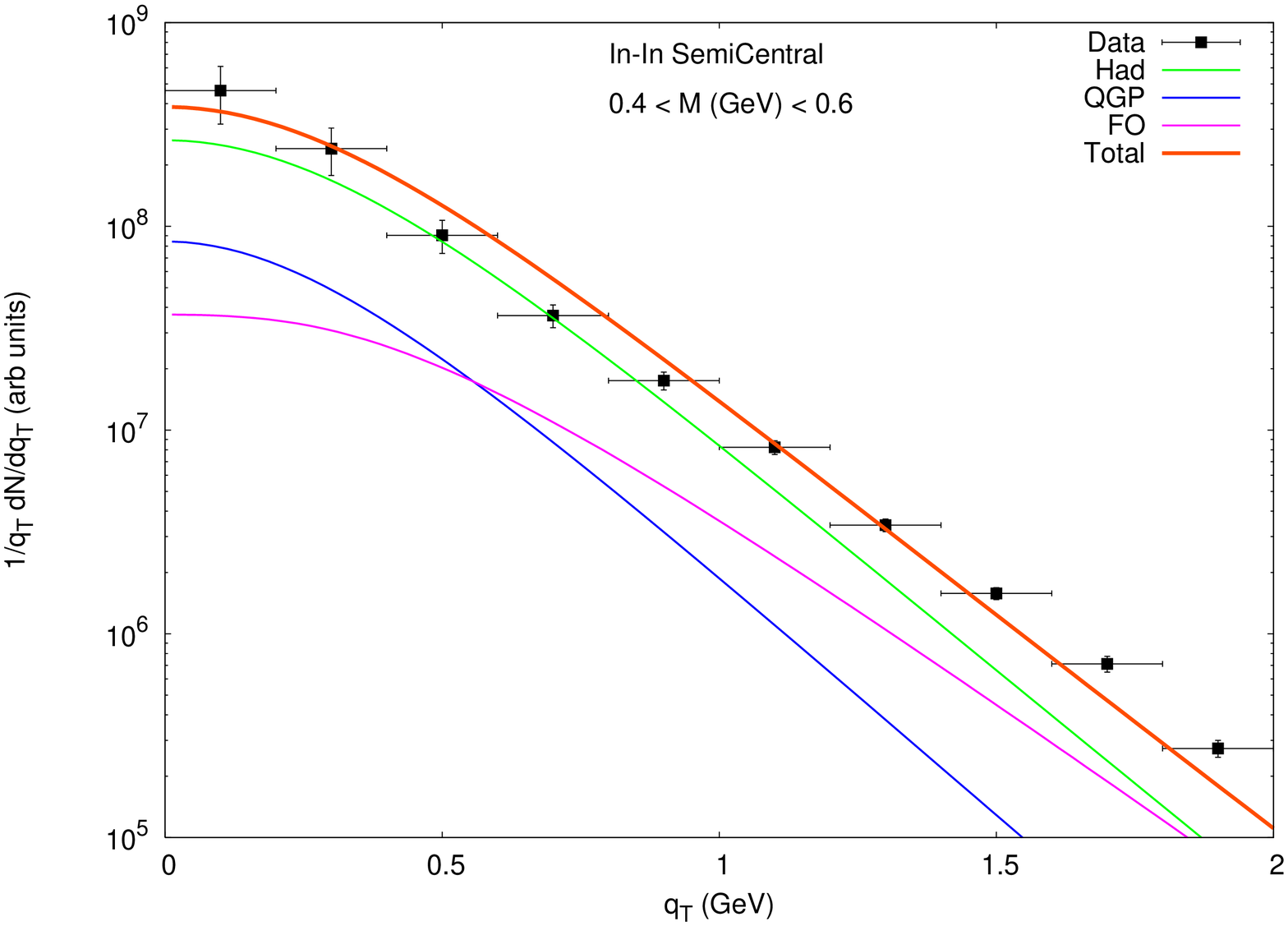,width=\textwidth}
\end{minipage}
\begin{minipage}[]{0.48\linewidth}
\epsfig{file=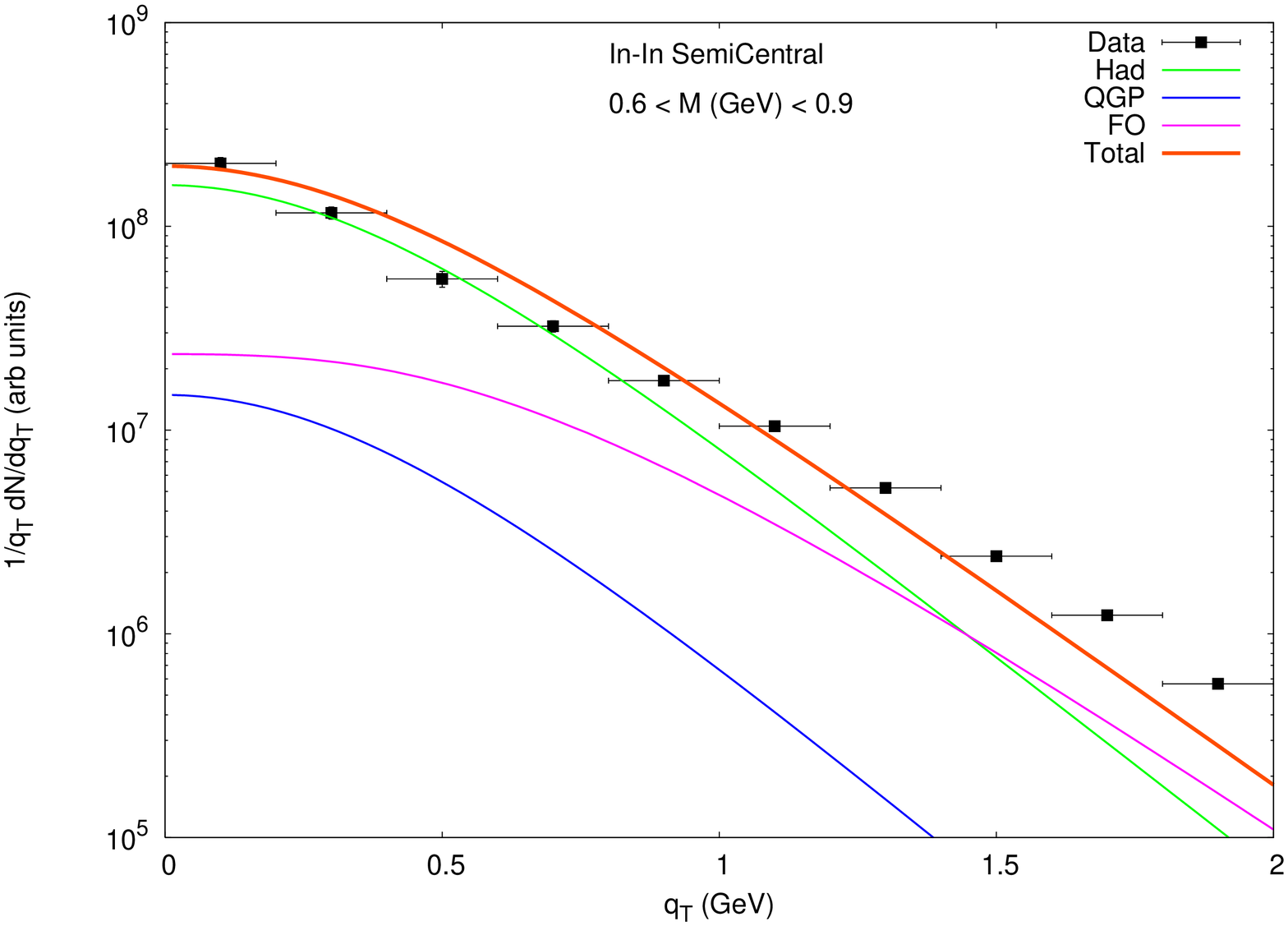,width=\textwidth}
\end{minipage}
\hfill
\begin{minipage}[]{0.48\linewidth}
\epsfig{file=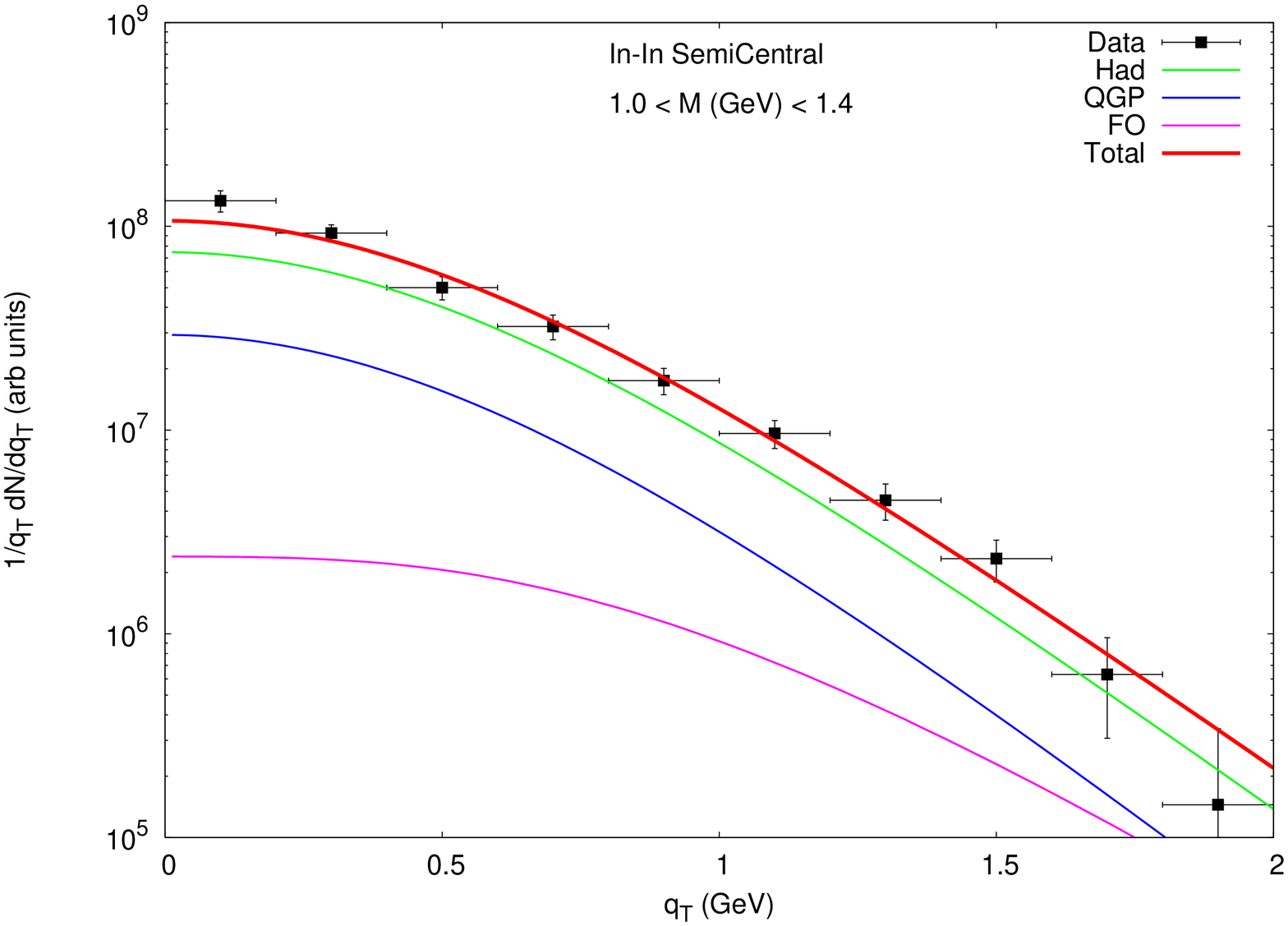,width=\textwidth}
\end{minipage}
\caption{\it NA60 dimuon 
  spectra~\cite{Arnaldi:2006jq,Damjanovic:2007qm,Arnaldi:2007ru} 
  compared to calculations~\cite{Dusling:2006yv,Dusling:2007kh}
  employing thermal rates from the chiral reduction 
  approach~\cite{Steele:1997tv} and perturbative $q\bar{q}$ 
  annihilation, folded over a hydrodynamic expansion for
  semicentral {\rm In(158~AGeV)-In} collisions, supplemented with 
  free $\rho$-meson decays after thermal freezeout. Upper left: 
  $M$-spectra; other panels: $q_t$ spectra in three mass bins.}
\label{fig_na60-DZ}
\end{figure}

\begin{figure}[!t]
\begin{minipage}{0.48\linewidth}
  \epsfig{file=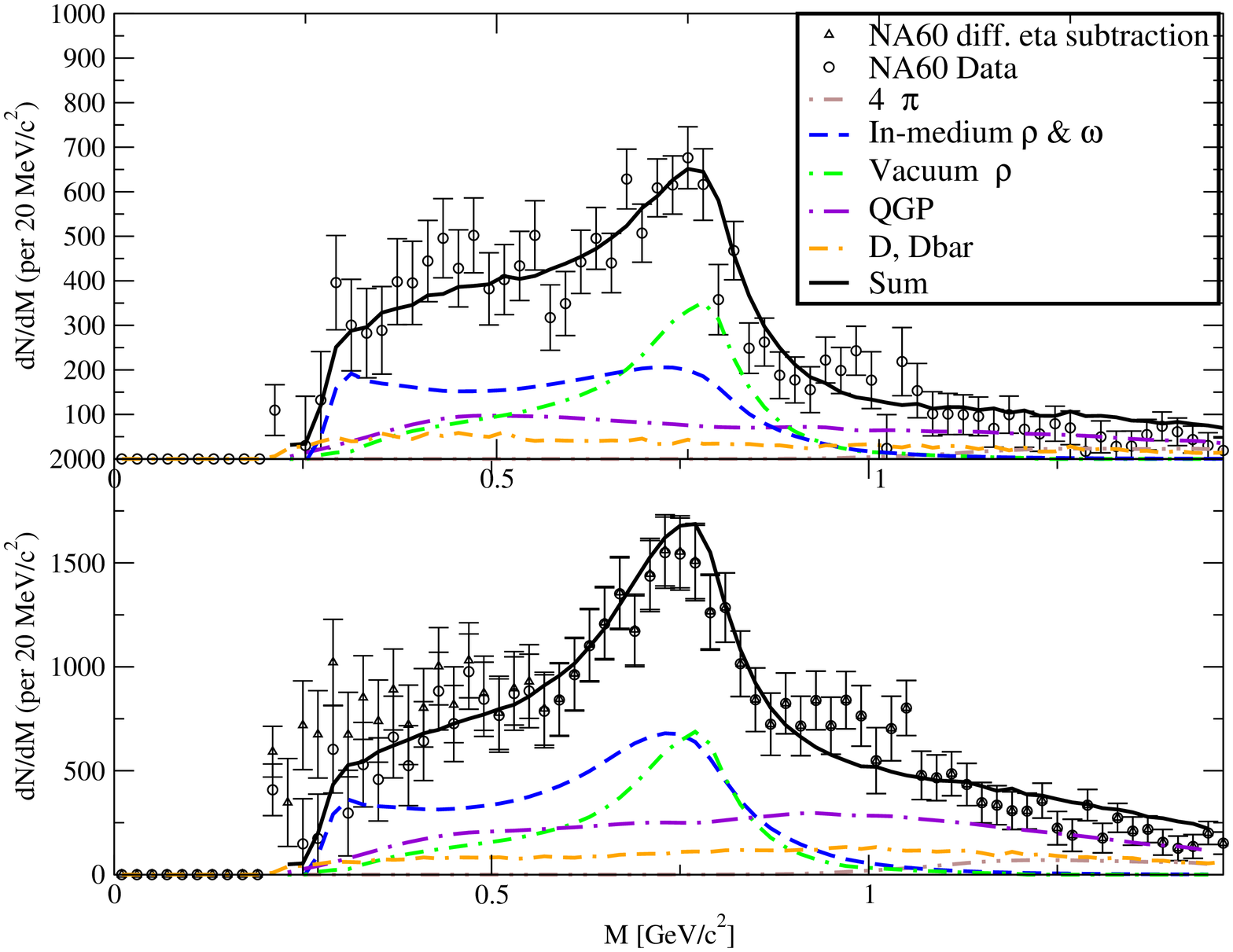,width=\textwidth}
\end{minipage}
\hfill
\begin{minipage}{0.48\linewidth}
\epsfig{file=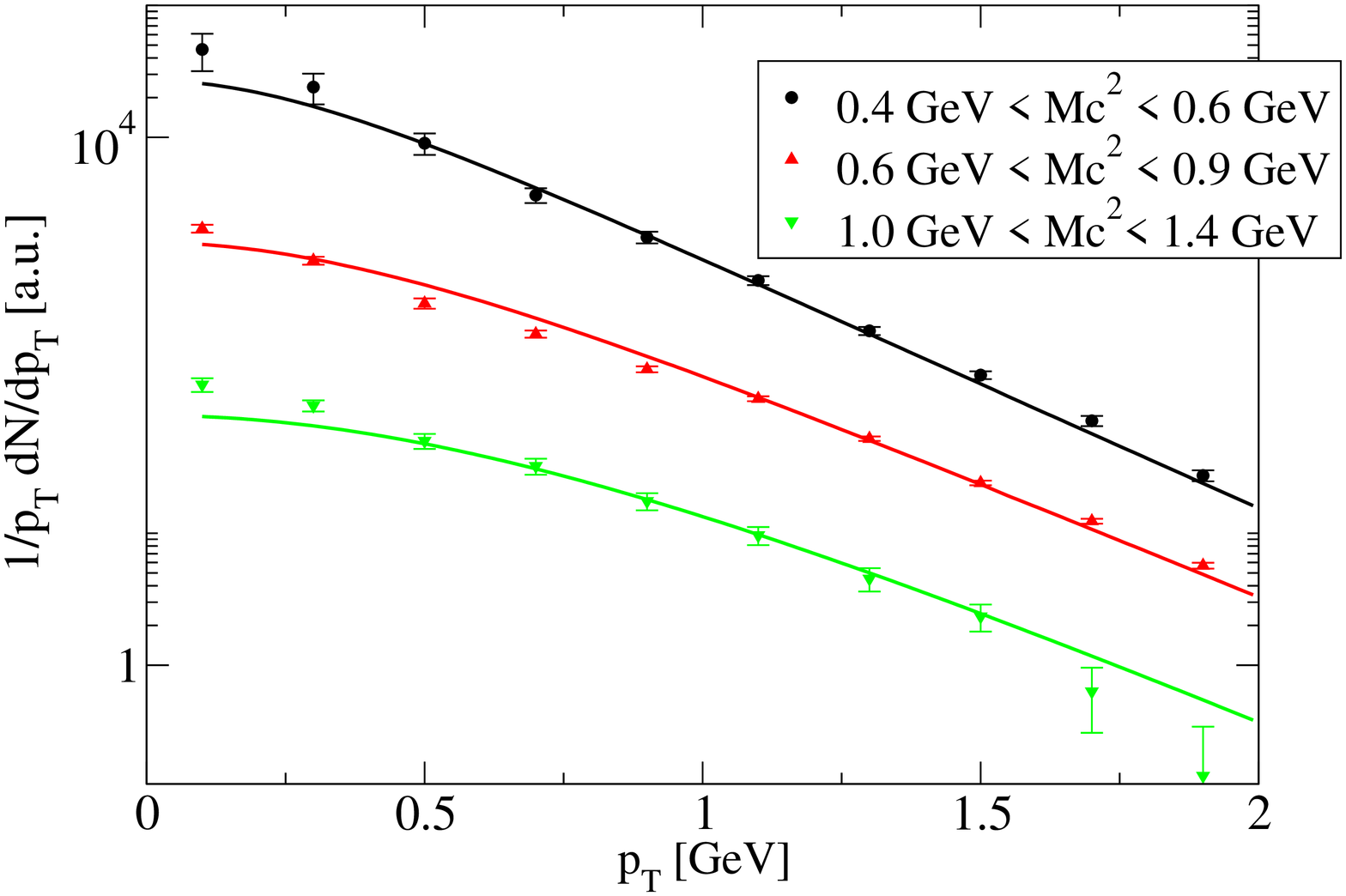,width=\textwidth}
\end{minipage}
\caption{\it NA60 dimuon 
  spectra~\cite{Arnaldi:2006jq,Damjanovic:2007qm,Arnaldi:2007ru} compared
  to calculations~\cite{Ruppert:2007cr} employing thermal rates with 
  in-medium $\rho$ and $\omega$ spectral functions~\cite{Eletsky:2001bb} 
   and free 4$\pi$ annihilation in hadronic matter,
  as well as $q\bar{q}$ annihilation in the QGP~\cite{Renk:2002md}, 
  folded over a thermal fireball expansion for semicentral 
  {\rm In(158\,AGeV)-In} collisions, supplemented with free $\rho$-meson 
  decays after thermal freezeout. Left panel: $M$-spectra for all 
  $q_t$ (bottom) and for $q_t$$>$1\,GeV (top); right panel: 
  $q_t$ spectra in three mass bins.}
\label{fig_na60-RR}
\end{figure}

Theoretical predictions~\cite{Rapp:2004zh} of the low-mass excess 
spectra utilizing the in-medium $\rho$ spectral function of 
Ref.~\cite{Rapp:1999us} (cf.~Sec.~\ref{sssec_hmbt}) showed good 
agreement with the first data of 
NA60~\cite{Arnaldi:2006jq}. More complete calculations including QGP 
radiation (as in Ref.~\cite{Rapp:1999zw} but with hard-thermal loop 
resummed rates~\cite{Braaten:1990wp}), in-medium $\omega$ and 
$\phi$ decays~\cite{Rapp:2000pe}, $4\pi$-like annihilation (relevant 
at intermediate mass)~\cite{vanHees:2006ng}, as well as primordial 
$\rho$ and Drell-Yan (DY) contributions (relevant at high $q_t$), 
are summarized in Fig.~\ref{fig_na60-HR}. 
In connection with a slight update of the fireball model (larger
acceleration implying smaller lifetime), the resulting description of
the NA60 invariant-mass spectra is quite satisfactory over the entire 
range, including projections onto low ($q_t$$<$0.5\,GeV) and high
($q_t$$>$1.0\,GeV). In-medium $\omega$ and $\phi$ contributions are
rather localized in mass, while QGP and DY radiation are at the 
10-15\% level at masses below 1\,GeV. The in-medium plus freezeout
(FO) $\rho$ contributions~\cite{Rapp:2004zh} remain the dominant source
confirming the notion that the NA60 low-mass data probe the in-medium 
$\rho$ spectral function. This is also borne out of the 
acceptance-corrected $q_t$-spectra where, for $M$$<$1\,GeV the 
$\rho$ contribution prevails up to momenta of 
$q_t$$\simeq$~1\,GeV.\footnote{The experimental 
$q_t$-spectra in Figs.~\ref{fig_na60-HR}, \ref{fig_na60-DZ},
\ref{fig_na60-RR} are not absolutely normalized; the theoretical 
$q_t$-spectra in Fig.~\ref{fig_na60-HR} are normalized using the 
$M$-spectra at low $q_t$; however, whereas the experimental $M$-spectra 
for $q_t$$>$1\,GeV are reasonably reproduced, the $q_t$-spectra for 
$M$=0.4-0.6\,GeV are underestimated for $q_t$$>$1\,GeV.} 
At $q_t$$>$1\,GeV DY and primordial $\rho$-mesons become an increasingly 
important source, but the data for $M$=0.4-0.6\,GeV and 0.6-0.9\,GeV 
cannot be fully accounted for. These discrepancies are less pronounced 
for central In-In collisions, and may possibly be resolved by a stronger
transverse expansion within the constraints of the hadronic spectra
(we return to this question below).
At masses $M$=1-1.4\,GeV, the most significant sources are hadronic 
emission from multi-pion states (e.g., $\pi$-$a_1$, $\rho$-$\rho$ or 
$\pi$-$\omega$ annihilation), QGP and $D\bar D$ decays. The hadronic
contribution is significantly enhanced (by maximally a factor of 
$\sim$2 around $M$$\simeq$1\,GeV) due to the effects of chiral 
mixing~\cite{vanHees:2006ng} (recall Sec.~\ref{sssec_mix}), which
currently cannot be discriminated by the data. The $q_t$ spectra for
$M$=1-1.4~GeV are well described over the entire momentum range.

Fig.~\ref{fig_na60-DZ} summarizes the results of hydrodynamic
calculations~\cite{Dusling:2006yv,Dusling:2007kh} based on hadronic
emission rates within the chiral-reduction
approach~\cite{Steele:1996su,Steele:1997tv,Steele:1999hf}
(cf.~Sec.~\ref{sssec_mix}), freezeout $\rho$ mesons (including the
proper $\gamma$ factor relative to thermal radiation) and perturbative
$q\bar{q}$ annihilation in the QGP (pQGP).
The overall structure of the NA60 mass spectrum is roughly reproduced
(cf.~upper left panel of Fig.~\ref{fig_na60-DZ}), but the $\rho$
resonance figuring into the EM correlator lacks significant in-medium
broadening, despite the reduction in peak strength due to the mixing
effect (the agreement improves for semiperipheral and peripheral 
collisions~\cite{Dusling:2006yv}). The freezeout-$\rho$ contribution
compares quite well with the one in the upper left panel of
Fig.~\ref{fig_na60-HR} which includes a broadening but also occurs at 
higher pion density (recall the discussion in Sec.~\ref{sssec_sources}). 
The level of the pQGP contribution is very similar to the fireball model 
of Refs.~\cite{vanHees:2006ng,vanHees:2007th} in Fig.~\ref{fig_na60-HR}. 
As in Refs.~\cite{vanHees:2006ng,vanHees:2007th} the hadronic  
contribution at $M$$>$1\,GeV is based on a fit to the EM correlator in 
vacuum, but the mixing effect is less pronounced in the virial scheme, 
leading to a slightly smaller contribution in the dilepton spectrum 
(possibly also due to the absence of pion chemical potentials). The 
shapes of the $q_t$ spectra (local slopes) of all 3 contributions
displayed in Fig.~\ref{fig_na60-DZ} (in-medium hadronic, freezeout 
$\rho$ and QGP) agree well with the fireball calculations of 
Refs.~\cite{vanHees:2006ng,vanHees:2007th} as demonstrated in a 
direct comparison in Ref.~\cite{Damjanovic:2007qm}. This suggests 
good consistency of the fireball and hydrodynamic evolution.  

In Ref.~\cite{Ruppert:2007cr} a thermal fireball expansion
(cf.~Tab.~\ref{tab_fireball}, middle column) has been applied to compute 
dimuon spectra utilizing in-medium $\rho$ and $\omega$ spectral functions 
(based on empirical scattering amplitudes on pions and nucleons, recall 
solid lines in Fig.~\ref{fig_Arho-had})~\cite{Eletsky:2001bb}, 
vacuum 4$\pi$ annihilation (with both charged~\cite{Lichard:2006kw} 
and neutral pions), as well as QGP rates based on the quasiparticle 
model of Ref.~\cite{Renk:2002md}, cf.~Fig.~\ref{fig_na60-RR}. 
The overall shape and magnitude of the mass spectra
is rather well reproduced, except close to the dimuon threshold
where the importance of baryon effects is apparently underestimated
(the underlying $\rho$ spectral function at $T$=150\,MeV shows little 
variation between baryon densities of $\varrho_B$=0.5\,$\varrho_0$
and $\varrho_0$~\cite{Ruppert:2007cr}). The $q_t$ spectra can 
be reasonably well
described without contributions from DY or primordial $\rho$'s, which 
differs from the hydrodynamic (DZ)~\cite{Dusling:2006yv,Dusling:2007kh}
and HR-fireball~\cite{vanHees:2006ng,vanHees:2007th} results,
cf.~Ref.~\cite{Damjanovic:2007qm} (recall that the RR fireball model 
describes NA60 pion spectra over the entire $p_T$ range by thermal 
emission); part of this discrepancy is due to the slightly
larger expansion velocity and the square-root radial profile of
the transverse rapidity, cf.~Tab.~\ref{tab_fireball}.   
Another significant difference concerns the magnitude of the QGP
contribution, which is by a factor of $\ge$2 larger in 
Ref.~\cite{Ruppert:2007cr} than in
Refs.~\cite{vanHees:2006ng,Dusling:2006yv}. Part of this discrepancy
is due to the quasiparticle QGP EoS employed in 
Ref.~\cite{Ruppert:2007cr}, which entails larger temperatures 
(including $T_0$) at given fireball volume. It is also related to the 
prevailing role of QGP radiation for $M$$\ge$1\,GeV.  


\begin{figure}[!b]
\begin{minipage}[]{0.49\linewidth}
\epsfig{file=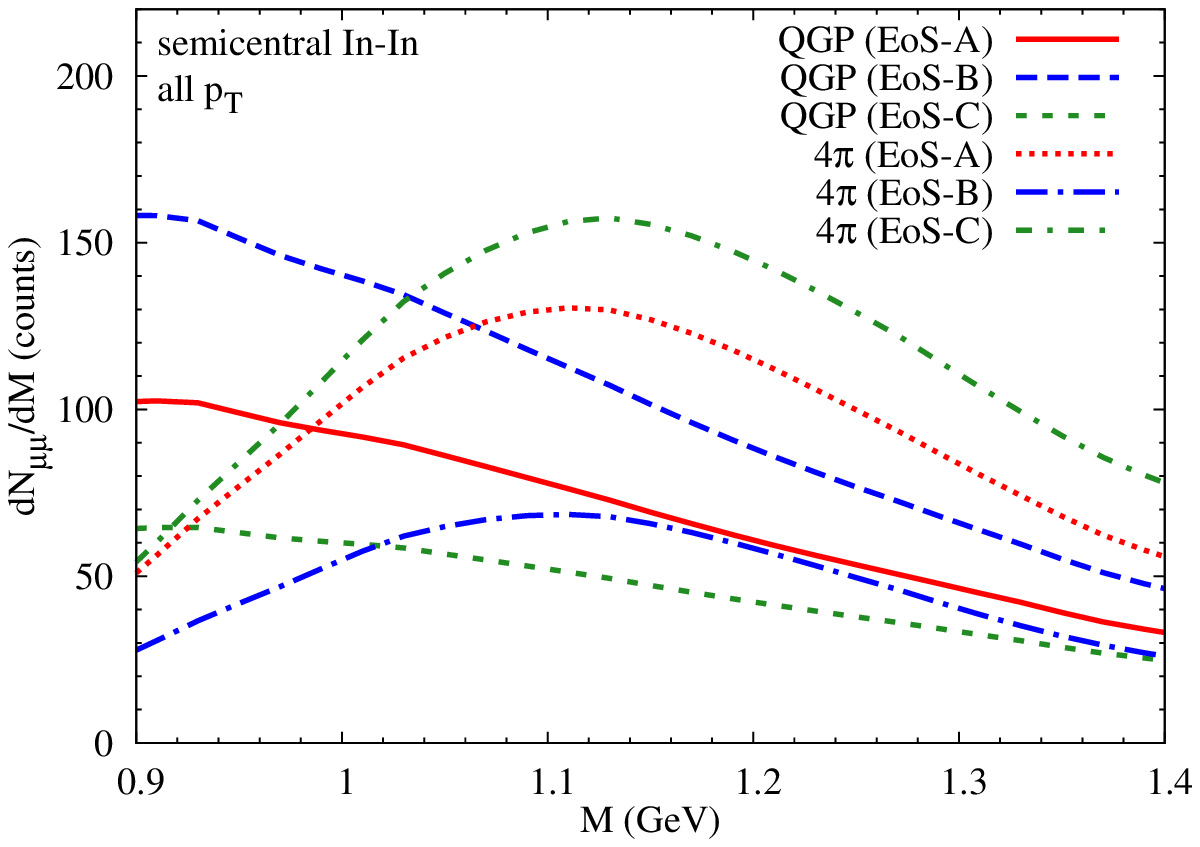,width=\textwidth}
\end{minipage}
\begin{minipage}[]{0.49\linewidth}
\epsfig{file=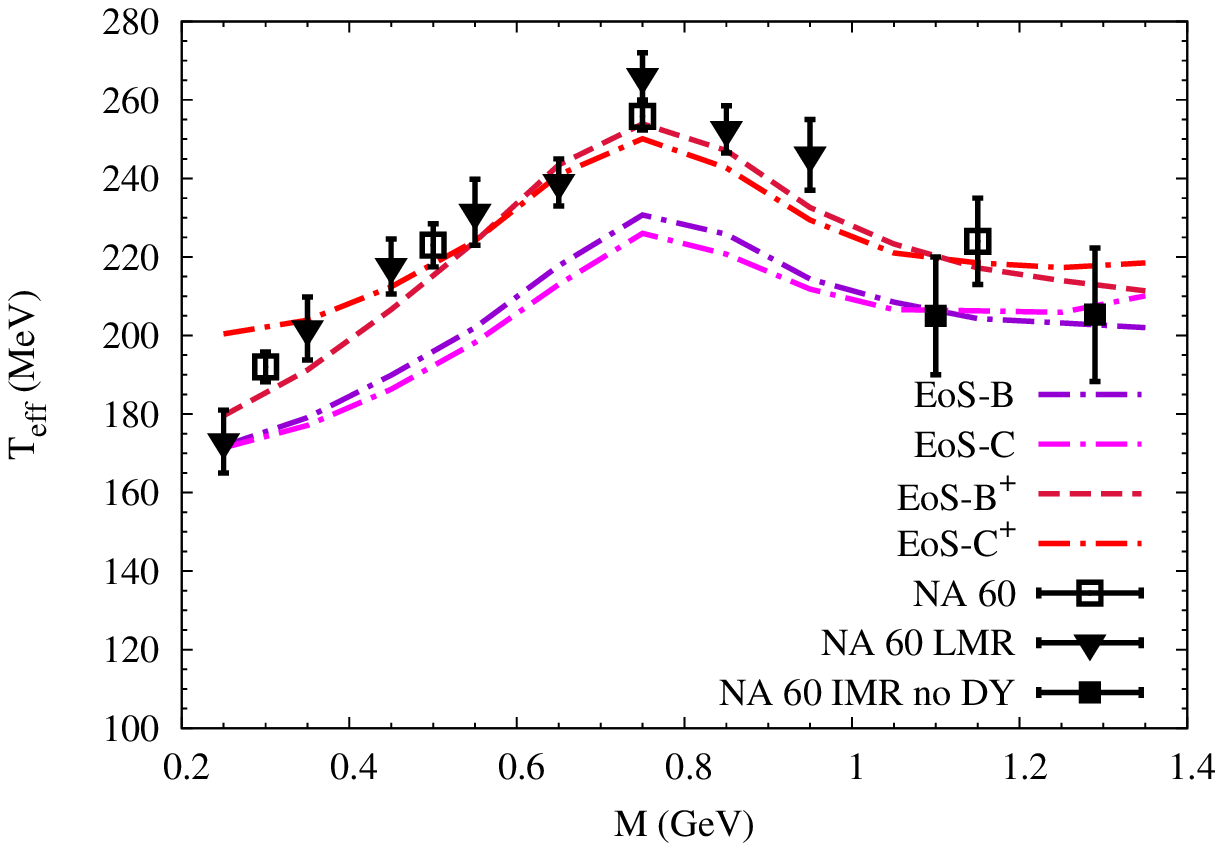,width=\textwidth}
\end{minipage}
\caption{\it Sensitivity of dilepton spectra to critical and 
chemical-freezeout temperature~\cite{vanHees:2007th}. Left panel: QGP 
and hadronic (4$\pi$) radiation at intermediate mass for 
$T_c$=$T_{\mathrm{ch}}$=175\,MeV (EoS-A), 
$T_c$=$T_{\mathrm{ch}}$=160\,MeV (EoS-B) and 
($T_c,T_{\mathrm{ch}}$)=(190,160)\,MeV (EoS-C); right panel: 
slope parameters in $q_t$-spectra for EoS-B and EoS-C (lower curves) 
and for 15\% increased fireball expansion 
(EoS-B$^+$ and EoS-C$^+$, where the latter additionally includes 
$\omega$ $t$-channel exchange in $\pi\rho\to\pi e^+e^-$ reactions;
all without Drell-Yan contribution);
data are from Refs.~\cite{Damjanovic:2007qm,Arnaldi:2007ru}.} 
\label{fig_eos-comp}
\end{figure}
The sensitivity of the NA60 data to the critical and chemical freezeout
temperatures has been elaborated in Ref.~\cite{vanHees:2007th},
by varying $T_c$ from 160-190~MeV and $T_{\rm ch}$ from 160-175~MeV
(keeping the fireball expansion parameters fixed), representing current 
uncertainties in lattice QCD~\cite{Cheng:2006qk,Aoki:2006br} and 
thermal model fits~\cite{BraunMunzinger:2003zd,Becattini:2005xt}. With 
``quark-hadron" duality in the thermal dilepton rates~\cite{Rapp:1999us}
in this temperature regime (at {\em all} masses, 
cf.~Fig.~\ref{fig_dlrates}), the invariant-mass spectra turn out to 
be remarkably {\em in}sensitive to these 
variations~\cite{vanHees:2007th} (duality of the QGP and hadronic 
emission for $M\lsim 1.5$~GeV close to $T_c$ is not realized in the 
rates underlying the calculations of 
Refs.~\cite{Dusling:2006yv,Ruppert:2007cr}). However, the partition of 
QGP and hadronic (4$\pi$) emission at intermediate masses changes 
appreciably from hadron-gas dominated spectra for $T_c$$\ge$175\,MeV 
to QGP dominated ones for $T_c$=160\,MeV, cf.~left panel of 
Fig.~\ref{fig_eos-comp}. In the latter case, the 
smaller value for $T_{\rm ch}$=160\,MeV implies smaller chemical 
potentials in the hadronic phase. This is part of the reason for the 
reduction in hadronic emission, but also leads to a larger freezeout 
temperature by about 15\,MeV (recall the discussion in 
Sec.~\ref{sssec_sources} and right column in Tab.~\ref{tab_fireball}). 
This, in turn, helps in the description of the transverse-momentum
spectra at $q_t$$>$1\,GeV. However, an additional increase in the
transverse fireball acceleration by 15\% seems to be required to achieve 
quantitative agreement with the effective slope parameters as displayed 
in the right panel of Fig.~\ref{fig_eos-comp}. It remains to be checked
whether this can be consistent with a more complete set of hadronic spectra 
in In(158\,AGeV)-In collisions.   

\begin{figure}[!t]
\begin{minipage}[]{0.325\textwidth}
\epsfig{file=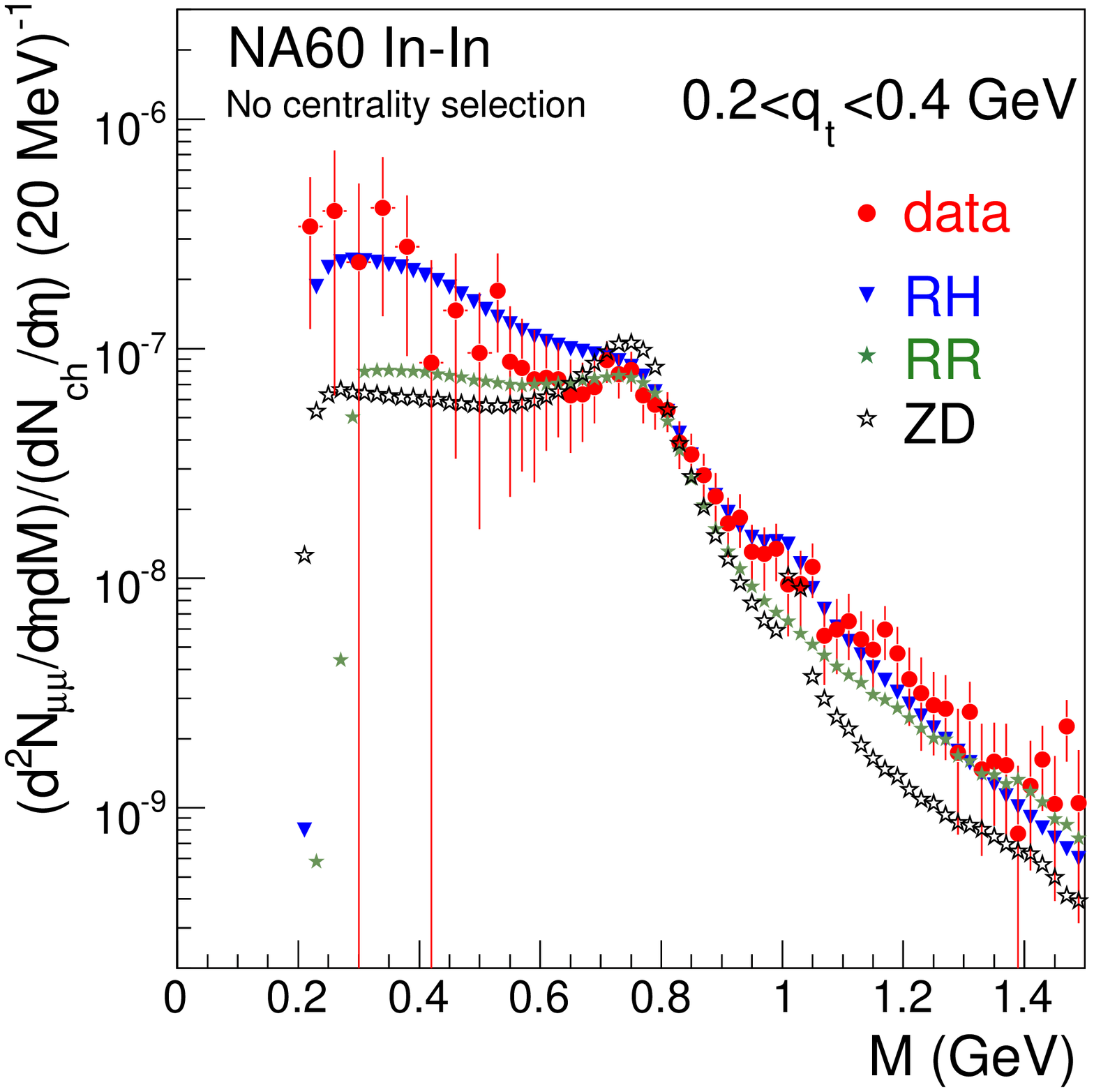,width=0.99\textwidth}
\end{minipage}
\hspace{0.1cm}
\begin{minipage}[]{0.325\textwidth}
\epsfig{file=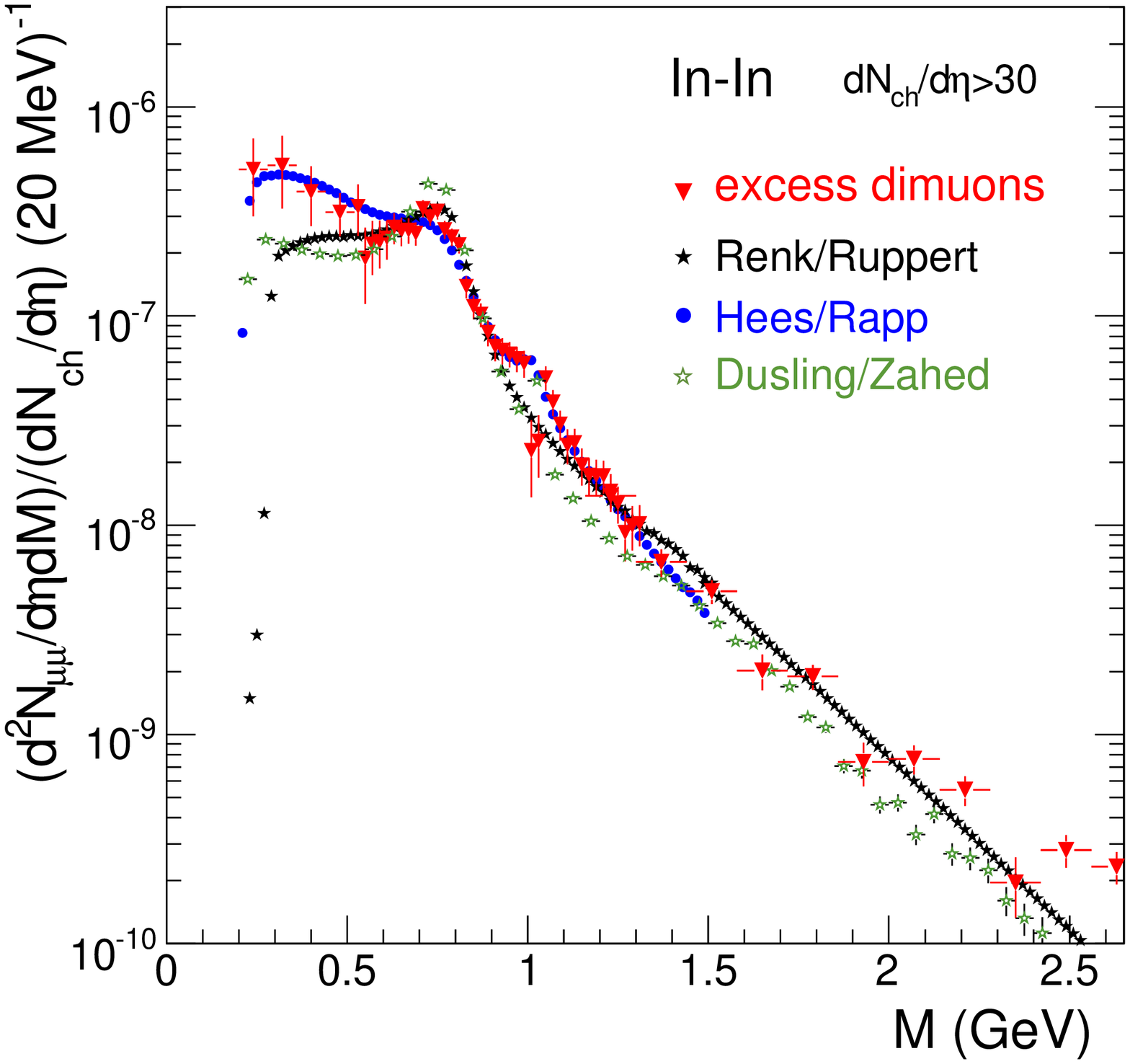,width=0.99\textwidth}
\end{minipage}
\begin{minipage}[]{0.325\textwidth}
\epsfig{file=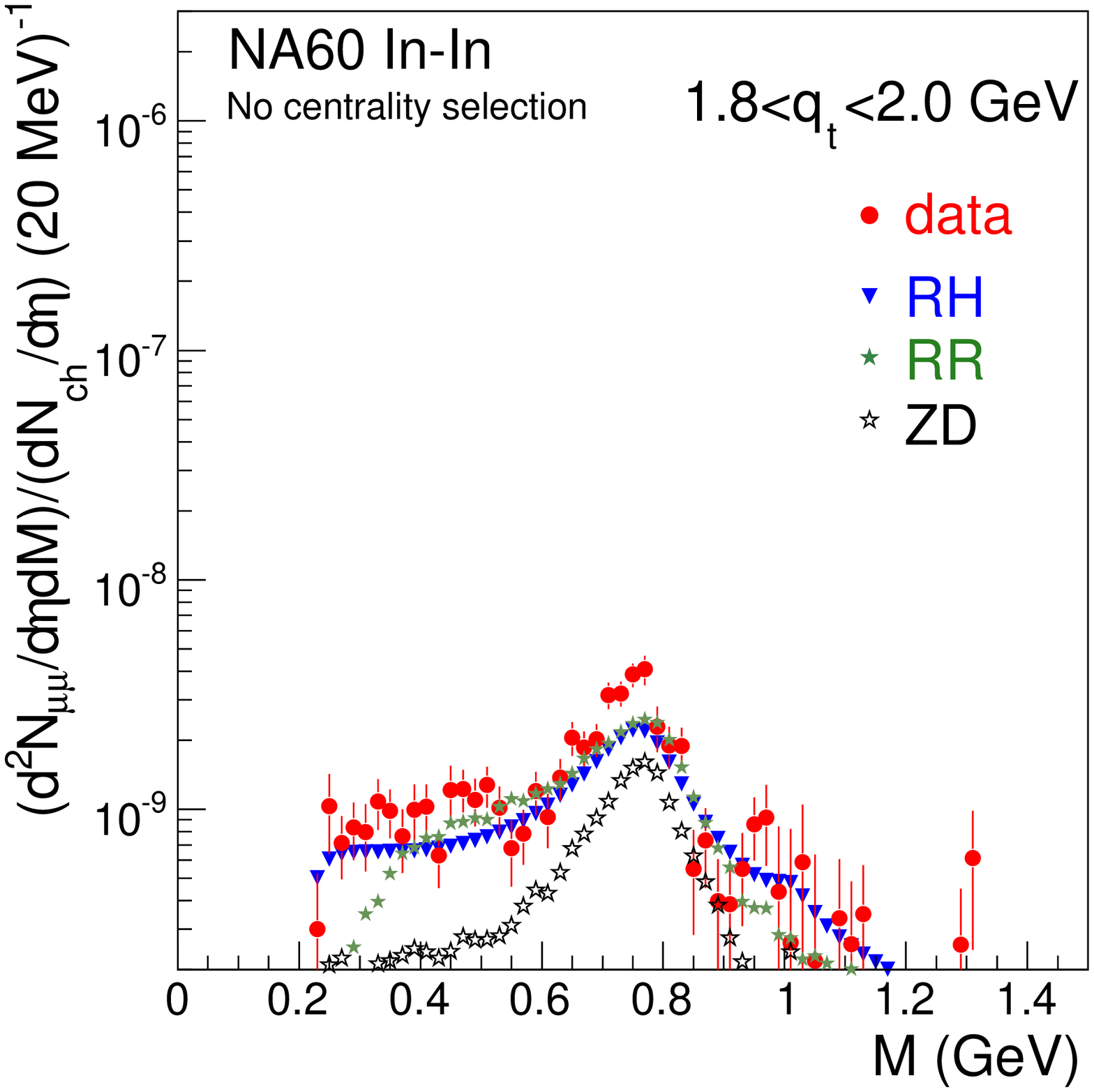,width=0.99\textwidth}
\end{minipage}
\caption{\it Comparison of acceptance-corrected excess dimuon mass spectra 
 (for 0.2$<$$q_t$/GeV$<$0.4 (left), $q_t$$>$0.2~GeV (middle) and 
  1.8$<$$q_t$/GeV$<$2.0 (right)) in minimum-bias {\rm In(158\,AGeV)-In} 
 collisions~\cite{Damjanovic:2008ta,Arnaldi:2008er,Arnaldi:2008fw}
 to model predictions for semicentral In-In of RH (EoS-A)~\cite{vanHees:2007th},
 RR~\cite{Ruppert:2007cr} and ZD~\cite{Dusling:2007kh}, normalized to
 the average $N_{\rm ch}$ of the data.}
\label{fig_M-acc-corr}
\end{figure}
A comparison the three model calculations discussed above to 
acceptance-corrected mass spectra in minimum-bias 
In(158\,AGeV)-In~\cite{Damjanovic:2008ta} in Fig.~\ref{fig_M-acc-corr}
reiterates the importance
of baryon-driven medium effects~\cite{vanHees:2007th} at low $M$ and 
low $q_t$, as well as the lack of high-$q_t$ yield in the
$\rho$-mass region and below for Refs.~\cite{Dusling:2007kh} and
\cite{vanHees:2007th} with EoS-A. The latter improves when increasing
the fireball expansion as in the right panel of Fig.~\ref{fig_eos-comp}.    
Also note that comparing minimum-bias data to calculations at 
an average $N_{\rm ch}$ underestimates the theoretical contributions 
which scale with $N_{\rm coll}$ (DY and primordial $\rho$'s).

\subsubsection{CERN-SPS II: CERES/NA45 and WA98}
\label{sssec_na45-wa98}
The refinements in the analysis of the NA60 dimuon spectra (fireball 
evolution and additional sources) have been rechecked against 
existing and updated EM data at the SPS.

\begin{figure}[!tb]
\begin{minipage}{0.5\textwidth}
\epsfig{file=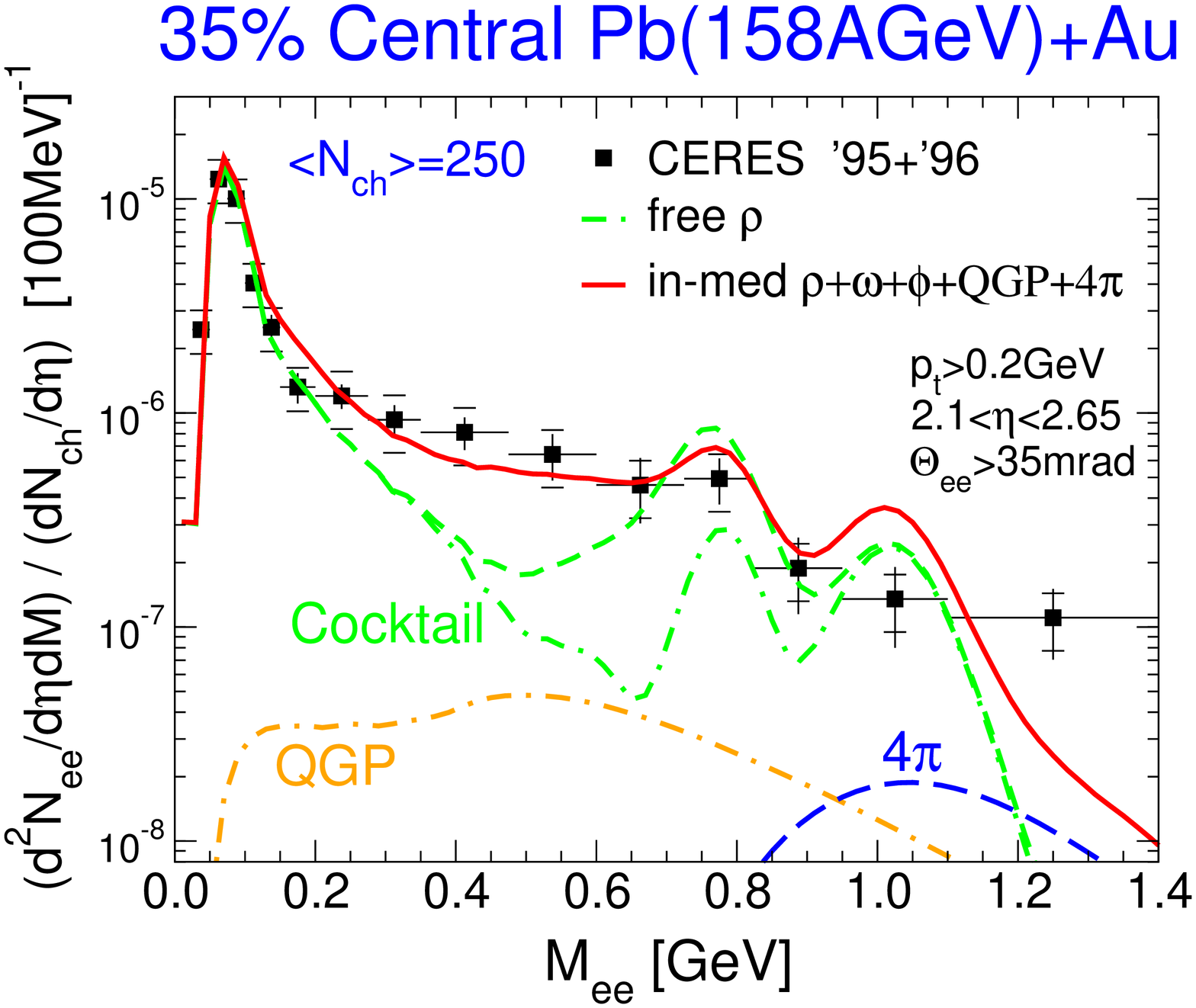,width=0.95\linewidth,
height=0.27\textheight}
\end{minipage}
\begin{minipage}{0.5\textwidth}
\vspace{0.0cm}
\epsfig{file=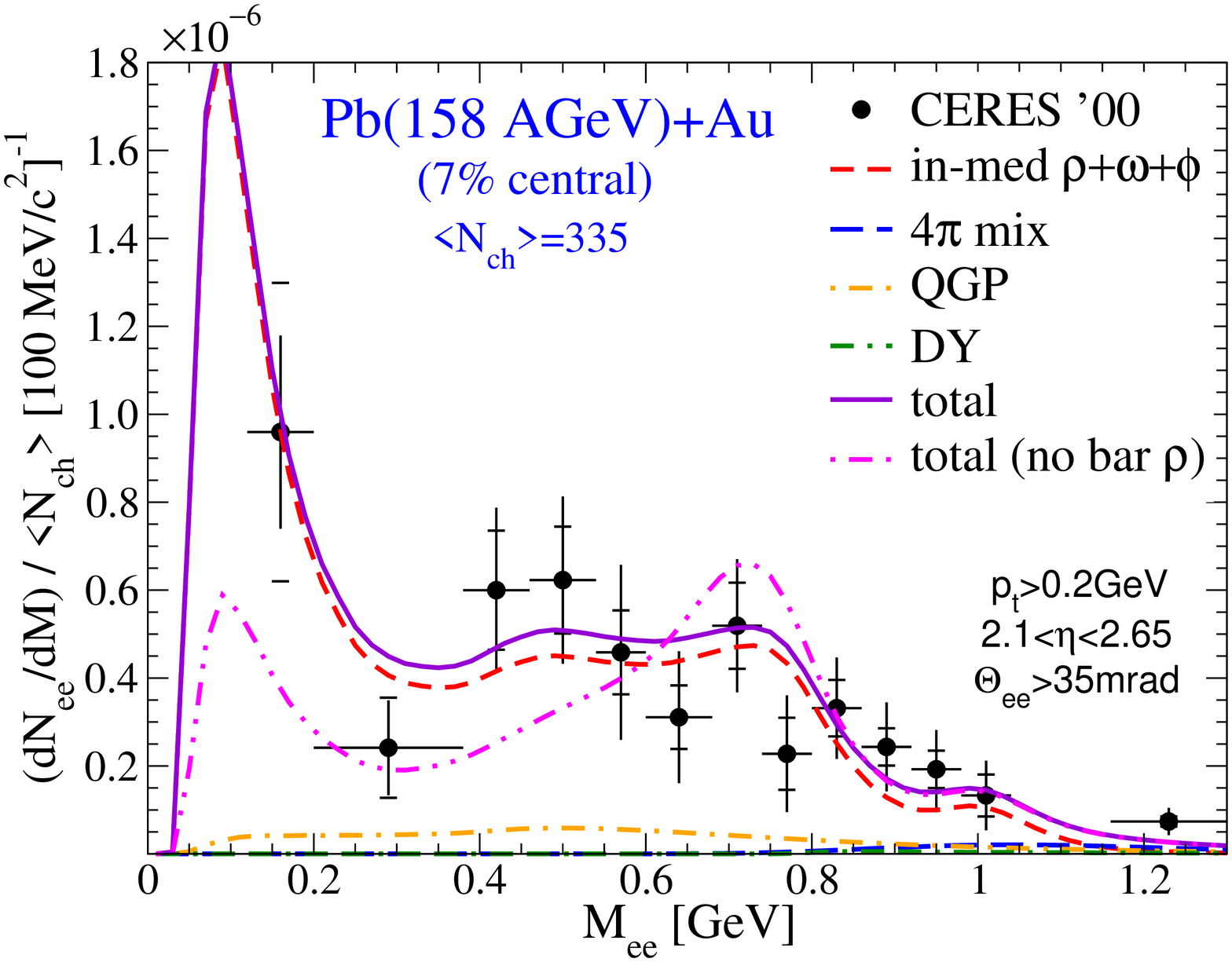,width=0.97\linewidth}
\end{minipage}
\caption{\it CERES/NA45 data for $e^+e^-$ invariant-mass spectra in 
  {\rm Pb(158\,AGeV)-Au} collisions at the 
  SPS~\cite{Agakichiev:2005ai,Adamova:2006nu}. The left
  panel corresponds to semicentral collisions including the
  contribution from long-lived hadron decays after freeze-out
  (``cocktail'', dash-dotted line), while the right panel shows 
  ``excess" spectra for central collisions. The theoretical 
  predictions are based on in-medium $\rho$ spectral 
  functions~\cite{Rapp:1999us} supplemented by $\omega$ and $\phi$ 
  decays, as well as Drell-Yan and 4$\pi$ 
  annihilation~\cite{vanHees:2007th} (as for the NA60 data, 
  see Fig.~\ref{fig_na60-HR}).}
\label{fig_ceres}
\end{figure}
The updated calculations of Ref.~\cite{vanHees:2007th} 
agree well with the combined '95/'96 CERES dielectron data (left panel 
of Fig.~\ref{fig_ceres}). For the 2000 data (right panel of 
Fig.~\ref{fig_ceres}), the cocktail-subtracted excess spectra in central
Pb-Au corroborate the main findings of the NA60 data, i.e., a 
quantitative agreement with the in-medium $\rho$ of 
Ref.~\cite{Rapp:1999us} and the predominance of baryon effects. 
The longer lifetime 
of the fireball in central Pb-Au (factor $\sim$2 relative to In-In) 
reduces the uncertainties due to $\rho$-meson cocktail contributions. 
In addition, dielectrons enable access to very low masses, where the 
'00 CERES data may bear a first hint of a large enhancement as predicted
by hadronic many-body theory.

\begin{figure}[!tb]
\begin{minipage}{0.5\textwidth}
\vspace{0.1cm}
\epsfig{file=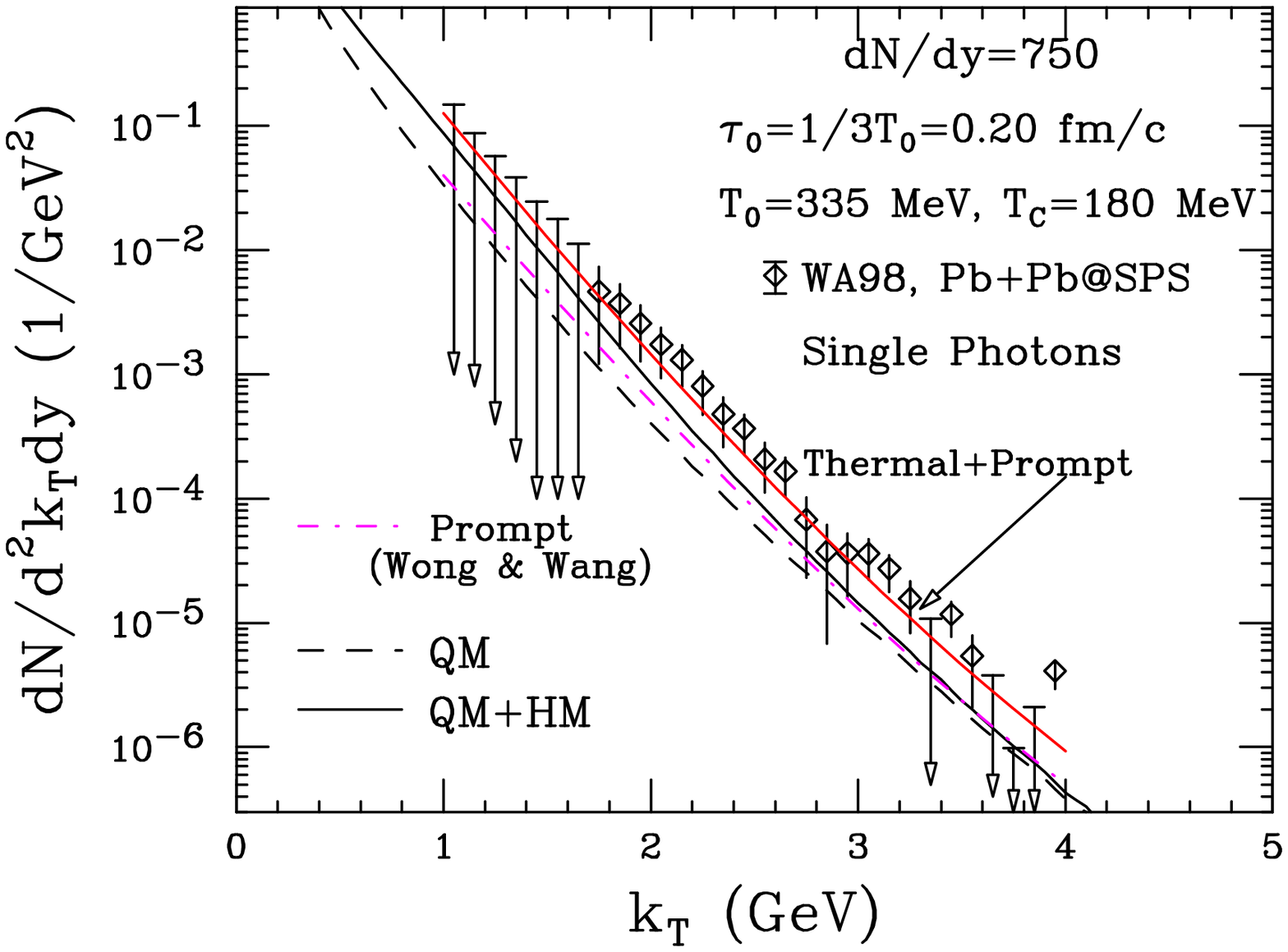,width=0.94\linewidth,height=0.24\textheight}
\end{minipage}
\begin{minipage}{0.5\textwidth}
\hspace{-0.4cm}
\epsfig{file=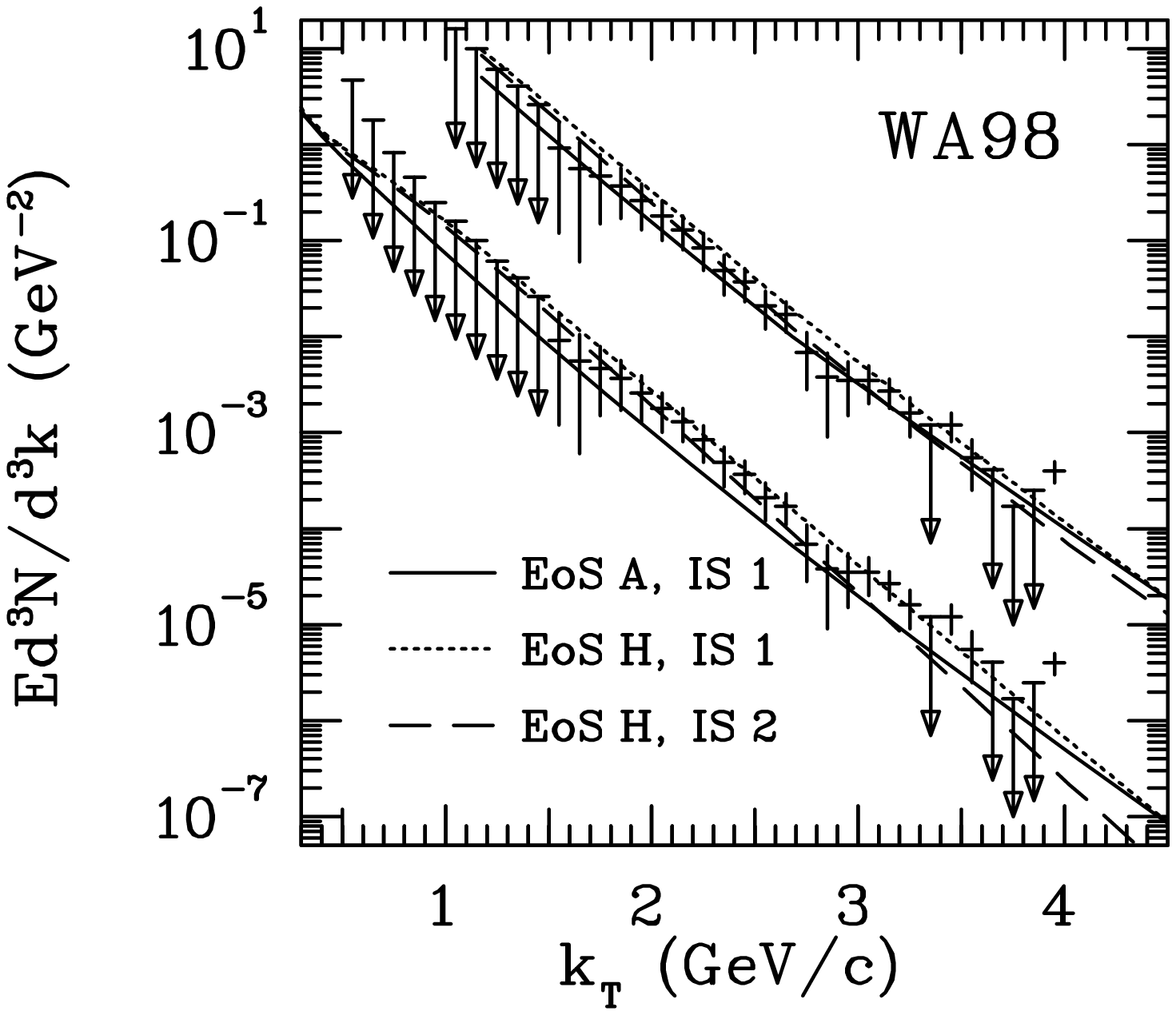,width=0.98\linewidth,height=0.24\textheight}
\end{minipage}
\begin{minipage}{0.5\textwidth}
\vspace{0.2cm}
\hspace{0.2cm}
\epsfig{file=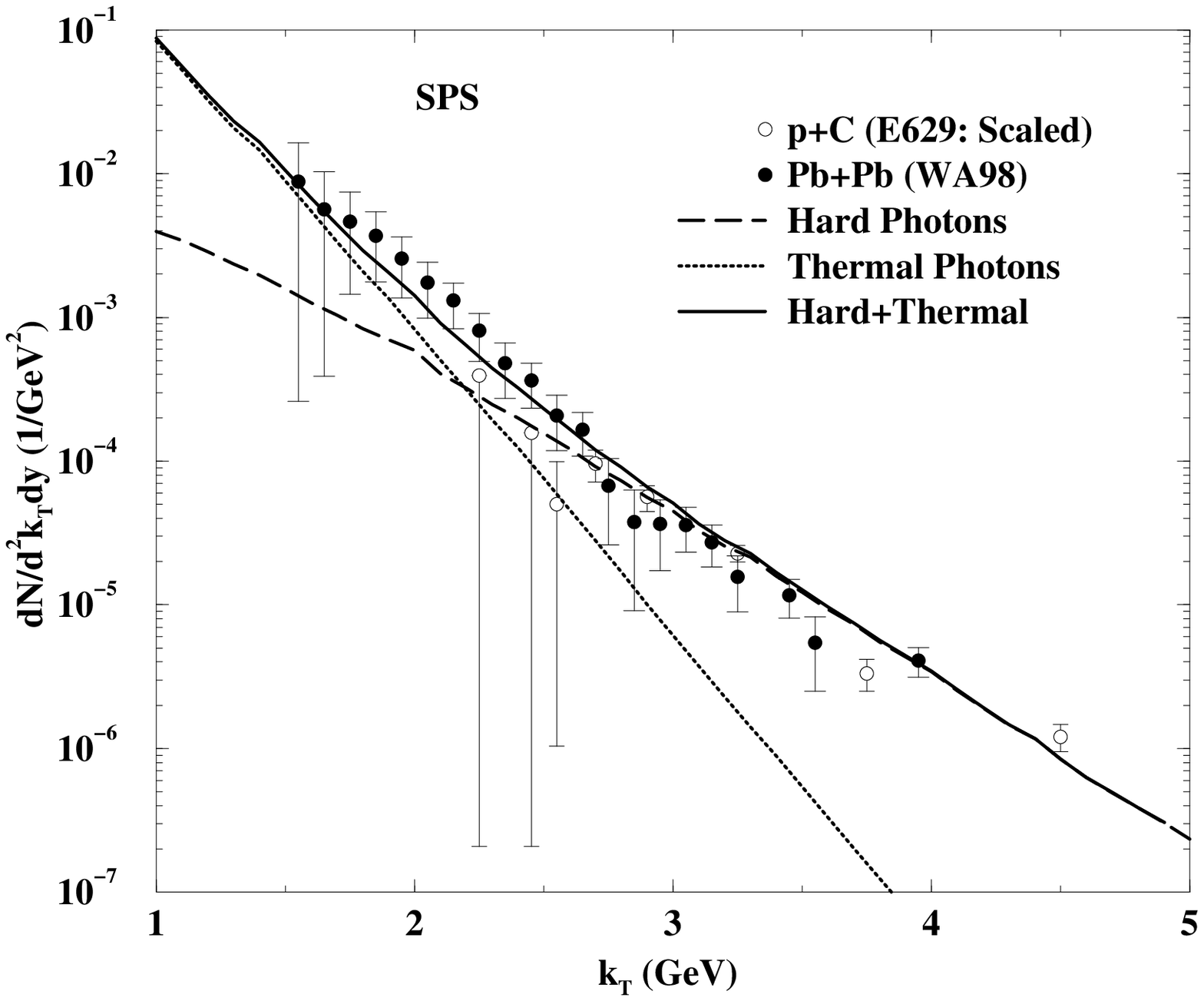,width=0.92\linewidth,height=0.24\textheight}
\end{minipage}
\begin{minipage}{0.5\textwidth}
\vspace{0.1cm}
\hspace{0.2cm}
\epsfig{file=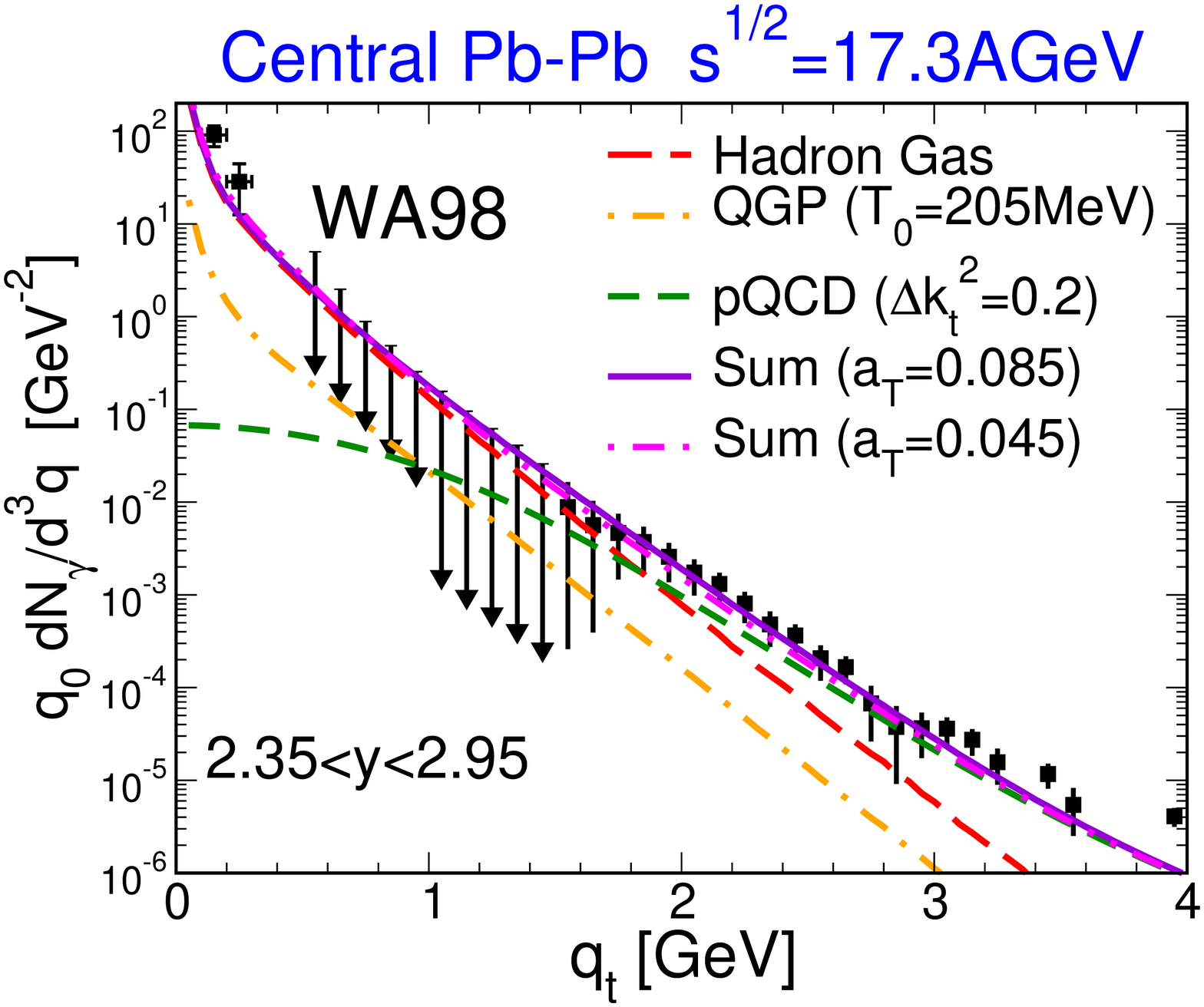,width=0.92\linewidth,height=0.27\textheight}
\end{minipage}
\label{wa98}
\caption{\it Direct photon spectra in central {\rm Pb(158~AGeV)-Pb} collisions
  at the SPS as measured by the WA98 collaboration~\cite{Aggarwal:2000th,
  Aggarwal:2003zy}. The upper panels contain theoretical 
  calculations~\cite{Srivastava:2000pv,Huovinen:2001wx} with a moderate 
  primordial pQCD component and a hot initial state 
  ($\tau_0$=0.2-0.33\,fm/$c$; in the upper left panel 
  the lower curves are without primordial emission), while the calculations 
  in the lower panels~\cite{Alam:2003gx,Turbide:2003si} contain primordial 
  emission with Cronin effect and a larger thermalization time 
  ($\tau_0$$\simeq$1\,fm/$c$).} 
\label{fig_wa98}
\end{figure}
As emphasized in Sec.~\ref{ssec_rates}, (very) low-mass dilepton rates
are intimately related to thermal photon spectra. In 
Ref.~\cite{Turbide:2003si} the in-medium $\rho$ spectral function of 
Ref.~\cite{Rapp:1999us} has been carried to the photon point and 
convoluted over the same fireball expansion as before; when supplemented 
with $t$-channel exchange reactions, QGP emission and primordial (hard) 
photons constrained by $p$-$A$ data, the resulting $q_t$ spectra are 
consistent with WA98 photon spectra, see lower right  panel of 
Fig.~\ref{fig_wa98}; the updated fireball evolution barely affects the 
total spectra. The contributions from the lightlike $\rho$ are 
prevalent up to $q_t$$\simeq$1\,GeV (cf.~Fig.~\ref{fig_phrate}), after 
which $t$-channel processes takes over.
Primordial photons outshine the combined thermal yield (hadronic+QGP)
for $q_t$$\gsim$2\,GeV. This is nicely consistent with the calculations of
Ref.~\cite{Alam:2003gx}, see lower left panel in Fig.~\ref{fig_wa98}. 
In earlier calculations of Refs.~\cite{Srivastava:2000pv,Huovinen:2001wx} 
the thermal yield is significantly larger, due to an increased QGP 
contribution caused by a short formation time of 
$\tau_0$=0.2-0.33\,fm/$c$ with associated peak temperatures of up to 
$T_0$=335\,MeV (for $\tau_0$=1\,fm/$c$~\cite{Turbide:2003si,Alam:2003gx}  
average initial temperatures are slightly above $\bar T_0$=200\,MeV).
Even for this upper estimate of QGP emission\footnote{At 
SPS energy, with a Lorentz contraction of $\gamma$$\simeq$9 for the 
incoming nuclei, the time for full nuclear overlap is ca.~0.8\,fm/$c$.}, 
the latter is smaller than the hadronic one for momenta 
$q_t$$\le$1.5-2~GeV, and the pQCD photons are at the $\sim$40\% level
of the combined thermal contribution at $q_t$$\simeq$2\,GeV.

Similar conclusions arise from theoretical 
analyses~\cite{Rapp:1999zw,Gallmeister:1999dj,Kvasnikova:2001zm} of 
intermediate-mass dimuon spectra (1.5$\ge$$M_{\mu\mu}$/GeV$<$3)
in Pb-Pb collisions at SPS~\cite{Abreu:2000nj}: unless the initial
temperature significantly exceeds $T_0$=250\,MeV, the thermal 
contribution falls below primoridial sources (DY) at masses and 
transverse momenta beyond $M,q_t$$\simeq$1.5-2\,GeV. This is
fully confirmed by the recent NA60 intermediate-mass dilepton
spectra~\cite{Arnaldi:2008er}. 

\subsubsection{Future Dilepton Measurements}
\label{sssec_future}
Dilepton programs will be pursued with high priority over a
wide range of collision energies. The large enhancement observed in a
low-energy (40\,AGeV) run at SPS~\cite{Adamova:2003kf} is in line
with the prediction of hadronic-many body theory that medium effects
caused by baryons play a leading role~\cite{Rapp:2002tw}. 
This trend continues down to 
much lower bombarding energies of 1-2\,AGeV. However, at these
energies recent transport calculations suggest that the low-mass 
enhancement, which could not be explained by hadronic in-medium 
effects~\cite{Bratkovskaya:1999xx}, is related to primordial $N$-$N$ 
Bremsstrahlung~\cite{deJong:1996ej}, as well as 
$\Delta\to Ne^+e^-$ and $\eta$ Dalitz 
decays~\cite{Bratkovskaya:2007jk,Schumacher:2006wc}. A better sensitivity
to medium effects appears to be in the $\rho$-$\omega$ mass region, where
the (lack of) yield indicates a strong broadening of the vector
resonances~\cite{Shekhter:2003xx,Bratkovskaya:2007jk,Santini:2008pk}.    

At the high-energy frontier, first RHIC data~\cite{Afanasiev:2007xw} find
a large $e^+e^-$ signal especially in the mass region around 
$M$$\simeq$0.3~GeV. The excess is concentrated at low $q_t$ and in
central collisions, and cannot be explained by current in-medium 
spectral functions. It is tempting to speculate that the excess is
caused by the formation of a disoriented chiral condensate 
(DCC), as pion-DCC annihilation shares the above 
features~\cite{Kluger:1997cm}. However, the magnitude of this dilepton
source cannot easily compete with hadronic medium effects, unless
the DCC domains are rather large and abundant. In this case, footprints
of the DCC should be visible in other observables (e.g., $\pi^0$ and
$p_T$ fluctuations). Precision measurements within the RHIC-II 
program will be of crucial importance here~\cite{David:2006sr}.     

Finally, dilepton data will play a critical role in the CBM experiment
at the future GSI facility (FAIR). In the planned energy regime,
$E_{\rm lab}$=10-40\,AGeV, one envisages the largest nuclear compression
and thus maximal baryon density, ideally suited to scrutinize the current
understanding of medium effects. An extra benefit could be the occurence 
of a critical point or a true mixed phase at a first-order transition,
with extended fireball lifetimes further enhancing the dilepton signal. 

\subsection{Critical Appraisal}
\label{ssec_appraisal}
In this section we evaluate the current
status of determining the in-medium vector spectral functions
(focusing on the $\rho$ meson) and the implications for chiral 
restoration.

\begin{figure}[!tb]
\center
\begin{minipage}{0.5\textwidth}
\epsfig{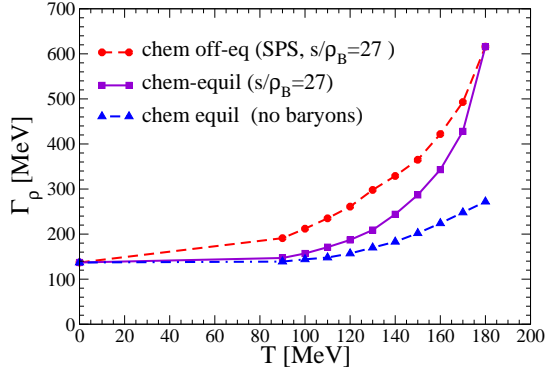}
\end{minipage}
\caption{\it Temperature dependence of the $\rho$ width as evaluated
in hadronic many-body theory~\cite{Rapp:1999us} 
at fixed entropy per baryon representative for heavy-ion collisions
at full SPS energy (dashed and solid line: with and without meson chemical
potentials). For the dash-dotted line, medium effects due to baryons
are switched off.}
\label{fig_gam-emsf}
\end{figure}

Calculations of $\rho$-meson spectral functions based on effective 
chiral Lagrangians coupled with many-body techniques agree on a strong 
broadening with small (positive) mass shifts. At normal nuclear matter 
density, one finds an increase in width of 
$\Delta\Gamma_\rho$$\simeq$250\,MeV with an 
estimated error of $\sim$30\%, i.e., the vacuum width almost triples. 
The question whether the parameters in the effective Lagrangian 
are subject to in-medium changes requires further input. In the 
vector-manifestation scenario, reduced bare masses
and coupling constants are inferred from a matching of the correlators
to an operator product expansion (OPE) at spacelike momenta governed 
by the in-medium reduction of the condensates.
However, it turns out that, within current uncertainties, the 
softening of the $\rho$ spectral function as imposed by the OPE at 
nuclear matter density is fully accounted for by the 
broadening due to hadronic many-body effects. More accurate tests
of this assertion, especially at higher densities/temperatures,
will require a more precise determination of the in-medium condensates 
on the OPE side of the QCD sum rule. 
The predicted broadening is supported by several recent experiments 
where dilepton spectra have been measured with impressive precision: 
at JLAB, photoproduction data off mid-size nuclei find a $\rho$ 
broadening of $\Delta\Gamma_\rho^{\rm NUC}$$\simeq$70-100\,MeV 
without significant mass shift, consistent with many-body effects at 
about half nuclear density and 3-momenta of $\sim$1-2\,GeV. 
The NA60 dilepton 
spectra in central In-In collisions exhibit an {\em average} $\rho$ 
width of $\bar\Gamma_\rho^{\rm HIC}$$\simeq$400\,MeV
i.e., an additional broadening of  
$\Delta\bar\Gamma_\rho^{\rm HIC}$$\simeq$250\,MeV. 
Typical kinetic freeze-out conditions at SPS energies are 
$(\rho_B^{\rm fo},T_{\rm fo})$$\simeq$(0.3$\rho_0$,120MeV). With 
initial temperatures of $T_0$$\simeq$200\,MeV (as suggested by 
``effective" slope parameters in the $q_T$ spectra for $M$$>$1\,GeV, 
as well as direct photon spectra in Pb-Pb), the average $\rho$ width
thus reflects the medium at an average temperature of 
$\bar T$$\simeq$150\,MeV (the growing 
fireball 3-volume ``biases" low-mass dilepton radiation to more dilute
stages). This implies that the $\rho$ width approaches its mass when 
the system moves toward the (pseudo) critical temperature, 
$\Gamma_\rho$($T$$\to$$T_c$)\,$\to m_\rho$, i.e., the 
resonance ``melts" (see also Ref.~\cite{Dominguez:1990hn}). 
Inspection of the theoretical predictions 
for the width of the $\rho$ as extracted from the vector spectral 
function corroborates this conclusion, cf.~Fig.~\ref{fig_gam-emsf}. The 
circumstantial ``duality" of hadronic and partonic EM emission rates 
close to $T_c$ lends robustness to the pertinent predictions for 
dilepton spectra in heavy-ion collisions as they become independent 
on details of the evolution model, in particular of the treatment of 
the phase transition region. The excess radiation at intermediate mass,
with its rather soft emission characteristics in $q_t$, as well as 
direct photon spectra, further consolidate the origin of thermal 
radiation from around $T_c$.    

A no less challenging task is to connect the above findings to the 
chiral transition. It has recently been argued that the statistical 
operator consisting of a (free) hadron resonance gas (HRG) is capable 
of reproducing several features of lattice QCD computations of the 
equation of state (EoS) until rather close to the (pseudo-) critical 
temperature~\cite{Karsch:2003zq,Ejiri:2005wq}. Beyond $T_c$ the lattice 
EoS levels off, reflecting quark-gluon degrees of freedom, while the HRG 
EoS diverges (Hagedorn catastrophe). The $\rho$ melting offers a 
microscopic explanation for this transition: under moderate conditions, 
the {\em interacting} HRG physics drives the $\rho$ broadening
to an extent which justifies the use of well-defined quasi-particle 
states in the statistical operator. With further increasing temperature
and density, resonance overlap in the $\rho$ spectral function drives 
it to a continuum shape {\em with a strength resembling a weakly 
interacting $q\bar q$ pair}, i.e., the resonance strength in the 
statistical operator converts into partonic 
strength. The phenomenon of overlapping resonances merging into a 
perturbative $q\bar q$ continuum is, of course, well known from 
the $e^+e^-$ annihilation cross section into hadrons above 
$M$$\simeq$1.5~GeV. It is suggestive that the thermal medium provides
the necessary phase space for low-mass resonances which, via their 
mutual ``mixing" in different hadronic correlators, ``restore" 
quark-hadron duality down to $M$$\to$0, implying chiral restoration.
To quantify this picture the evaluation of chiral order parameters
is mandatory. It is tempting to speculate that the rather sharp increase
of the $\rho$ width close to the expected critical temperature 
(especially in chemical equilibrium as realized in lattice QCD,
represented by the solid line in Fig.~\ref{fig_gam-emsf}) is 
signaling the chiral transition. QCD sum rules remain a valuable tool 
if the $T$ (and $\mu_B$) dependence of the quark and gluon condensates 
can be made more precise. 
Ideally, the latter are determined from first-principle lattice QCD 
calculations.  Possibly the most promising approach, which has been 
little exploited thus far, are chiral (or Weinberg) sum rules. Their 
use hinges on the in-medium axialvector spectral function.  
The latter is much more difficult to constrain due to a principal
lack of experimental information, encoded in either 3-pion or 
$\pi$-$\gamma$ final states. This stipulates the importance of
calculating the axialvector correlator in chiral models. In connection
with a realistic vector correlator and lattice-QCD input on the
in-medium condensates, the explicit realization of chiral restoration
can be investigated. First efforts in this direction have been 
undertaken~\cite{Rapp:2002pn,Harada:2005br,Struber:2007bm,Harada:2008hj}, 
but a full treatment 
including quantitative $V$ and $A$ spectral functions, even in the 
vacuum, is currently lacking.







\section{Conclusions}
\label{sec_concl}

Medium modifications of hadronic spectral functions play a key role 
in the diagnosis of hot/dense strongly interacting matter and 
its condensate structure. Experimentally, the most promising approach 
is dilepton spectroscopy which directly probes the vector spectral 
function of the hadronic medium. For the $\rho$-meson, which dominates 
the low-mass vector channel, effective hadronic theories largely agree 
on a strong broadening of the resonance, with little mass shift. 
Baryon effects prevail over those induced by mesons, and the
predicted modifications in cold nuclear matter are compatible with 
QCD sum rules at finite density. 
Intense experimental efforts over the last $\sim$15 years have 
culminated to a new level of precision which broadly confirms the 
theoretical expectations:
production experiments off ground-state nuclei find an increase of the
$\rho$ width by $\sim$80\,MeV, while the effect in heavy-ion collisions 
at the SPS is by a factor of $\sim$3 larger. Part of this difference is 
due to the access to the low-momentum regime in the heavy-ion 
measurements. It is therefore highly desirable to push the sensitivity 
of the nuclear experiments to low 3-momenta where significantly 
larger medium effects are predicted.
The average $\rho$ width extracted in heavy-ion collisions suggests that
the $\rho$ resonance ``melts" close to the expected phase boundary, in
agreement with extrapolations of hadronic models. This is a first
explicit evidence that melting resonances are involved in the transition 
from hadronic to quark degrees of freedom. Modern quark-model 
calculations could provide complementary insights when approaching 
$T_c$ from above. 
Unquenched lattice QCD computations of the vector correlator would 
undoubtedly set valuable benchmarks and possibly 
shed light on the conjecture that the width is connected to order 
parameters of chiral symmetry restoration.     
In addition, information on quark condensates and pion decay constant(s) 
below $T_c$ can be connected to hadronic vector and axialvector spectral 
functions utilizing Weinberg sum rules.  
The synergy of hadronic and quark models with first-principle lattice
QCD computations, augmented by quantitative applications to experiment 
at current and future facilities, opens exciting perspectives to improve 
our knowledge about the chiral transition in hot/dense QCD matter and
the generation of luminous mass in the Universe.

\section*{Acknowledgments}
We thank H.J.~Specht and S.~Damjanovic for valuable discussions and
suggestions. 
RR has been supported by a U.S.~National Science Foundation CAREER 
Award under grant PHY-0449489, and by the A.\,v.\,Humboldt-Foundation
through a Bessel Research Award.


\end{document}